\documentclass[namedreferences]{solarphysics}

\usepackage[hyperref,optionalrh]{spr-sola-addons} % For Solar Physics 
\usepackage{graphicx}    % For eps figures, newer & more powerfull
\usepackage{color}      % For color text: \color command
\usepackage{breakurl}    % For breaking URLs easily trough lines
\usepackage{threeparttable}

\renewcommand{\vec}[1]{{\mathbfit #1}}

\newcommand{\rmd}{ {\ \mathrm d} }

\begin{document}

\begin{article}

\begin{opening}

\title{Helioseismic Investigation of Modeled and Observed Supergranule Structure}

\author{K.~\surname{DeGrave}$^{1}$\sep
    J.~\surname{Jackiewicz}$^{1}$
    %M.~\surname{Author-c}$^{2}$
    }
\runningauthor{K. DeGrave and J. Jackiewicz}
\runningtitle{Helioseismic Investigation of Modeled and Observed Supergranule Structure}

  \institute{$^{1}$ New Mexico State University, Department of Astronomy 1320 Frenger Mall, Las Cruces, NM 88003, USA\\
           email: \url{degravek@nmsu.edu} email: \url{jasonj@nmsu.edu}\\ 
       %$^{2}$ Second affiliation
           %email: \url{e.mail-c} \\
       }

\begin{abstract}
The subsurface structure of an ``average" supergranule is derived from existing HMI pipeline time-distance data products and compared to the best helioseismic flow model detailed in \citet{duvall2012}. We find that significant differences exist between them. Unlike the shallow structure predicted by the model, the average HMI supergranule is very extended in depth, exhibiting horizontal outflow down to $7$--$10$~Mm, followed by a weak inflow reaching a depth of $\sim20$~Mm below the photosphere. The maximal velocities in the horizontal direction for the average supergranule are much smaller than the model, and its near-surface flow field RMS value is about an order of magnitude smaller than the often-quoted values of $\sim250-350$~$\rm{m\,s^{-1}}$ for supergranulation. Much of the overall HMI supergranule structure and its weak flow amplitudes can be explained by examining the HMI pipeline averaging kernels for the near-surface inversions, which are found to be very broad in depth, and nearly identical to one another in terms of sensitivity along the $z$-direction. We also show that forward-modeled travel times in the Born approximation using the model (derived from a ray theory approach) are inconsistent with measured travel times for an average supergranule at any distance. Our findings suggest systematic inaccuracies in the typical techniques used to study supergranulation, confirming some of the results in \citet{duvall2012}.
\end{abstract}
\keywords{Helioseismology; Supergranulation; Interior, Convective Zone}
\end{opening}
%-------------------------------------------------

\section{Introduction}
   \label{intro}
Supergranulation has been an active area of solar research now for several decades. Since its initial discovery by \citet{hart1954}, extensive work has been done to characterize these structures at the photosphere \citep[e.g.][among others]{hart1956,simon1964,worden1976,hathaway2000,hathaway2002,hirzberger2008,rieutord2008,roudier2014,williams2014}. However, many open questions still remain regarding their subsurface flow structure and vertical extent through the convection zone. To better characterize the subsurface properties of supergranulation, the methods of local helioseismology must be employed.

One of the first studies of supergranulation from a helioseismic standpoint was carried out by \citet{kosovichev1997}, where large-scale flow structures were observed through time-distance inversions down to a depth of $5$~Mm below the photosphere. Follow-up work by \citet{duvall1998} suggested that supergranules exhibit a radial outflow down to a depth of $\sim8$~Mm, followed by a deeper inflow. These findings were based on two-dimensional spatial correlations of near-surface inversions with flows recovered at various depths. Subsequent inversions by \citet{zhao2003} also reported the presence of an inflow at depths of $9$--$12$~Mm. A later analysis by \citet{jackiewicz2008} found that supergranulation persists down to at least $\sim5$~Mm, but it was determined that the signal at larger depths is completely dominated by random noise and is therefore irretrievable with short data sets of duration $\sim24$~hrs. Additionally, \citet{woodard2007} found that inversions below depths of $4$~Mm were infeasible due to high levels of noise.

To quantify the accuracy of these results, some efforts have been made to validate current time-distance techniques. \citet{zhao2007} performed the first time-distance analysis of large-scale numerical simulation data from \citet{benson2006}. It was determined that it was not possible to accurately recover horizontal flows below a depth of $\sim3$~Mm, and near-surface inversions for the vertical flow component were actually anticorrelated with those of the simulation, likely due to cross-talk effects. Recently, \citet{degrave2014} performed a full time-distance analysis of two $\sim1$~day magnetohydrodynamic quiet-Sun simulations \citep[for simulation details, see][]{rempel2014}, each containing large-scale supergranule-sized flows, and found that recovered flow coherence was lost at depths exceeding $3$--$5$~Mm. In fact, inversions performed at depths of $7$--$9$ Mm actually suggested the presence of spurious flows that were not present in the simulation. This result has since been supported by \citet{svanda2015}. Inversions by \citet{degrave2014} for the vertical flow component were also found to be unreliable at all depths, even after trying to minimize the effects of cross-talk.

It is becoming more apparent that recovering the subsurface flow properties of individual supergranules through helioseismic inversions at depths greater than 5~Mm is quite difficult due to high levels of random noise. To this end, there have been several attempts to study these features from a statistical viewpoint. Averages can be taken over many thousands of features to help boost the signal-to-noise ratio \citep{birch2006}. To date, these analyses of an ``average" supergranule have relied on identifying regions of strong outflow located in near-surface divergence (or nearly equivalently, point-to-annulus travel-time difference) maps. These near-surface averages have then been used to more accurately measure surface properties \citep[e.g.][]{duvall2010} or to constrain subsurface flow models \citep{duvall2012,svanda2012,duvall2014} of supergranulation.
%To date, no study has been carried out to examine the subsurface properties of an average supergranule through direct helioseismic inversions at depths greater than about $1$~Mm.

\begin{figure}[t!]
\begin{center}$
\begin{array}{c}
\includegraphics[width=0.98\linewidth,clip=]{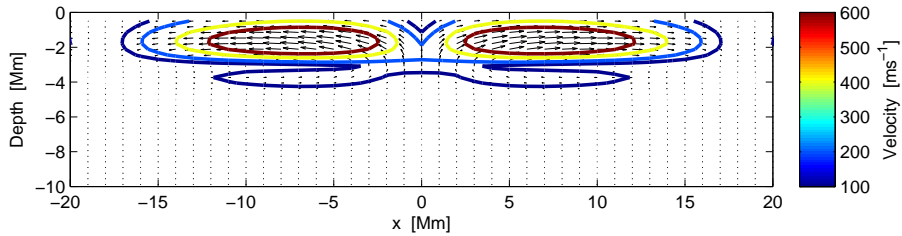}
\end{array}$
\end{center}
\caption{A cut in depth at $y=0$ through the $\rm{DH}_2$ supergranular flow field, where arrows denote the vectors flow in the $v_x-v_y$ plane. The longest arrow represents about $700~\rm{m\,s^{-1}}$. The contour lines mark the $100$, $200$, $400$, and $600$~$\rm{m\,s^{-1}}$ scalar velocity $|\vec{v}|$ levels corresponding to the color bar.}
\label{fig:model}
\end{figure}

Recently, \citet{duvall2012} derived a series of supergranule flow models based on averaging features measured using large-distance separation ($\Delta=2$--$24^\circ$) point-to-annulus travel-time difference maps. It was observed that the measured travel-time differences at the center of the average supergranule approached a nearly-constant 5.1~sec for the largest $\Delta$ values. The idea is that waves propagating over these large distances penetrate so deeply through the convection zone that their paths are nearly vertical when leaving and approaching the solar surface, and are therefore likely to be mostly sensitive to the vertical component of the supergranular flow field. \citet{duvall2012} argue that the current value for the surface vertical velocity component of supergranulation, $v_z(z=0)\approx10$~$\rm{m\,s^{-1}}$ \citep{duvall2010,hathaway2002}, is too small to account for this $5.1$~sec travel-time measurement. Therefore, the vertical flow field must increase in magnitude and peak below the surface at some depth. Constrained by the $10~\rm{m\,s^{-1}}$ surface $v_z$ measurement and assuming a Gaussian profile in the vertical direction, along with a mass-conserving flow field, they find the model that best approximates the data (referred to as $\rm{DH}_2$ hereafter) has the following properties:
\\\\
$\rm{DH}_2$:\\
$v_z(z=0)=10~\rm{m\,s^{-1}}$ and $v_z^{\rm{max}}(z=-2.3~\mathrm{Mm})=240~\mathrm{m\,s^{-1}}$, $\mathrm{FWHM}_z=2.1~\mathrm{Mm}$\\
$v_{\rm{h}}(z=0)=138~\mathrm{m\,s^{-1}}$ and $v_{\rm{h}}^{\rm{max}}(z=-1.62~\mathrm{Mm})=700~\mathrm{m\,s^{-1}}$
\\\\
for the vertical ($v_z$) and horizontal ($v_{\rm{h}}$) flow components, respectively. Figure~\ref{fig:model} shows a cut through the $\rm{DH}_2$ supergranule ($v_x,v_z$) flow component.
%, with contours marking the $100$, $200$, $400$, and $600~\rm{m\,s^{-1}}$ scalar velocity ($v=\sqrt{v_x^2+v_y^2+v_z^2}$) levels. 
Such a shallow feature might appear quite surprising in light of the various works discussed previously.

% Figure

The goal of this work is to test the \citet{duvall2012} flow model against an ``average" supergranule derived using existing helioseismic data products provided by the Joint Science Operations Center (JSOC) Data Resource Management System (DRMS) for the HMI time-distance analysis pipeline \citep[for details about these data products, see][]{couvidat2012,zhao2012}. This will allow us to directly examine the subsurface properties of an average supergranule at depth to compare to the results from these novel measurements. In Section~\ref{data}, we briefly describe the HMI data and their properties. In Section~\ref{alg}, we explain the algorithm employed to identify and average supergranules in these data. Section~\ref{results} describes the resulting HMI supergranule in terms of travel-time and flow field averages. Forward modeling comparisons between various supergranule models in the Born approximation are made in Section~\ref{forward}. Concluding remarks are made in Section~\ref{conclusions}, and a short discussion of supergranule mass conservation is given in Appendix~\ref{mass}. To avoid confusion, we note that the short-hand notation for the various flow features used throughout this work is given in Table~\ref{shtab}.

\begin{table}
\caption{Short-hand notation for the various flow features described throughout this paper.}
\label{shtab}
\begin{tabular}{lcc} % centered columns (4 columns)
\hline
feature & abbreviation\\
\hline
\multicolumn{1}{l}{HMI average supergranule }       & HMI\\
%\multicolumn{1}{l}{D\&H (2012) Model 1}     & $\rm{DH}_1$\\
\multicolumn{1}{l}{D\&H (2012) Model 2}     & $\rm{DH}_2$\\
%\multicolumn{1}{l}{D\&H (2012) Model 3}     & $\rm{DH}_3$\\
\multicolumn{1}{l}{Rempel average supergranule}    & $\rm{RSG}_1$\\
\multicolumn{1}{l}{Rempel Model}          & $\rm{RSG}_2$\\
\multicolumn{1}{l}{Best-fit Model (short-distance)}          & $\rm{BFM_1}$\\
\multicolumn{1}{l}{Best-fit Model (large-distance)}          & $\rm{BFM_2}$\\
\hline
\end{tabular}
\end{table}

\begin{table}
\caption{Parameters for each of the eight HMI phase-speed data filters \citep[adapted from][]{zhao2012}.}
\label{pstab}
\begin{tabular}{c c c c} % centered columns (4 columns)
\hline %inserts double horizontal lines
filter No. & annulus range & phase-speed & FWHM\\
& [heliographic degree] & [$\rm{km\,s^{-1}}$] & [$\rm{km\,s^{-1}}$] \\
\hline
1 & 0.54--0.78 & 14.87 & 4.37\\
2 & 0.78--1.02 &17.49 & 4.37\\
3 & 1.08--1.32 & 21.43 & 5.47\\
4 & 1.44--1.80 & 28.83 & 9.40\\
5 & 1.92--2.40 & 36.48 & 5.91\\
6 & 2.40--2.88 & 40.62 & 5.17\\
7 & 3.12--3.84 & 47.33 & 8.29\\
8 & 4.08--4.80 & 55.94 & 8.95\\
%9 & 5.04--6.00 & 64.95 & 9.07\\
%10 & 6.24--7.68 & 74.35 & 9.72\\
%11 & 7.68--9.12 & 83.67 & 8.92\\
\hline %inserts single line
\end{tabular}
\end{table}

\section{Data} %%%%%%%%%%%%%%%%%%%%%%%%%%%%%%%%%%%%%%%%
   \label{data}
To briefly summarize, the HMI pipeline divides each raw, full-disk, 8~hr Doppler velocity time series into twenty-five smaller patches, each spanning $30^\circ\times30^\circ$ and centered at $0^\circ$, $\pm24^\circ$, and $\pm48^\circ$ in solar latitude and longitude. These data are tracked and remapped using Postel's projection, and subsequently filtered in Fourier space using a series of Gaussian phase-speed filters (referred to as $\rm{td_1}$--$\rm{td_{8}}$ hereafter, parameters of which are given in Table~\ref{pstab}). Cross-covariances are measured from the filtered data in the point-to-annulus and quadrant configurations over a range of annuli radii ($\Delta$). Travel-time differences are then computed from the cross-covariances in the out-in (`oi', denoted $\delta \mathrm{t_{oi}}$), west-east (`we', denoted $\delta \mathrm{t_{we}}$), and north-south (`ns', denoted $\delta \mathrm{t_{ns}}$) geometries \citep{duvall1997} using two definitions: a ``least-squares" method defined by \citet{gb02}, and the other a Gabor wavelet fitting method \citep{kosovichev1997,couvidat2012}. For the current work, we are only interested in the \citet{gb02} `oi' measurements. Each travel-time map is of size $256\times256$ pixels, with horizontal grid spacing $\rmd x=\rmd y=0.12^\circ~\rm{pixel}^{-1}$. Travel times are then inverted for the horizontal flow components ($v_x,v_y$) in the Born approximation, centered around eight depths (0--1, 1--3, 3--5, 5--7, 7--10, 10--13, 13--17, 17--21~Mm). Travel-time maps computed using several additional filters of higher phase-speed, as well as inversion flow maps centered at several depths larger than the eight mentioned here are also available from JSOC. However, these have been deemed unreliable by the HMI pipeline team, and have not been fully tested. They are therefore not included in the subsequent analysis.

JSOC allows one to download travel-time and velocity maps measured within some user-defined time frame, centered at any of the twenty-five desired latitude, longitude pairs. Here, data were selected to cover the 32 days spanning 10 June 2010 -- 11 July 2010 \citep[i.e. the same time span analyzed by][]{duvall2012}. Only the nine patches centered at $0^\circ$, $\pm24^\circ$ were used to avoid the effects of foreshortening that one might encounter toward the limb \citep{zhao2012}. These areas overlap, some as many as four times. We also use only data computed from the first 8~hr time series of each date (00:00--07:59~UT). Since supergranules only live for $\sim1.5$~days on average, individual cells will have evolved significantly from one time series to the next, and we can effectively treat these as different features. In all, we are left with a total of $72$ travel-time maps per date (one map for each of the eight filters over the nine $30^\circ\times30^\circ$ patches), as well as $72$ flow maps for each of the three flow components.

To aid in the analysis, we also have available the line-of-sight (LOS) magnetograms covering this same 32 day period. These allow us to identify any areas of strong magnetic field that might be present in our data to avoid averaging supergranules near these regions.

\section{Supergranule Identification} %%%%%%%%%%%%%%
 \label{alg}

% Figure
\begin{figure}[t!]
\begin{center}$
\begin{array}{c}
\includegraphics[width=0.65\linewidth,clip=]{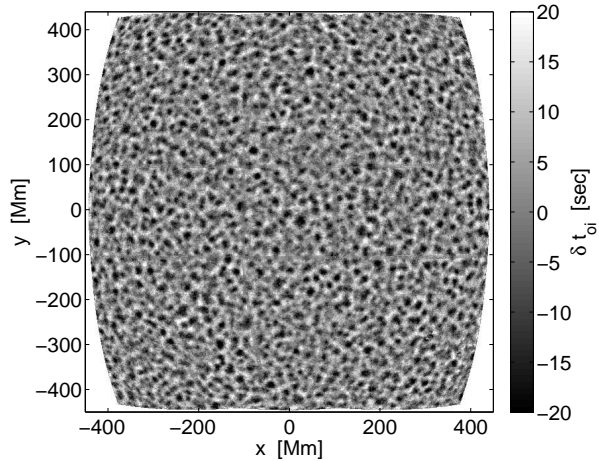}
\end{array}$
\end{center}
\caption{An example $\rm{td_1}$ `oi' travel-time map for the 10 June 2010 date. This results from combining the nine $30^\circ\times30^\circ$ patches centered at $0^\circ$, $\pm24^\circ$ and remapping onto a Cartesian grid. The color scale has been clipped to more easily see the regions of presumed large-scale outflow (negative `oi' travel time) which trace the positions of supergranule cells.}
\label{fig:ttoi}
\end{figure}

%This choice was based on some experimentation, which has shown that even in the case of a circularly-symmetric radial flow field, inverted flow maps (and therefore the corresponding divergence maps) will often be slightly skewed away from cell center, likely due to the inversion weights. This results in averages being taken about locations which are somewhat different than those located in the `oi' maps.

%rather than from smoothed divergence maps recovered through near-surface inversions \citep[e.g.][]{duvall2010}
% (i.e. they are on the $x,y$ grids given by Equations~1 and 2).
% Equation
%\begin{eqnarray}
%x &=& k^{\prime} \cos\phi \sin(\lambda-\lambda_0)\\
%y &=& k^{\prime} [\cos\phi_1\sin\phi - \sin\phi_1\cos\phi\cos(\lambda-\lambda_0)]\\
%c &=& \cos^{-1}[\sin\phi_1\sin\phi + \cos\phi_1\cos\phi\cos(\lambda-\lambda_0)]\\
%k^{\prime} &=& \frac{c}{\sin c}
%\label{eq:one}
%\end{eqnarray}
%Here, $\phi$ and $\lambda$ are the latitude and longitude of the data, $\phi_1$ and $\lambda_0$ are the latitude and longitude of the center of the projection, and $c$ is the angular distance from the center.
An algorithm was written to identify supergranules and subsequently average them about all cell-center locations. The supergranules are identified directly from the `oi' travel-time maps. The algorithm begins by reading in the $72$ travel-time maps for a given date. These ``raw" maps are in the Postel projection. To properly overlay the maps and average their overlapping regions, all travel times were first projected back into latitude and longitude coordinates. From here, maps corresponding to a particular phase-speed filter were then combined via interpolation onto a large Cartesian grid of spacing 1.4577~Mm~$\rm{pixel}^{-1}$. This results in a total of eight large travel-time maps (one for each phase-speed filter) for a particular date, an example of which is shown in Figure~\ref{fig:ttoi}.
% using Equations 5--7.
%\begin{eqnarray}
%\phi &=& \sin^{-1}\Big[\cos c \sin\phi_1 + \frac{y \, \sin c \cos\phi_1}{c}\Big] \\
%\lambda &=& \lambda_0 + \tan^{-1} \Big[\frac{x \, \sin c}{c \cos\phi_1\cos c - y \, \sin\phi_1\sin c}\Big] \\
%c &=& \sqrt{x^2 + y^2}
%\label{eq:two}
%\end{eqnarray}
% via Equations~8 and 9 \citep{thompson2006}, where $R_{\odot}$ is the radius of the Sun, and $B_0$ is the so-called solar B-angle.
%\begin{eqnarray}
%x &=& R_{\odot}\cos\phi \sin(\lambda - \lambda_{0}) \\
%y &=& R_{\odot} [\sin\phi \cos B_0 - \cos\phi \sin B_0 \cos(\lambda - \lambda_{0})]
%\label{eq:three}
%\end{eqnarray}

% Figure
\begin{figure}[t!]
\begin{center}$
\begin{array}{cc}
\includegraphics[width=0.48\linewidth,clip=]{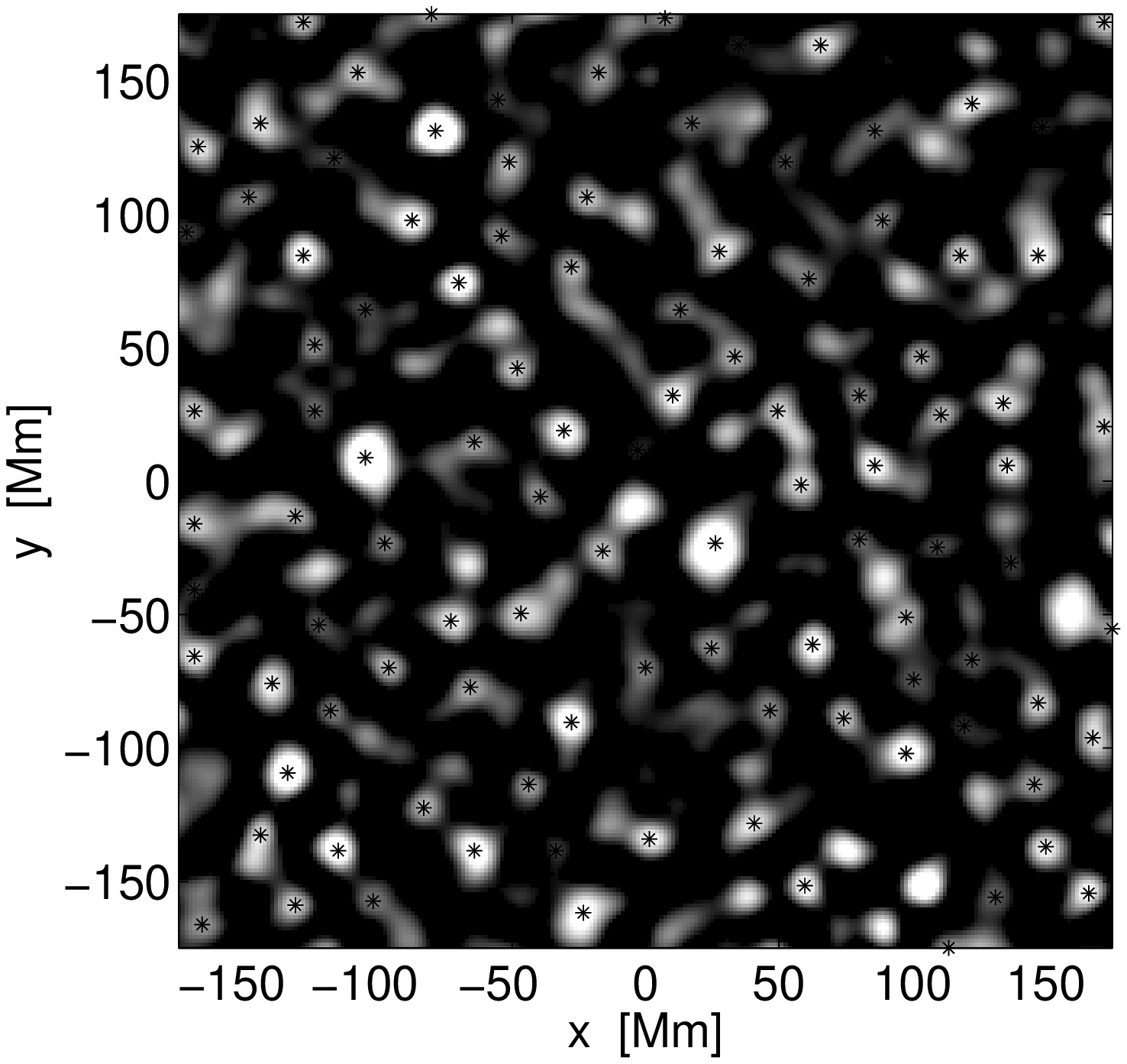} &
\includegraphics[width=0.48\linewidth,clip=]{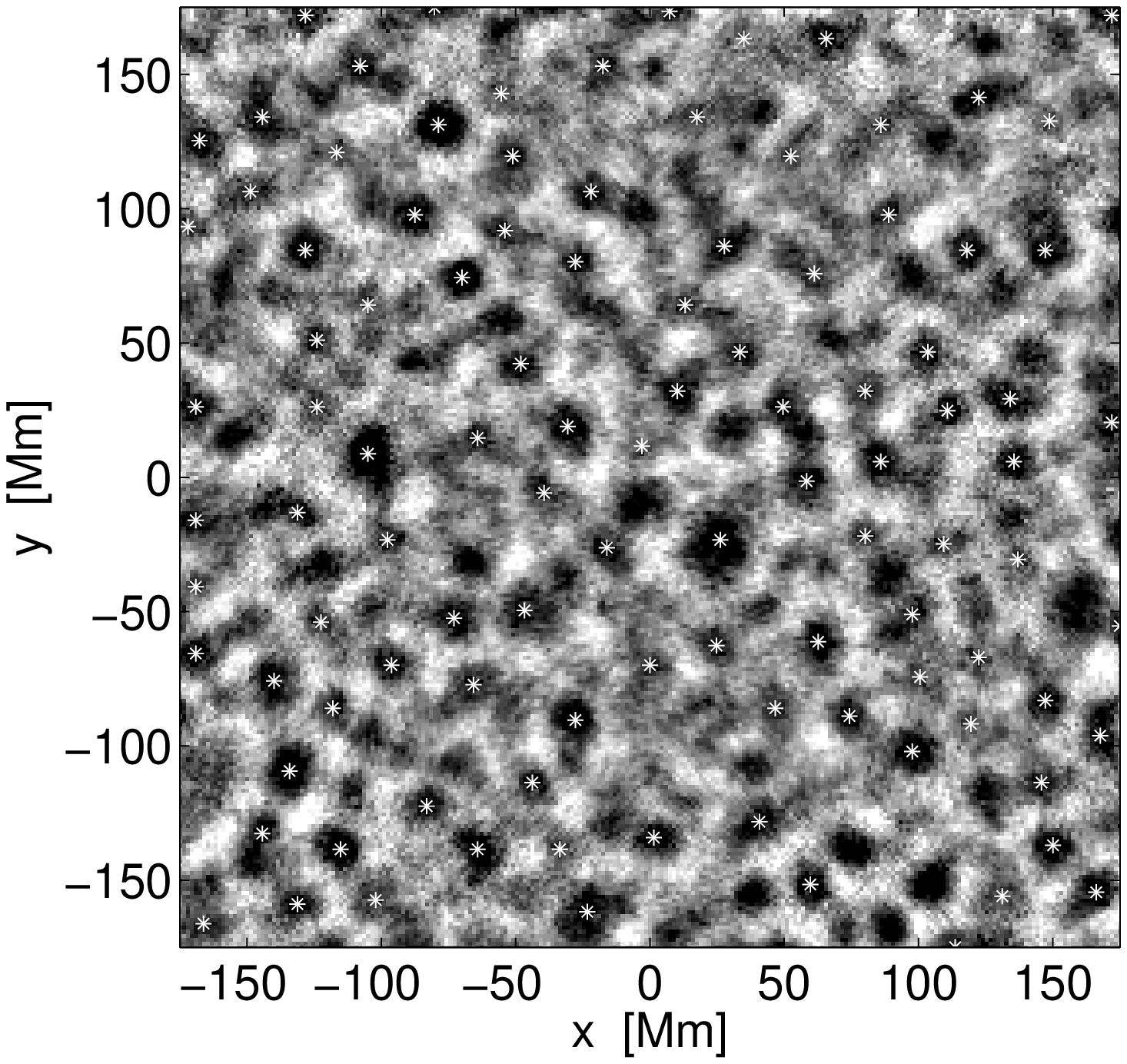}
\end{array}$
\end{center}
\caption{Left: A close-up of the central region of Figure~\ref{fig:ttoi} (in negative) after smoothing via convolution with a 2D Gaussian function of $\sigma=4$~Mm. Supergranule cell centers identified by our algorithm are marked with black asterisks. Right: The cell centers (now marked by white asterisks) plotted over the same area of the unsmoothed travel-time map.}
\label{fig:peaks}
\end{figure}

Following \citet{duvall2012}, the large $\rm{td_1}$ (lowest phase-speed) map was smoothed via convolution with a two-dimensional Gaussian function of $\sigma=4$~Mm, where the Gaussian $\rm{FWHM}=2\sigma\sqrt{2\ln2}$. This serves to smooth out most of the small-scale noise fluctuations while preserving the overall large-scale supergranule structures. 
%The smoothed map is treated as an array, and its pixels are stacked column-wise into a one-dimensional row vector. Once in this form, we employ a function to locate all `oi' travel-time minima (i.e. the areas of strongest divergence, presumably tracing supergranule cell-center locations) through a nearest-neighbor comparison. The indices of these minima are labeled \textit{peak set~$1$} and are saved. The array is then stacked row-wise into a one-dimensional column vector, and the process is repeated with the new minima indices saved as \textit{peak set~$2$}. The cell-center candidate positions are then obtained by computing the intersection of the two sets (i.e. the indices common to both \textit{peak set~$1$} and \textit{peak set~$2$}). 
A simple algorithm was developed that takes each $\rm{td_1}$ travel-time map and finds the coordinates of large negative signals vertically (column-wise) and horizontally (row-wise). It then determines the coordinates common to both searches, and discards candidate cells separated from one another by less than 23~Mm. Any candidate within 23~Mm of the domain edges is also removed to avoid the counting of partial cells. Lastly, regions of strong magnetic field ($B_{\rm{LOS}}>250~\rm{G}$) were located using the magnetograms, and any cells within these areas were eliminated. As an example, Figure~\ref{fig:peaks} (left) shows the central portion of Figure~\ref{fig:ttoi}, smoothed and made negative, with potential supergranule features shown by the lighter colors. The identified cell centers are marked by the black asterisks. Figure~\ref{fig:peaks} (right) shows the cell centers (now denoted by the white asterisks) plotted over the same portion of the corresponding unsmoothed travel-time map. The results look reasonable. Then, a cutout is taken centered at each identified position of radius $\sim50$~Mm and stacked and averaged. This is done for each travel-time map for each filter. It is then carried out for each of the inverted velocity maps using the same coordinates.

% Each of the eight travel-time maps is then averaged separately about the cell-center locations to produce an average supergranule feature for each filter. We note here that the combining of velocity maps follows this process exactly. The velocity maps are also averaged over the same cell-center locations identified in the large $\rm{td_1}$ travel-time maps.

On average, the algorithm identified 625 cells per date, for a total of $\sim20,000$ over the entire 32 day period. Though larger than the number ($\sim5,500$ on account of selecting only the largest ones) found by \citet{svanda2012}, this is substantially smaller than the number ($\sim 55,000$) averaged over by \citet{duvall2012}, since we are using data rather close to disk center. However, a sample of 20,000 cells should be sufficiently large for our purposes and serves to reduce the level of noise in the travel-time and flow maps by a factor of $\sqrt{\rm{\#~of~cells}}=142$.

% TT RESULTS
\section{Results}
	\label{results}

\subsection{Travel-Time Averages}\label{ttavg}
\begin{figure}[t!]
\begin{center}$
\begin{array}{ccc}
\includegraphics[width=0.275\linewidth,clip=]{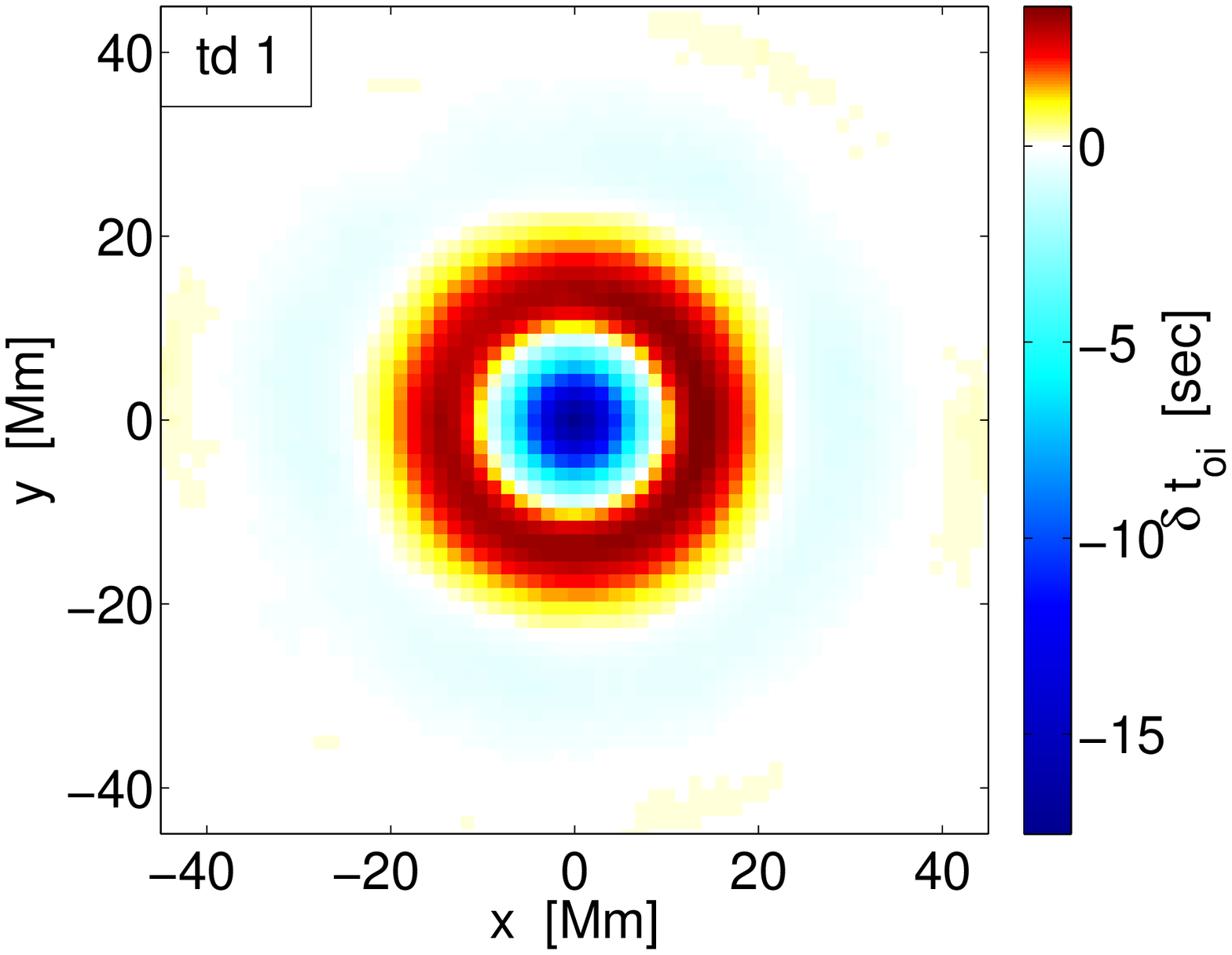} &
\includegraphics[width=0.275\linewidth,clip=]{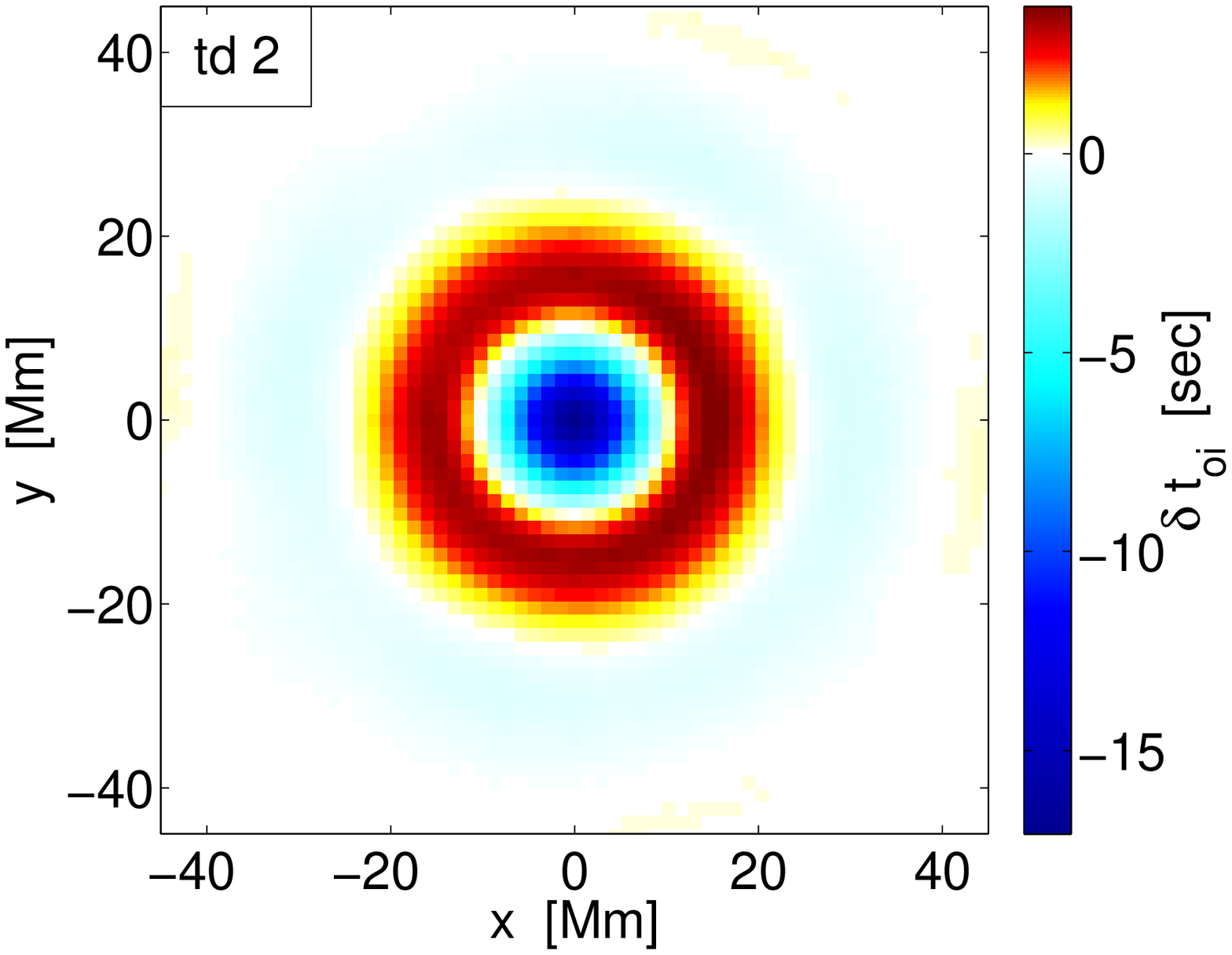} &
\includegraphics[width=0.275\linewidth,clip=]{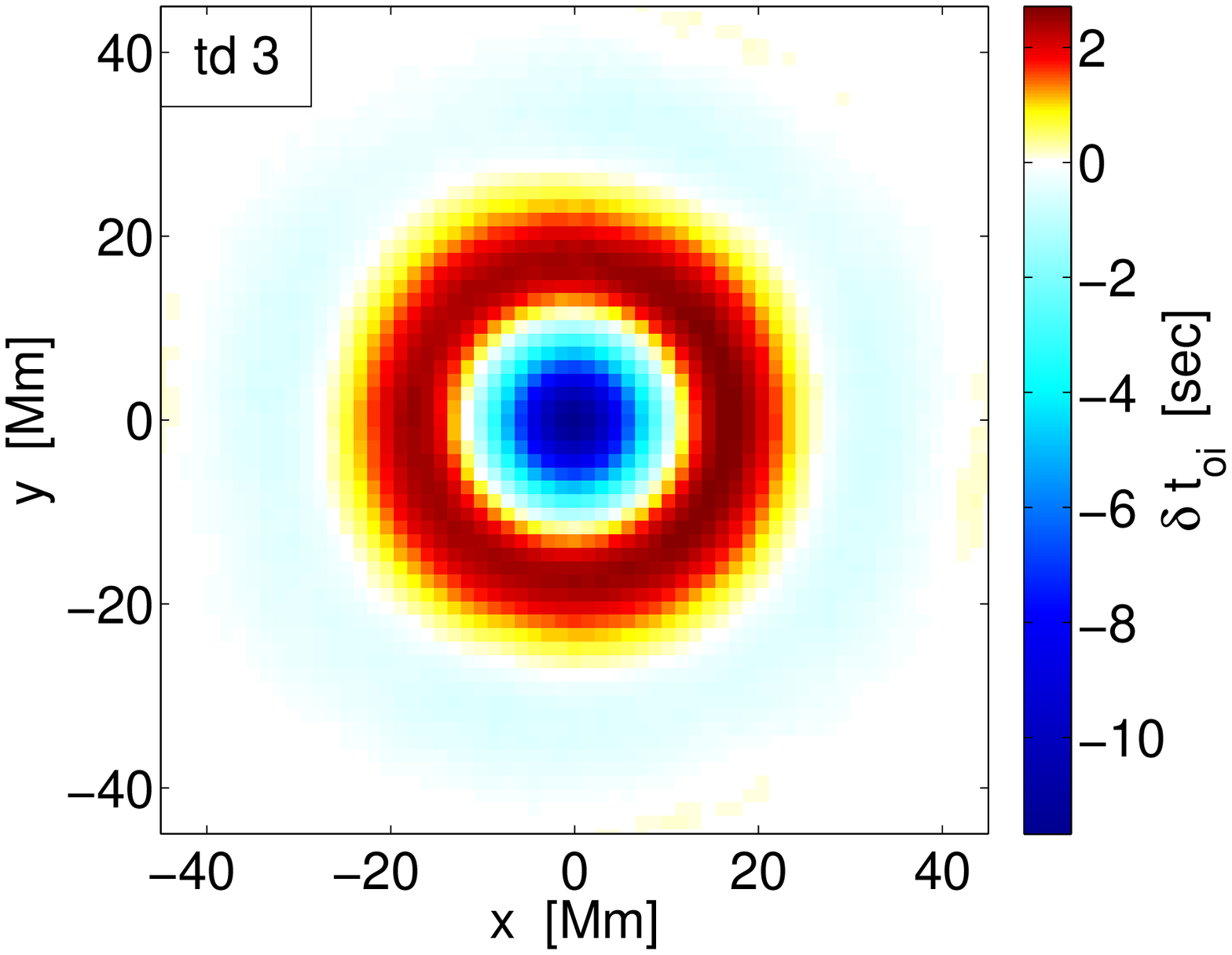} \\
\includegraphics[width=0.275\linewidth,clip=]{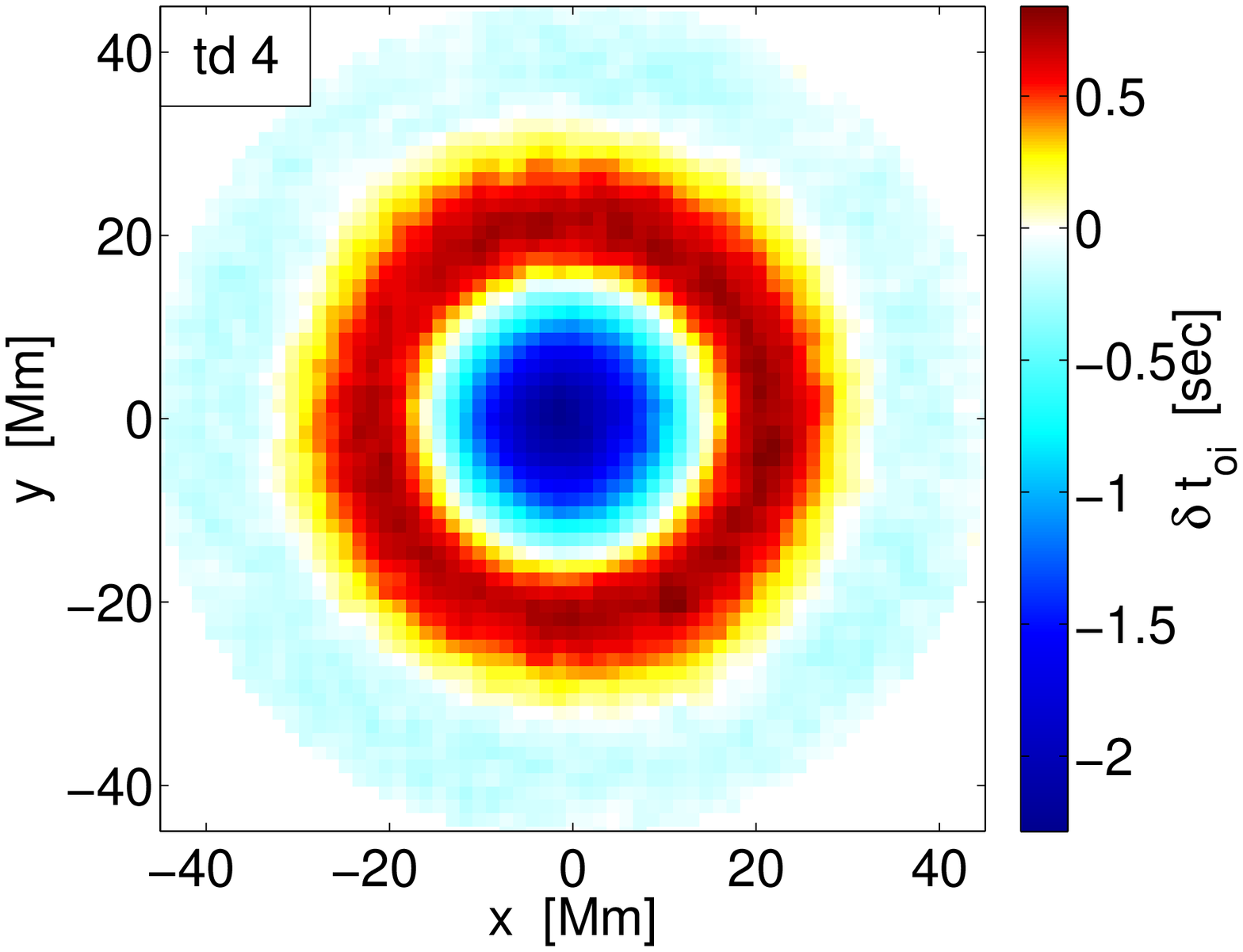} &
\includegraphics[width=0.275\linewidth,clip=]{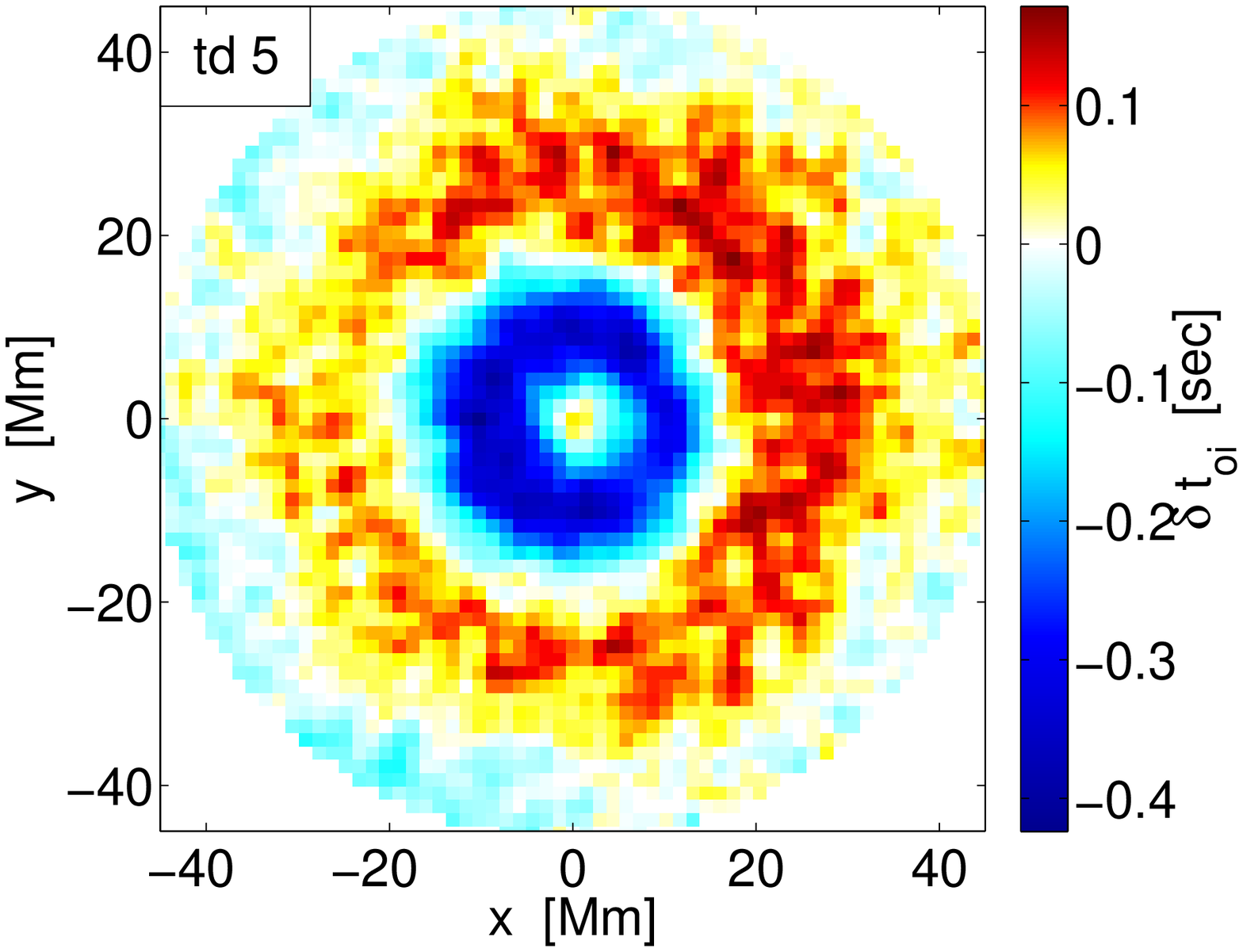} &
\includegraphics[width=0.275\linewidth,clip=]{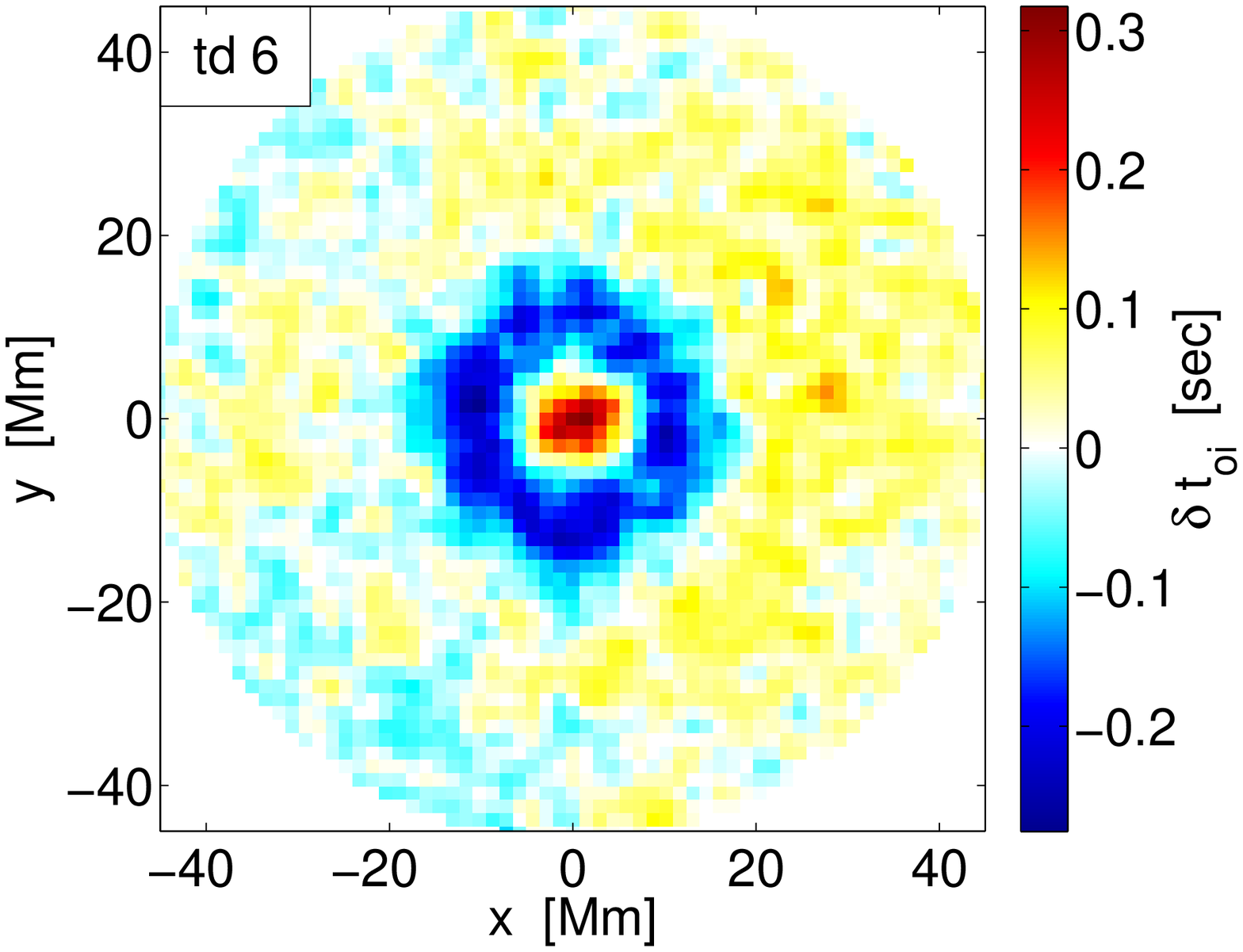} \\
\includegraphics[width=0.275\linewidth,clip=]{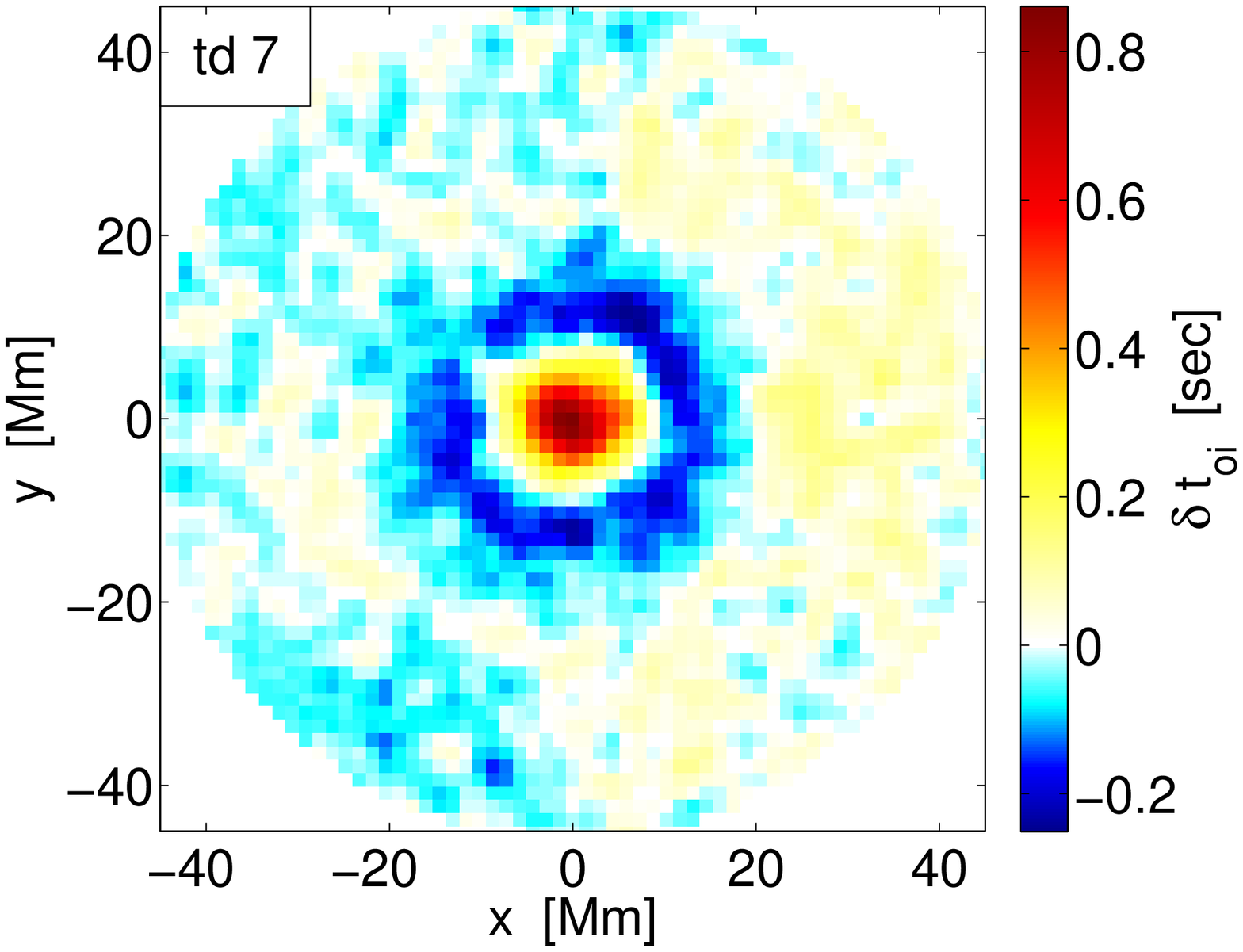} &
\includegraphics[width=0.275\linewidth,clip=]{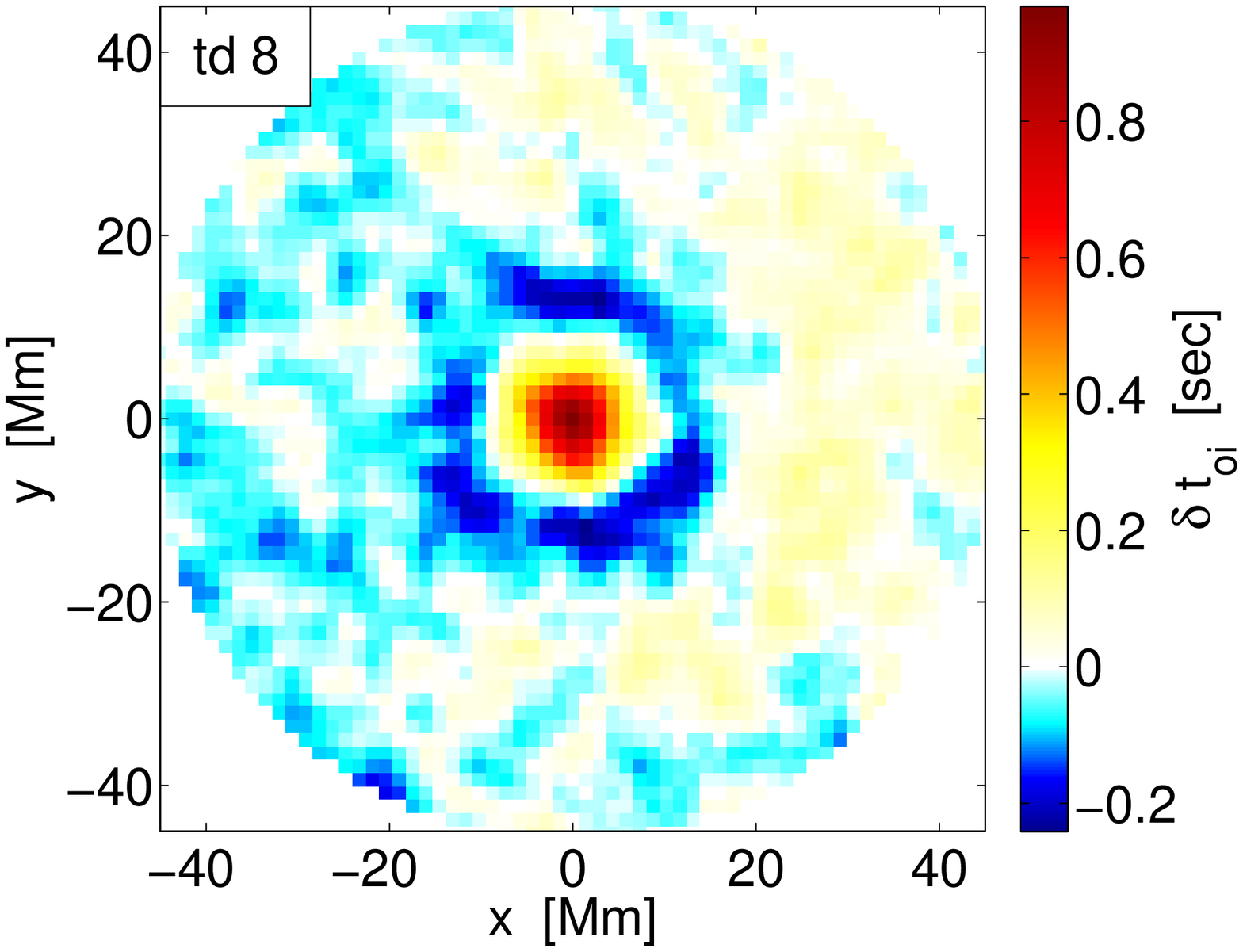}
\end{array}$
\end{center}
\caption{HMI `oi' travel-time maps for each of the eight phase-speed filters $\rm{td_1}$--$\rm{td_{8}}$ after averaging over $\sim20,000$ supergranule features. Filter number is shown in the upper left-hand corner of each panel with corresponding parameters given in Table~\ref{pstab}.}
\label{fig:ttcuts}
\end{figure}

Figure~\ref{fig:ttcuts} shows the resulting supergranule feature derived by averaging the `oi' travel-time maps about all $\sim20,000$ cell-center locations for each of the eight filters. Maps from filters $\rm{td_1}$--$\rm{td_4}$ show a well-defined circularly-symmetric negative travel time at cell center, surrounded by a positive ring arising from the contribution of adjacent supergranules. Looking to filter $\rm{td_5}$, some interesting structure begins to arise -- a small positive signal emerges at the center of the cell. In fact, all subsequent filters $\rm{td_6}$--$\rm{td_{8}}$ exhibit this same central positive `oi' travel time, which appears to rise in magnitude and peak $\sim1$~sec. A large (spatially) $\pm$ travel-time asymmetry in the east-west direction is observed in the ring surrounding the supergranule for phase-speed filters $\rm{td_4}$--$\rm{td_{8}}$, though its origin is unknown.

% Figure

Often, positive and negative `oi' travel times are interpreted as regions of inflow and outflow respectively. This assumption is well-warranted in the shallow subsurface layers where low phase-speed travel times are more significantly effected by horizontal flows (and where horizontal supergranular flows are likely largest). It is worth noting, however, that waves propagating to deeper layers (waves of higher phase-speed) travel along paths that are more vertical when leaving and approaching the solar surface, and are therefore likely to be more sensitive to the vertical flows within a cell. Filters $\rm{td_5}$--$\rm{td_{8}}$ isolate signal from waves which propagate quite deeply through the solar interior, with lower turning points ranging from $7$--$20$~Mm. It is therefore important not to necessarily interpret these deeper positive and negative `oi' measurements strictly as inflows and outflows. To see more clearly what is going on at these depths, it is necessary to examine the averaged flow maps.

% VEL RESULTS
\subsection{Velocity Averages}
	\label{velavg}
\begin{figure}[t!]
\begin{center}$
\begin{array}{ccc}
\includegraphics[width=0.275\linewidth,clip=]{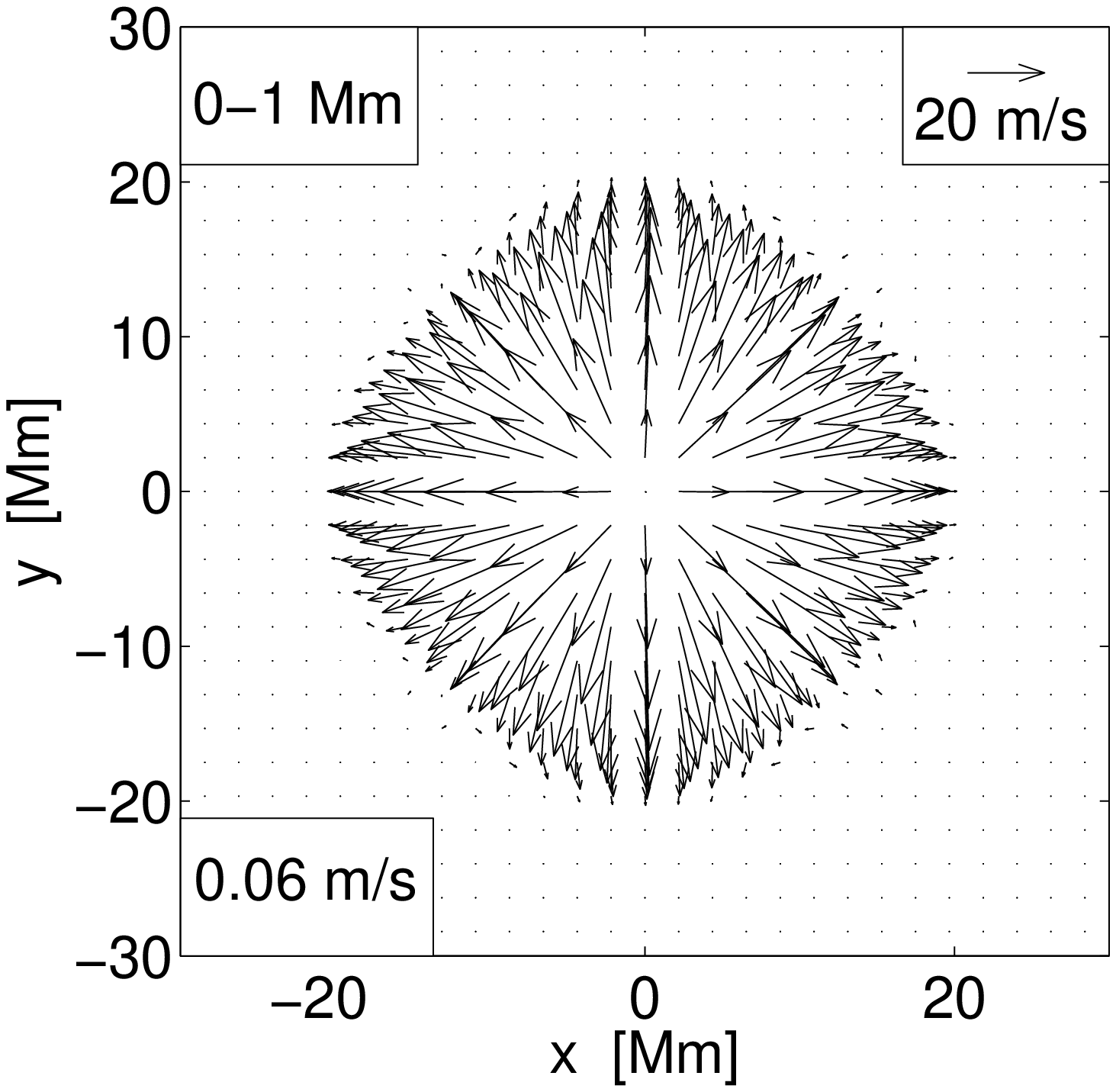} &
\includegraphics[width=0.275\linewidth,clip=]{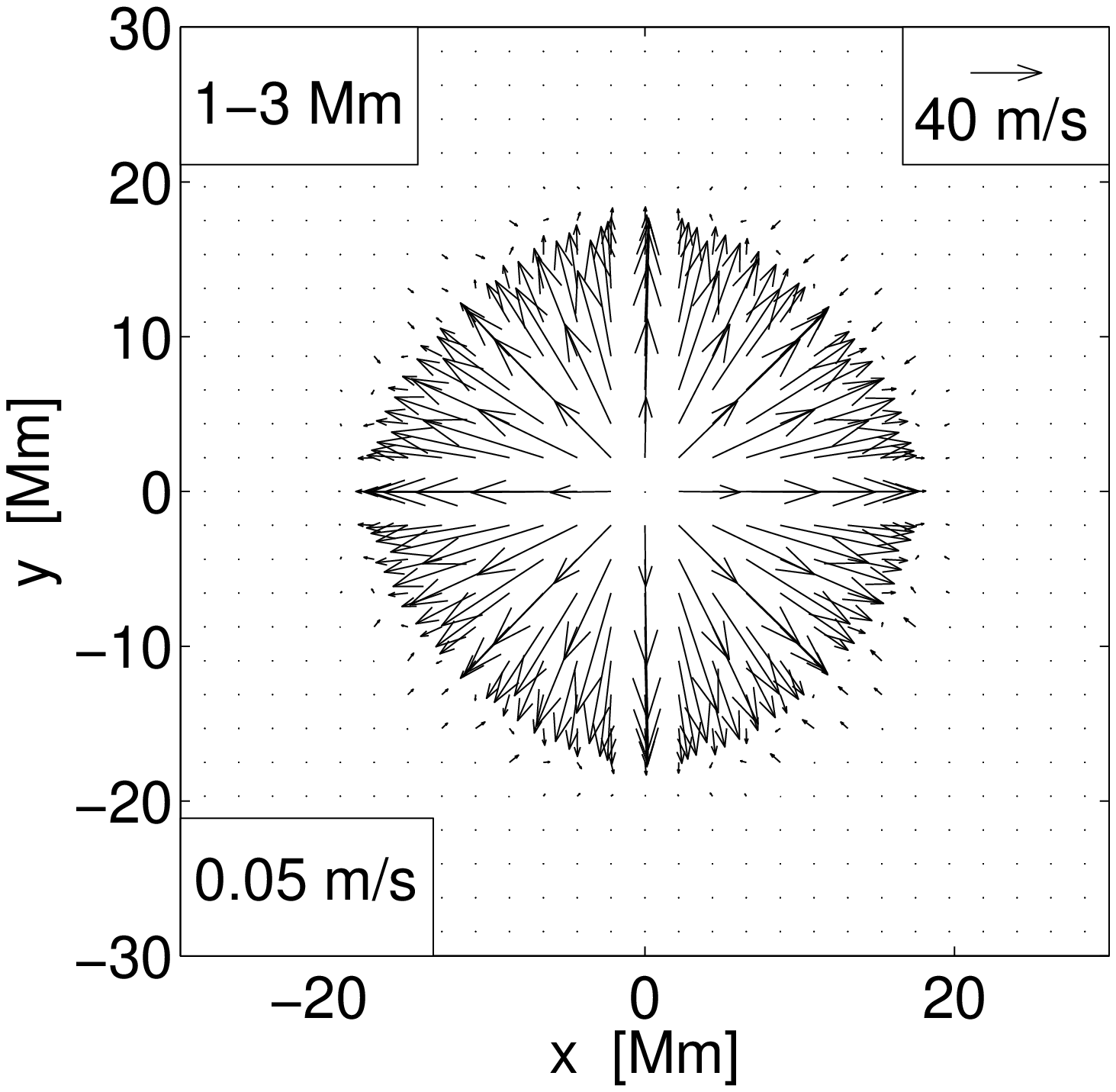} &
\includegraphics[width=0.275\linewidth,clip=]{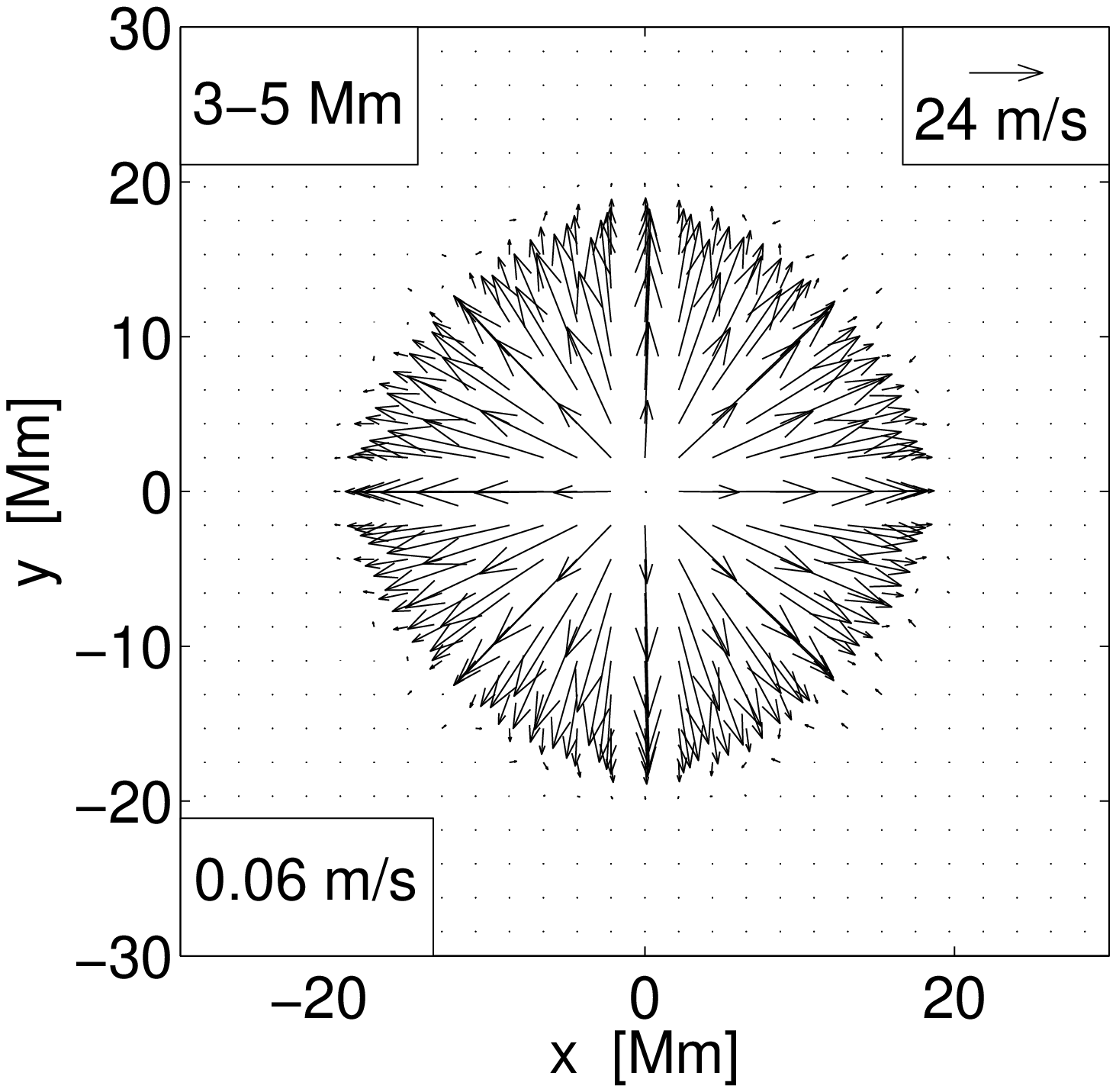} \\
\includegraphics[width=0.275\linewidth,clip=]{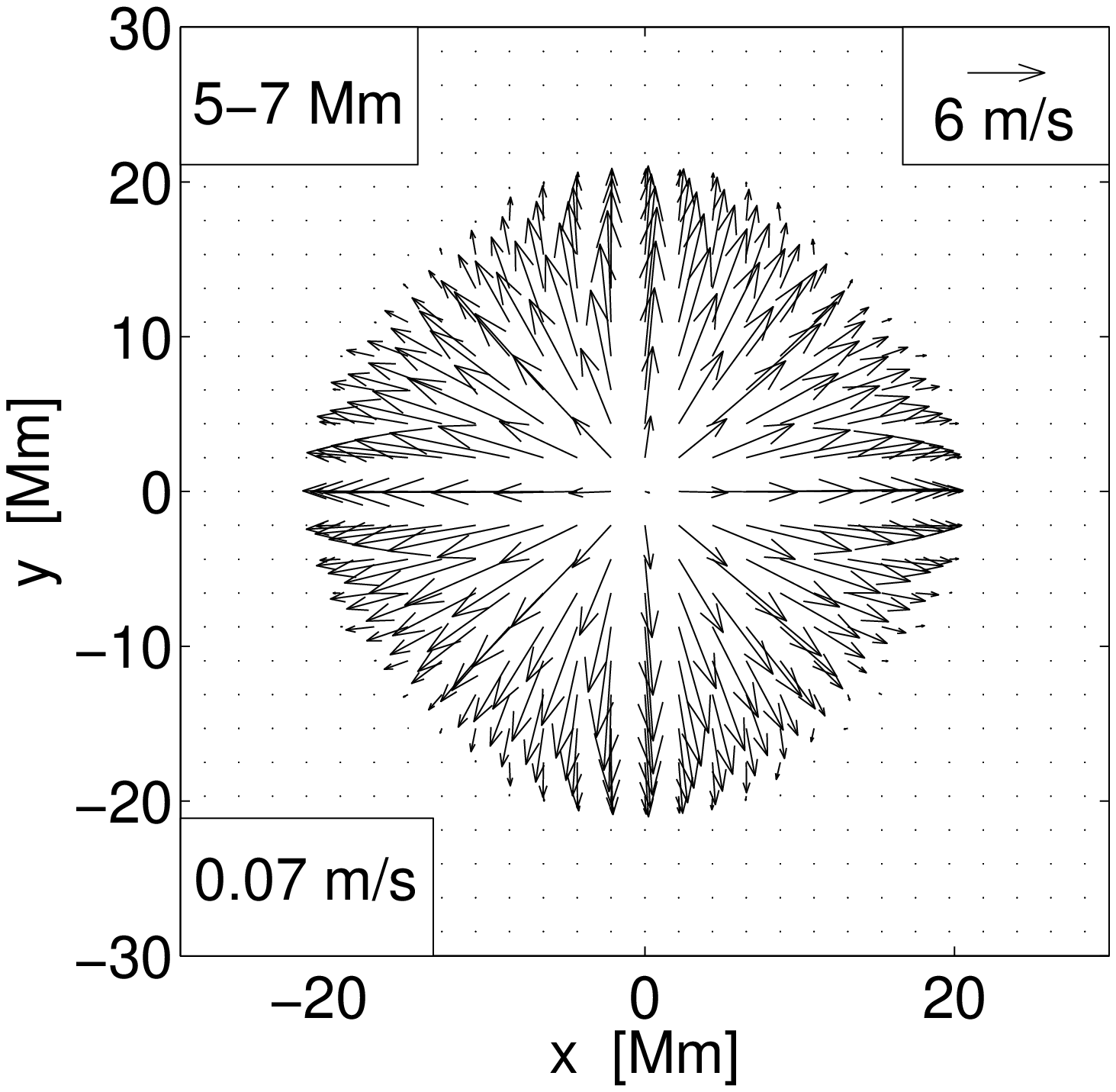} &
\includegraphics[width=0.275\linewidth,clip=]{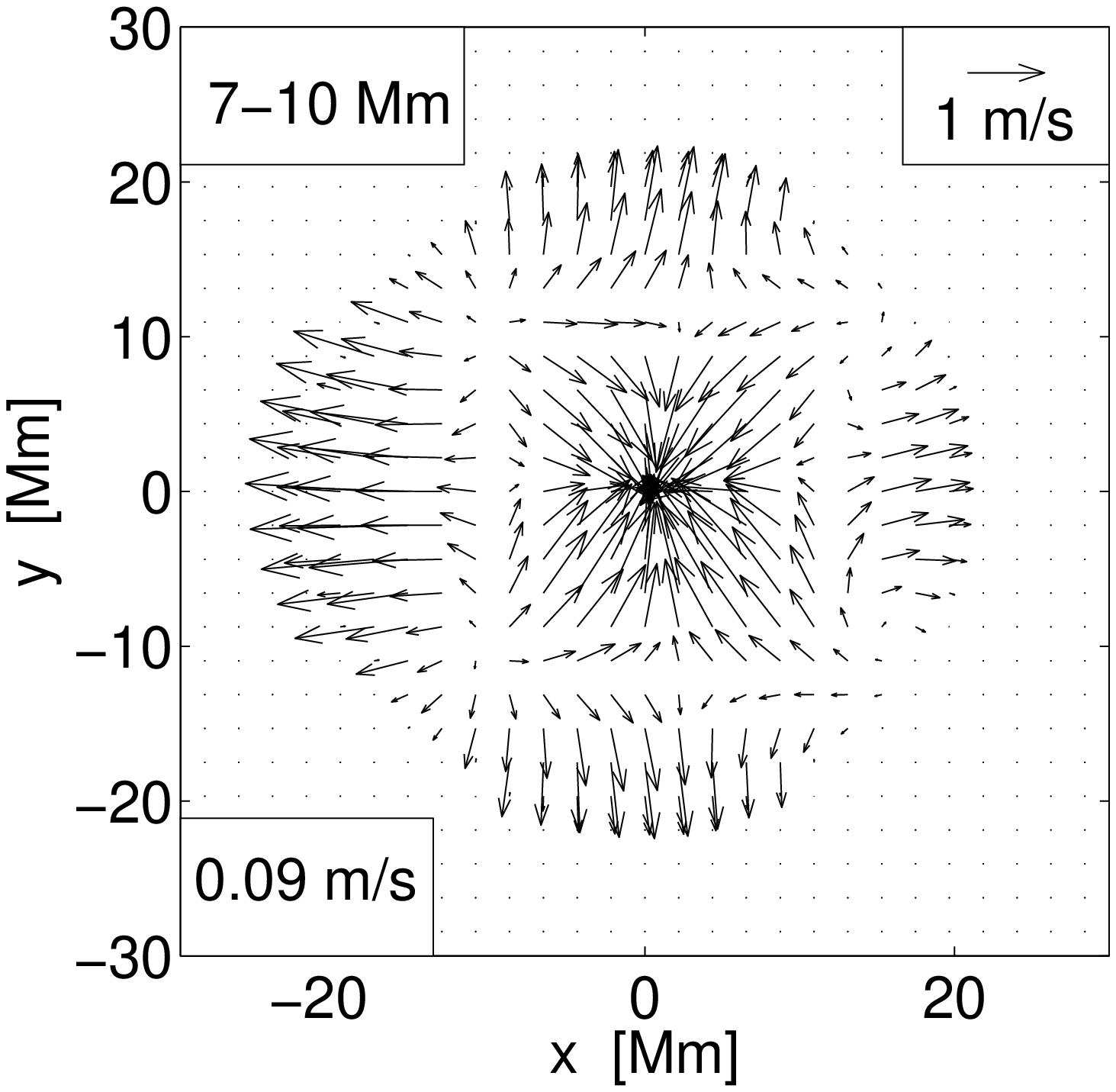} &
\includegraphics[width=0.275\linewidth,clip=]{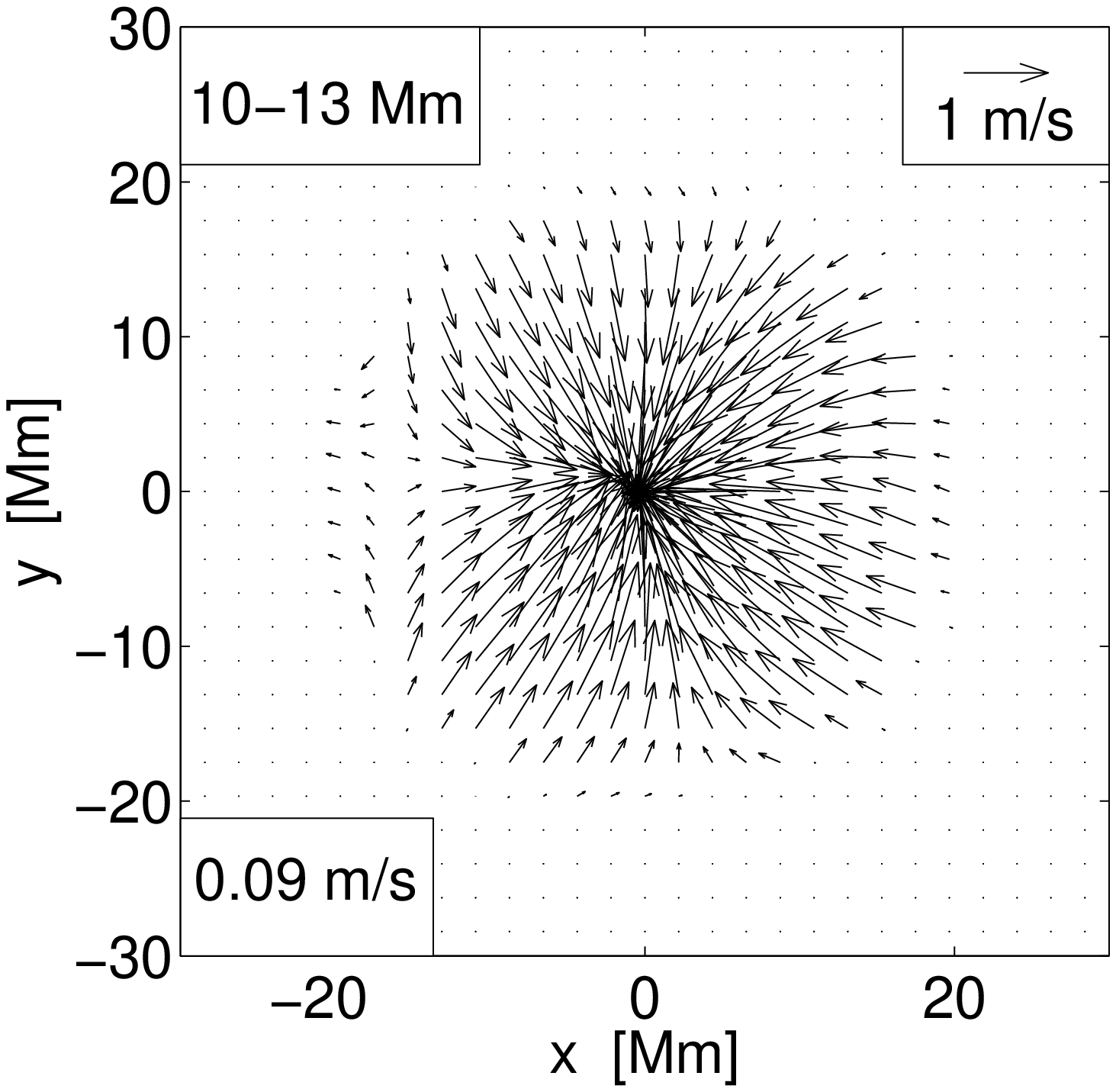} \\
\includegraphics[width=0.275\linewidth,clip=]{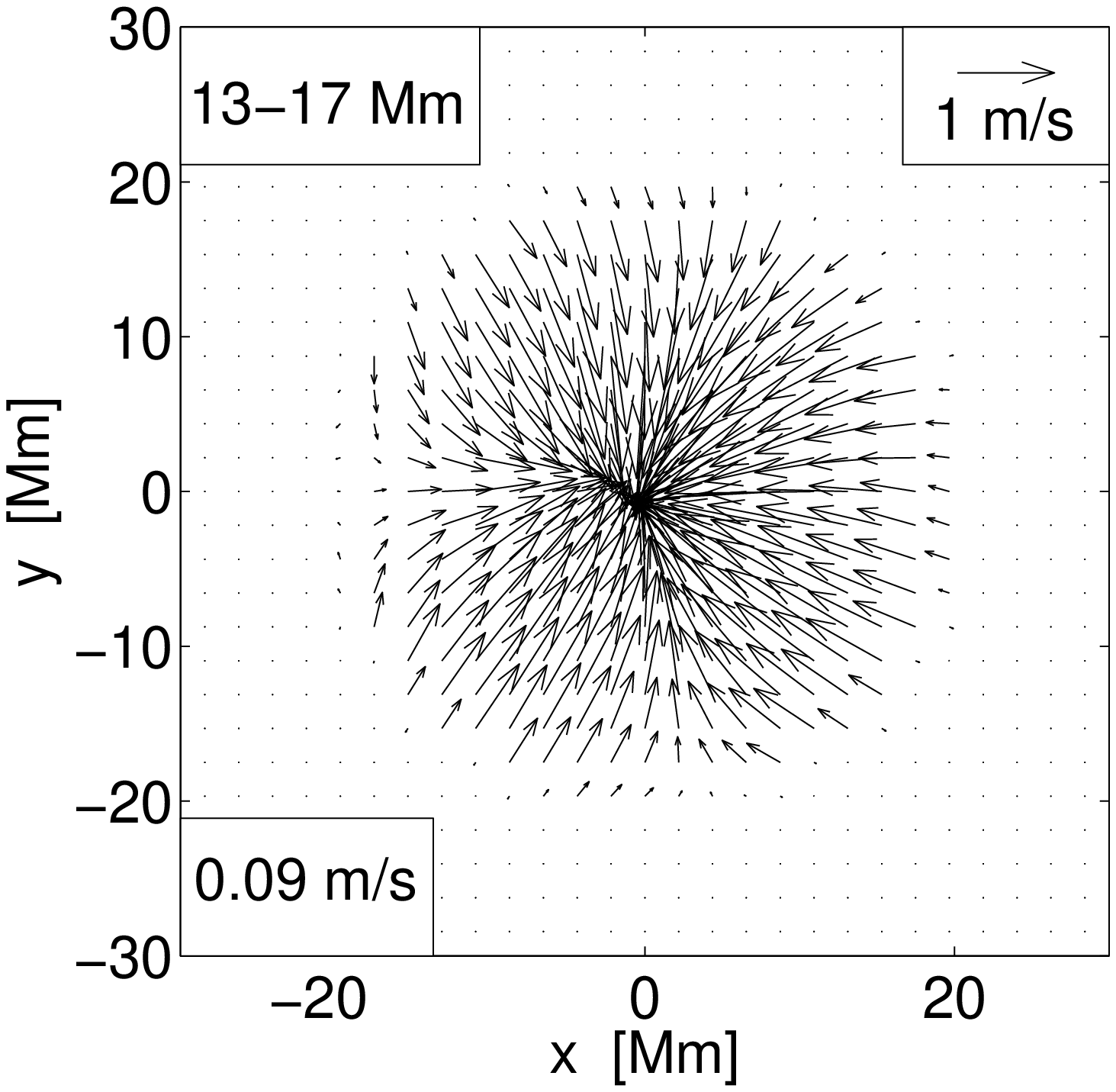} &
\includegraphics[width=0.275\linewidth,clip=]{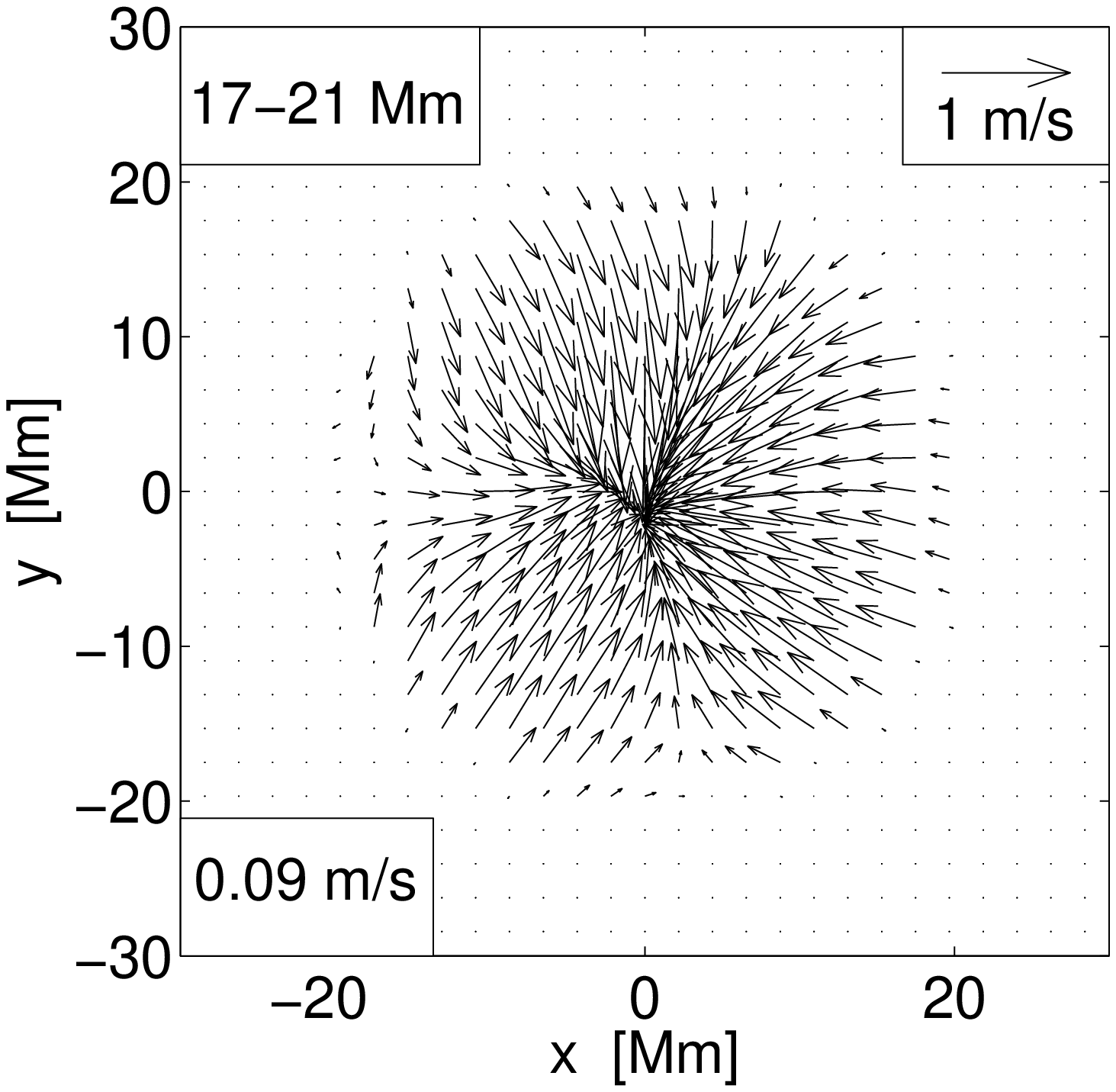}
\end{array}$
\end{center}
\caption{HMI ($v_x,v_y$) flow maps for each of the eight inversion depths defined in the text (shown in the upper left-hand corner of each panel) after averaging over $\sim20,000$ supergranule features. The RMS value of all non-zero flow elements in each flow map is shown in the upper right-hand corner of each panel, with the associated noise level given in the lower left-hand corner.}
\label{fig:velcuts}
\end{figure}

Figure~\ref{fig:velcuts} shows similar cuts taken through the inverted horizontal flow components of the average supergranule. The inversion depth for which each average was computed is shown in the upper left-hand corner of each panel. The root-mean-square (RMS) average of all non-zero flow elements in each map is shown in the upper right-hand corner, with the associated noise level given in the lower left-hand corner. The noise estimates for the first six inversion depths are given in Table~3 of \citet{zhao2012}, and have been corrected here to take into account the statistical averaging. As they are not explicitly stated in the table, the last two inversion depths have been assigned the same noise level as the deepest layer given in \citet{zhao2012}.

% Figure

The first four panels exhibit a nearly symmetric radial outflow down to a depth of $5$--$7$~Mm, roughly consistent with some of the earlier studies mentioned in Section~\ref{intro}. At a depth of $7$--$10$~Mm, the sign of the flow changes around cell center, and an inflow appears. As the remaining panels show, this extremely weak inflow is seen to subsist at all subsequent depths. Interestingly, the near-surface ($0$--$1$~Mm, panel~$1$) RMS flow amplitudes are substantially lower (by an order of magnitude) than the often-quoted values of $\sim250$--$350$~$\rm{m\,s^{-1}}$ for supergranulation \citep[e.g.][]{hart1954,simon1964,hathaway2002}.

\begin{figure}[t!]
\begin{center}$
\begin{array}{c}
\includegraphics[width=0.98\linewidth,clip=]{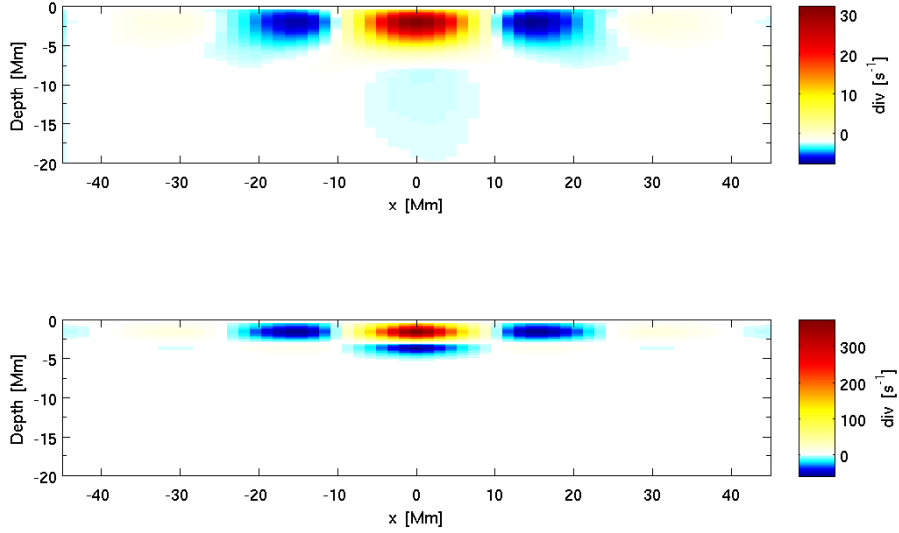}
\end{array}$
\end{center}
\caption{Cuts in depth at $y=0$ through the horizontal divergence ($\nabla_{\rm{h}} \cdot v_{\rm{h}}$) for the average HMI supergranule (top) and $\rm{DH}_2$ (bottom).}
\label{fig:divcut}
\end{figure}

Statistically averaging the flow field enables us to make direct comparisons with model $\rm{DH}_2$ of \citet{duvall2012}. The maps in Figure~\ref{fig:velcuts} were stacked at each central depth, and the horizontal divergence ($\nabla_{\rm{h}} \cdot v_{\rm{h}}$) was computed from the flows. Figure~\ref{fig:divcut} shows a cut in depth through the divergence at $y=0$ for both the HMI supergranule (top) and $\rm{DH}_2$ (bottom). The HMI divergence has been interpolated onto the $\rm{DH}_2$ grid for comparison. We find that the HMI cell exhibits an outflow which peaks at a depth of about $2$~Mm, and extends to approximately twice the depth of the $\rm{DH}_2$ outflow. The HMI supergranule shows the transition region at a depth of $\sim8$~Mm, with a weak extended inflow reaching down to $\sim20$~Mm. As expected from the vector map RMS values, the HMI divergence is an order of magnitude smaller than that of $\rm{DH}_2$.

% Figure

% Figure
%\begin{figure}[t!]
%\begin{center}$
%\begin{array}{c}
%\includegraphics[width=0.98\linewidth,clip=]{vz_slice_2d_two.eps}
%\end{array}$
%\end{center}
%\caption{Cuts in depth at $y=0$ through the $v_z$ flow component for the average HMI supergranule (top) and $\rm{DH}_2$ (bottom).}
%\label{fig:vzcut}
%\end{figure}

% Figure
\begin{figure}[t!]
\begin{center}$
\begin{array}{c}
\includegraphics[width=0.65\linewidth,clip=]{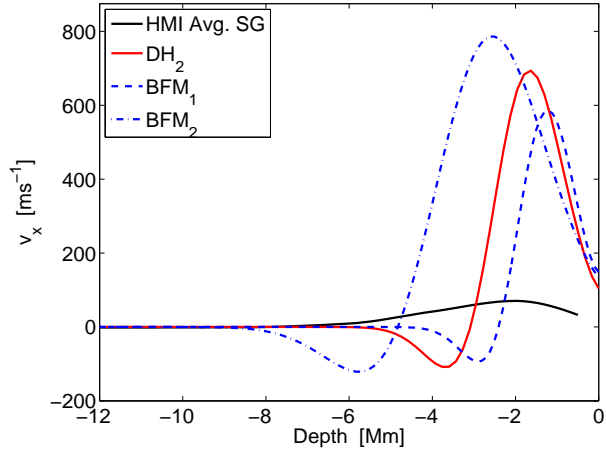}
\end{array}$
\end{center}
%\caption{One-dimensional cuts in depth through the average HMI supergranule and $\rm{DH}_2$ $v_x$ (left) and $v_z$ (right) flow components. Also shown are cuts through the best-fit flow models to all short and large-distance `oi' travel-time measurements ($\rm{BFM}_1$ and $\rm{BFM}_2$ respectively). These models are discussed in Section~\ref{bestfit}. All profiles are taken through the points of maximum surface $v_x$ and $v_z$ for the flow fields.}
\caption{One-dimensional cuts in depth through the average HMI supergranule and $\rm{DH}_2$ $v_x$. Also shown are cuts through the best-fit flow models to all short and large-distance `oi' travel-time measurements ($\rm{BFM}_1$ and $\rm{BFM}_2$ respectively). These models are discussed in Section~\ref{bestfit}. All profiles are taken through the points of maximum surface $v_x$ for the flow fields.}
\label{fig:1Dprofile}
\end{figure}

%Similar comparisons were also made for the vertical flow component. Figure~\ref{fig:vzcut} shows cuts in depth through $v_z$ at $y=0$ for the HMI (top) and $\rm{DH}_2$ (bottom). We find that the maximum amplitude of the HMI vertical velocity is an order of magnitude smaller than that of $\rm{DH}_2$, peaking at the surface rather than at a depth of $2.3$~Mm. The HMI $v_z$ RMS value at the surface is found to be $26$~$\rm{m\,s^{-1}}$, in line with the values found by \citet{hathaway2002}, \citet{duvall2010}, and \citet{svanda2012} for supergranulation. The HMI feature again exhibits a transition region, switching from upflow to downflow, though at a shallower depth than what was found in the divergence profile. This weak downflow extends to a depth of $\sim20$~Mm.

\begin{figure}[t!]
\begin{center}$
\begin{array}{c}
\includegraphics[width=0.98\linewidth,clip=]{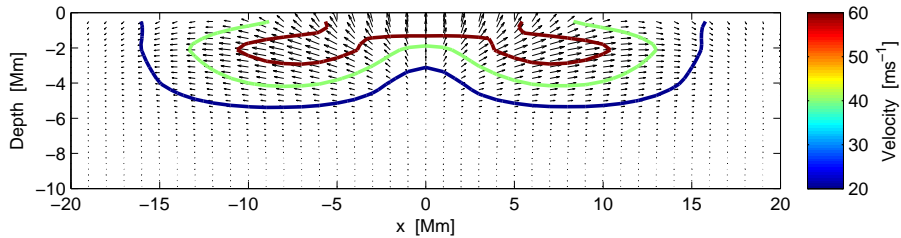}
\end{array}$
\end{center}
\caption{A cut in depth at $y=0$ through the average HMI supergranule ($v_x,v_z$) flow components. The contour lines denote the $20$, $40$, and $60$~$\rm{ms^{-1}}$ scalar velocity levels corresponding to the color bar.}
\label{fig:vxvz}
\end{figure}

\rm{Figure~\ref{fig:1Dprofile} shows one-dimensional depth cuts through the $v_x$ flow component for the HMI and $\rm{DH}_2$ supergranules. These profiles were taken through the points of maximum surface $v_x$}. Figure~\ref{fig:vxvz} shows a cut through the HMI supergranule in vector form. The contours mark the $20$, $40$, and $60~\rm{m\,s^{-1}}$ scalar velocity levels. This can be directly compared with $\rm{DH}_2$ in Figure~\ref{fig:model}.

%One sees the magnitude of the vertical flows are of the same order as the horizontal flows.

% Figure

\subsubsection{Weak Near-Surface Flow Amplitudes}
	\label{stdfilt}
\begin{figure}[t!]
\begin{center}$
\begin{array}{c}
\includegraphics[width=0.6\linewidth,clip=]{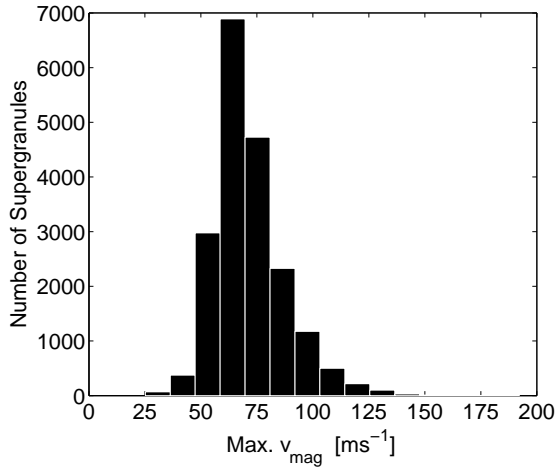}
\end{array}$
\end{center}
\caption{Histogram showing the maximum value of the azimuthally-averaged near-surface flow magnitude ($v_{\rm{mag}}=\sqrt{v_x^2+v_y^2}$) for every supergranule in our sample.}
\label{fig:hist_vmag}
\end{figure}

It could be argued that the small horizontal flow amplitudes observed near the surface for the HMI supergranule are simply a result of the statistical averaging procedure. If the supergranule cell centers have not been correctly aligned (say they are shifted by a pixel or two from one another), we might expect the overall flow amplitude to diminish significantly after combining $20,000$ features. To check if this was indeed the case, the near-surface horizontal flow magnitude ($v_{\rm{mag}}=\sqrt{v_x^2+v_y^2}$) was computed and averaged in an azimuth about cell center, individually for each supergranule in our sample. This gives an idea for what typical supergranule flow amplitudes are that is independent of the overall statistical averaging scheme. Figure~\ref{fig:hist_vmag} shows a histogram of the resulting maximum values for $v_{\rm{mag}}(r,z=0)$ for every supergranule. We find that the majority of HMI supergranules have maximum flow amplitudes in the range of $60$--$80$~$\rm{m\,s^{-1}}$, which is indeed significantly less than values quoted elsewhere.

It is also possible that since our analysis focused only on data within $\pm24^\circ$ of disk center, we were not truly ``capturing" the full strength of the supergranule horizontal flow field like we would if we were looking close to the solar limb \citep[e.g.][]{hathaway2002}. In addition to the twenty-five data patches discussed in Section~\ref{data}, each date also included another four patches centered at ($0,\pm68^{\circ}$ ) and ($\pm68^{\circ},0$) in latitude and longitude. Flow magnitudes were examined individually for the supergranules identified in these patches over several days and compared to the ones near disk center. We find no major quantitative differences in terms of overall flow amplitudes for the sample when looking closer to the limb.

% Figure

% Figure
\begin{figure}[t!]
\begin{center}$
\begin{array}{ccc}
\includegraphics[width=0.3\linewidth,clip=]{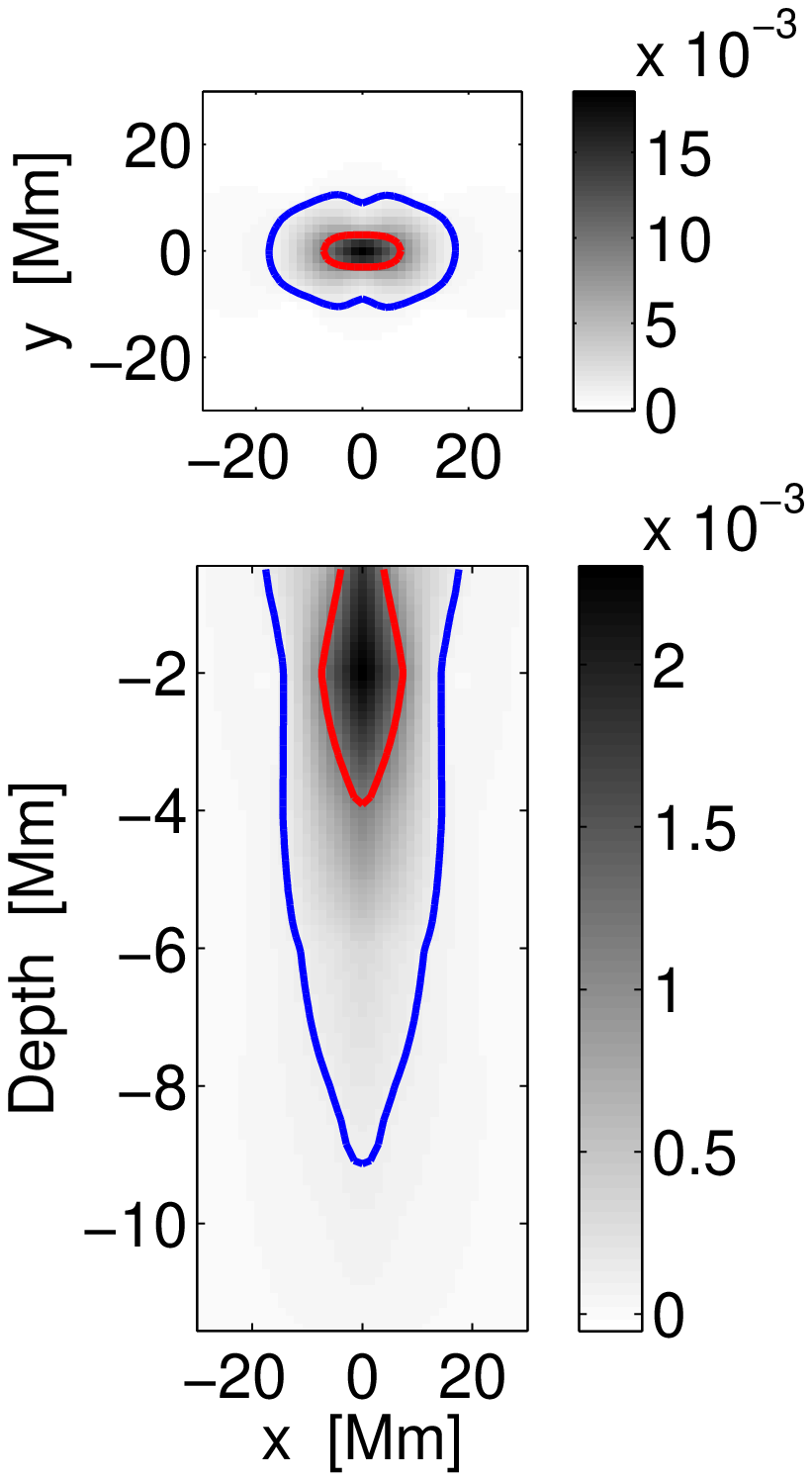}
\includegraphics[width=0.3\linewidth,clip=]{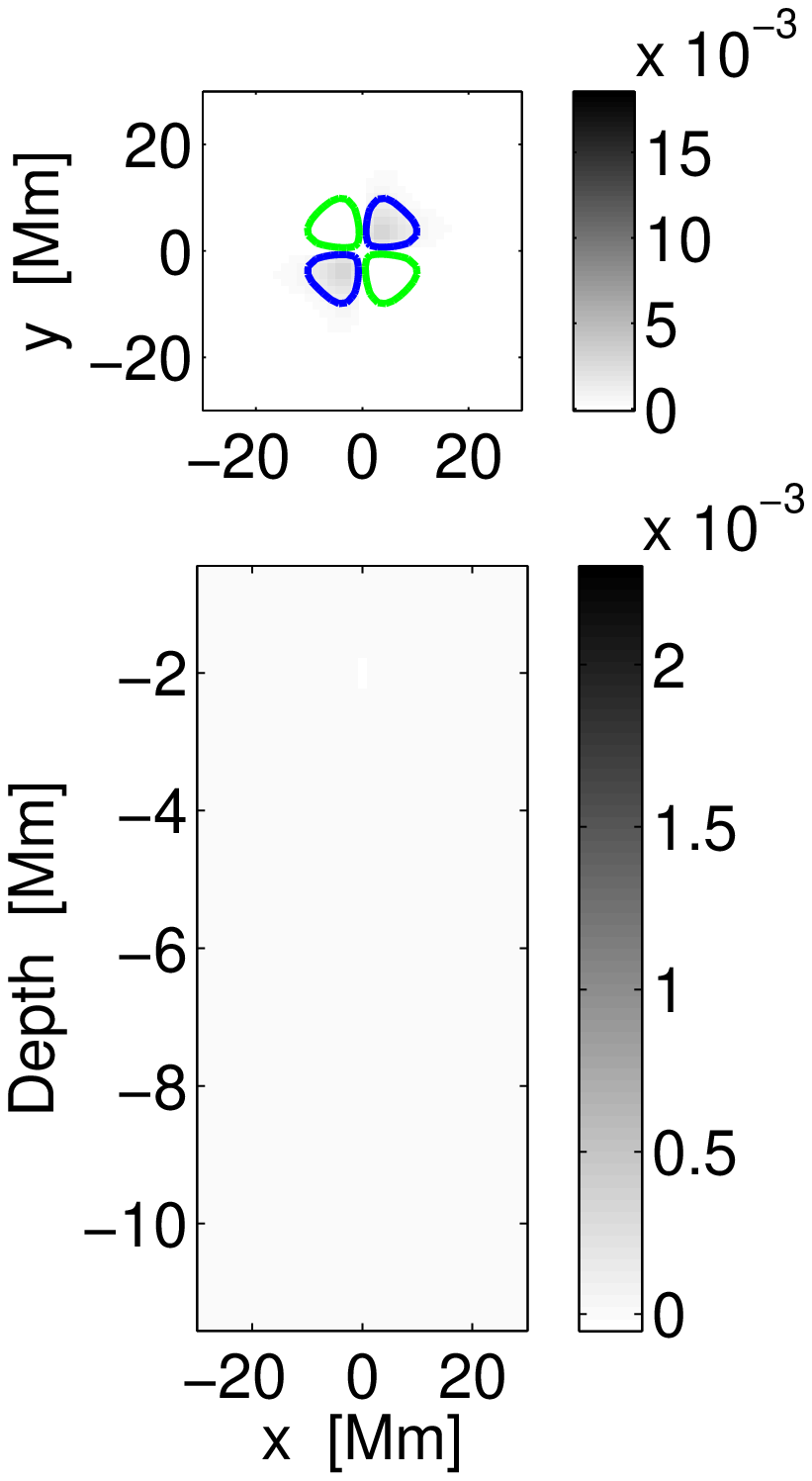}
\includegraphics[width=0.3\linewidth,clip=]{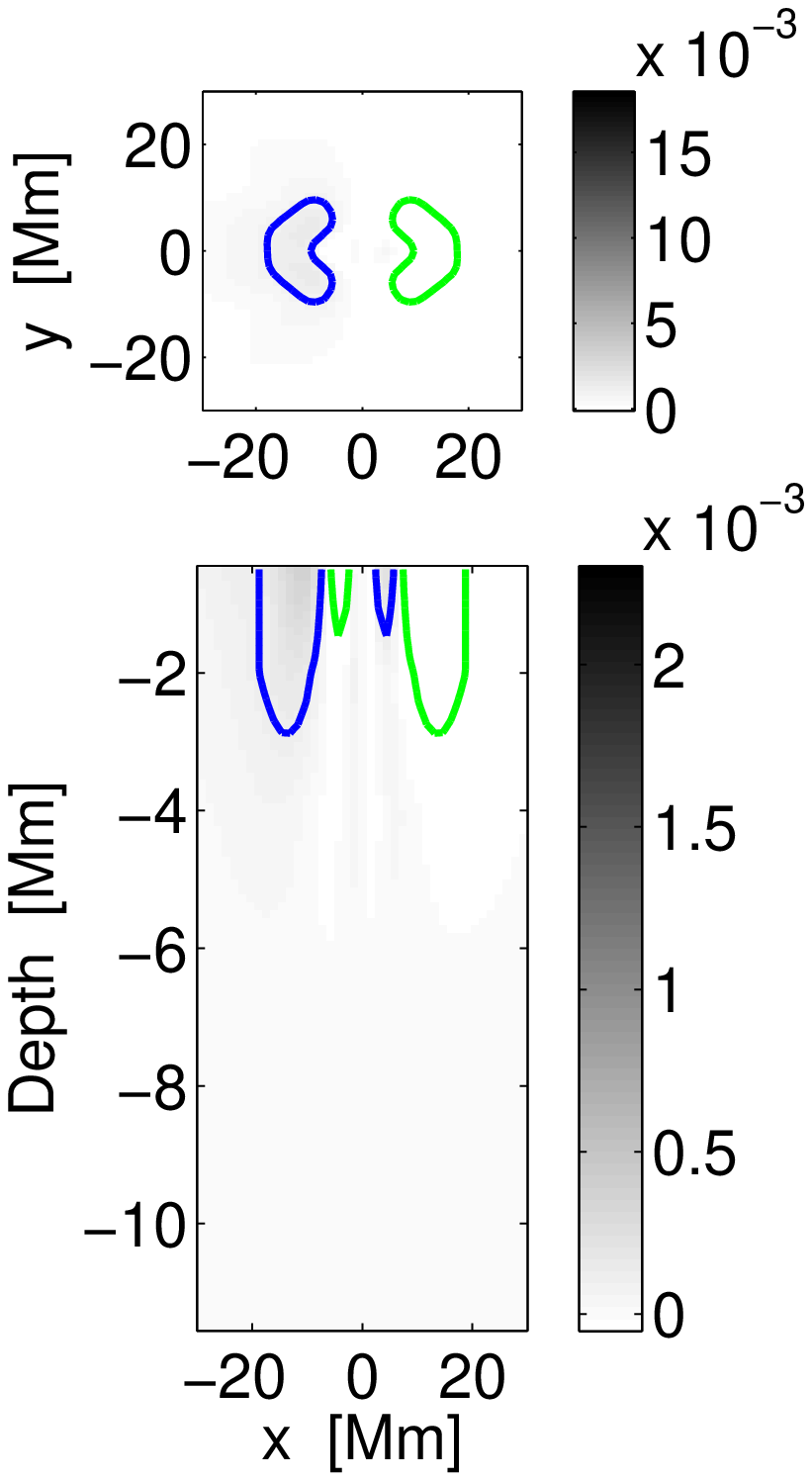}\\
%\hspace{-0.075\textwidth}\includegraphics[width=0.8\linewidth,clip=]{akern_cuts.eps}
\hspace{-0.075\textwidth}\includegraphics[width=0.25\linewidth,clip=]{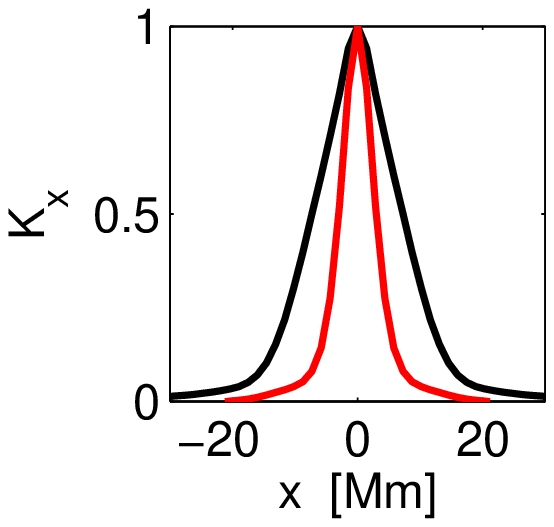}
\includegraphics[width=0.5275\linewidth,clip=]{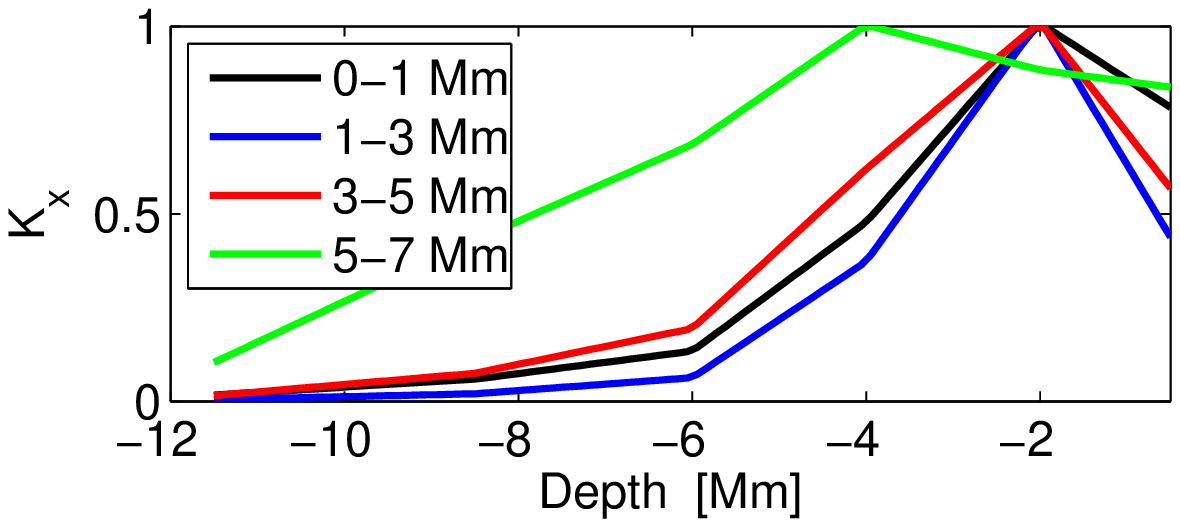}
\end{array}$
\end{center}
\caption{From left to right: Cuts through the $x$, $y$, and $z$ components of the HMI averaging kernel for the near-surface (0--1~Mm target depth) $v_x$ flow inversions. The square panels in the top row show each averaging kernel component after integrating over depth. The middle row panels show cuts in depth through the components along $y=0$. Red and black contour lines denote $50\%$ of the maximum value of the $x$-component of the kernel, respectively, while the blue and green curves mark the $\pm5\%$ contours respectively. The left-most bottom panel shows normalized 1D cuts through the depth-integrated $x$-component of the averaging kernel along $y=0$ (black curve) and $x=0$ (red curve). The right-most bottom panel shows a normalized 1D cut in depth through the $x$-component of the averaging kernel along $y=x=0$ (black curve). Also plotted here for comparison are 1D cuts in depth through the $x$-component of the HMI averaging kernels for inversions at depths of 1--3, 3--5, and 5--7~Mm (colored curves).}
\label{fig:ave_kern}
\end{figure}

Another factor contributing to the reduced flow amplitudes likely comes from the smoothing inherent in the inversion procedure itself, since what is recovered is essentially a convolution between the inversion averaging kernel and the flow field at some target depth. The resulting flow map therefore has a resolution approximately corresponding to the width of the averaging kernel (generally considered to be the averaging kernel FWHM in the case of a Gaussian-shaped function). In the best-case inversion scenario, the averaging kernel FWHM would be as narrow as possible, though this is generally not practical due to the resulting magnification of noise in the problem. As such, this constraint is often relaxed so that near-surface inversions typically have resolutions $\leq10$~Mm or so. \citet{degrave2014} and some further testing showed that the convolution of a Gaussian function of $\rm{FHMW}=10$~Mm with a diverging flow field with $250\,{\rm m\,s^{-1}}$ flows can reduce the amplitude by about $35\%$. Figure~\ref{fig:ave_kern} shows cuts through the $x$, $y$, and $z$ components of the HMI averaging kernel for the near-surface (0--1~Mm target depth) $v_x$ flow inversions. The kernel is found to peak at a depth of roughly $2$~Mm, and is quite broad in depth relative to inversions performed elsewhere \citep[e.g.][]{jackiewicz2008,svanda2012,degrave2014}. The effective horizontal and vertical resolutions of the kernel are roughly $10$~Mm and $4$~Mm respectively (bottom row figures), and so we would expect some influence from this factor. The effect was tested by convolving the averaging kernel with the model $\rm{DH_2}$ flow field. This resulted in an RMS $v_x$ flow amplitude that was $50\%$ lower than the actual model value at a depth of 0.5~Mm.

For comparison, additional cuts through the HMI averaging kernels for inversions centered at depths of 1--3, 3--5, and 5--7~Mm are also shown in the bottom right panel of Figure~\ref{fig:ave_kern}. We find that the three near-surface inversion averaging kernels are almost identical in terms of their sensitivity in the $z$-direction, and also the depth at which peak sensitivity occurs. This likely explains much of the observed correlation between the shallowest flow maps in Figure~\ref{fig:velcuts}.

Another likely effect is the use of standard phase-speed filters employed in the HMI time-distance pipeline (Table~\ref{pstab}). \citet{duvall2012} demonstrated that these rather narrow (in phase space) filters do not capture the full scattered wavefield from supergranules, causing travel times to be smaller than what would otherwise be found if wider filters were used. To this end, they advocate the use of very wide ($\Delta\ell=400$) phase-speed filters.

\section{Comparisons Through Forward Modeling}
	\label{forward}
The main observational result of the \citet{duvall2012} paper was the fact that measured `oi' travel times at the center of an average supergranule over the distance range $\Delta = 2$--$24^\circ$ approached a nearly-constant 5.1~sec for the largest $\Delta$ values. Travel times calculated for a series of models in the ray approximation were compared to these measured travel times, where it was determined that model $\rm{DH}_2$ provided the best fit to the data. Later work by \cite{duvall2014} also compared large-distance measured and ray-approximation travel times computed in the `we' and `ns' quadrant configurations, again concluding that $\rm{DH}_2$ adequately represented the data. The application of ray theory to their work was validated through favorable comparisons of ray-theory travel-time calculations and travel times measured using simulated data having a prescribed vertical flow field.

In this section, we further test this flow model through forward modeling in the Born approximation. Forward-modeled travel times are compared to measured ones using kernels produced using two very different sets of phase-speed filters, each of which is valid over a different, yet overlapping range of distances. The first is a short-distance range spanning $\Delta=0.5$--$4.5^\circ$, and will be collectively referred to as $\Delta_1$ hereafter. The second, a large-distance range spanning $\Delta=2$--$24^\circ$, will be referred to as $\Delta_2$.

% Narrow filter comparisons
\subsection{Short-Distance Born Comparison Using Standard HMI Phase-Speed Kernels}
	\label{stdfilt}
\begin{figure}[t!]
\begin{center}$
\begin{array}{c}
\includegraphics[width=0.6\linewidth,clip=]{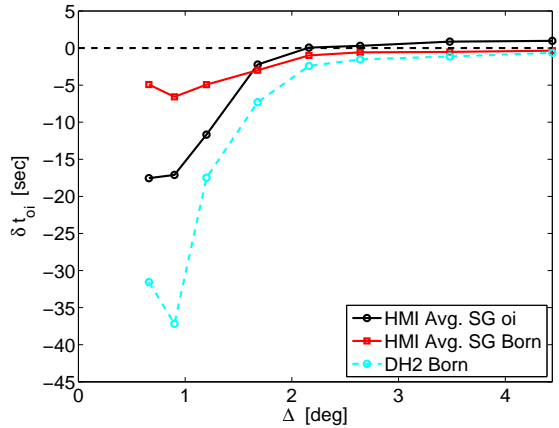}
\end{array}$
\end{center}
\caption{Comparing the statistically-averaged HMI supergranule `oi' travel times (HMI Avg. SG oi) to forward-modeled Born-approximation travel times for the average HMI supergranule (HMI Avg. SG Born) and model $\rm{DH}_2$. These forward-modeled travel times were computed by convolving each supergranule feature with the HMI time-distance pipeline Born kernels. All `oi' travel-time values were taken from cell center.}
\label{fig:hmifigure}
\end{figure}

To begin, the actual set of Born kernels used in the HMI time-distance pipeline was downloaded from JSOC. These kernels were produced using the set of standard HMI phase-speed filters whose parameters are given in Table~\ref{pstab}, and are valid over $\Delta_1$. These were convolved separately with the average HMI supergranule and $\rm{DH}_2$ flow fields to produce a series of forward-modeled travel-time maps. The central `oi' travel-time value from each map was plotted as a function of $\Delta$ for both features, as shown in Figure~\ref{fig:hmifigure}. Also plotted here for comparison are the measured travel-time values of the statistically-averaged HMI supergranule (those shown in Figure~\ref{fig:ttcuts}).

% Figure

It is apparent that the forward-modeled travel times of the average HMI supergranule are somewhat inconsistent with the corresponding statistically-averaged measurements over distances $\Delta < 2^{\circ}$, differing in amplitude by a factor of $3.5$ at the smallest $\Delta$ values. At larger distances, agreement is closer, though the measured travel times are small and positive, while the forward-modeled ones remain negative. At short distances, $\rm{DH}_2$ exhibits travel-time amplitudes that are larger (more negative) than those of the statistically-averaged measurements by a factor of $1.5$--$2$. The statistically-averaged measurements in this figure compare favorably to Figure~3 of \cite{duvall2012}, where travel times for an average supergranule were measured from HMI data after first applying a series of nominal filters similar to those used in the HMI pipeline.

The comparison between the measured HMI and forward-modeled travel times using the average supergranule flow field is essentially only a test of the inversion algorithm. To the extent that some of the inputs to the inversion, such as the travel-time noise covariance matrices and some of the constraint parameter values, are unknown, the agreement is not unreasonable.

% Figure
\begin{figure}[t!]
\begin{center}$
\begin{array}{cc}
\includegraphics[width=0.49\linewidth,clip=]{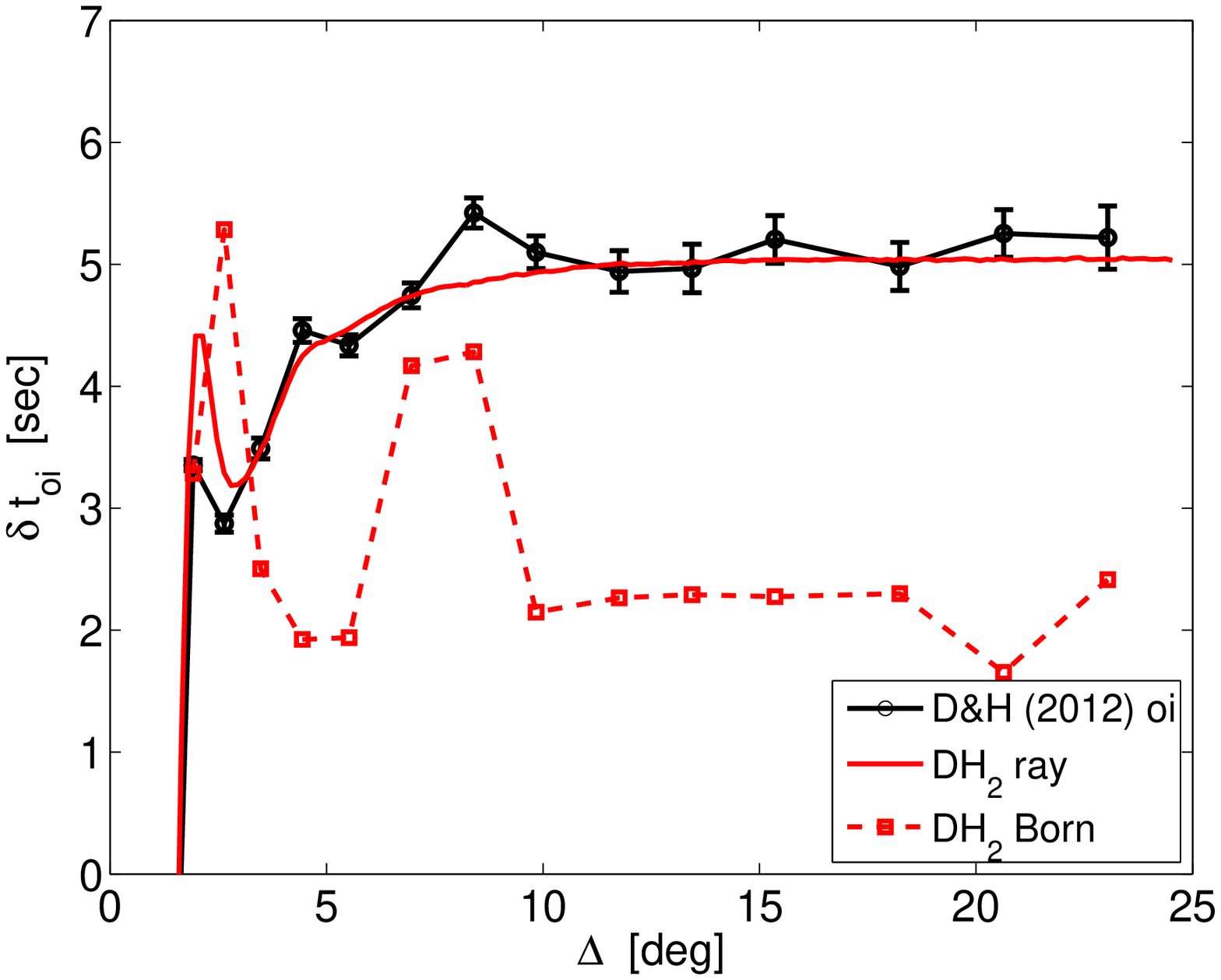}
\includegraphics[width=0.49\linewidth,clip=]{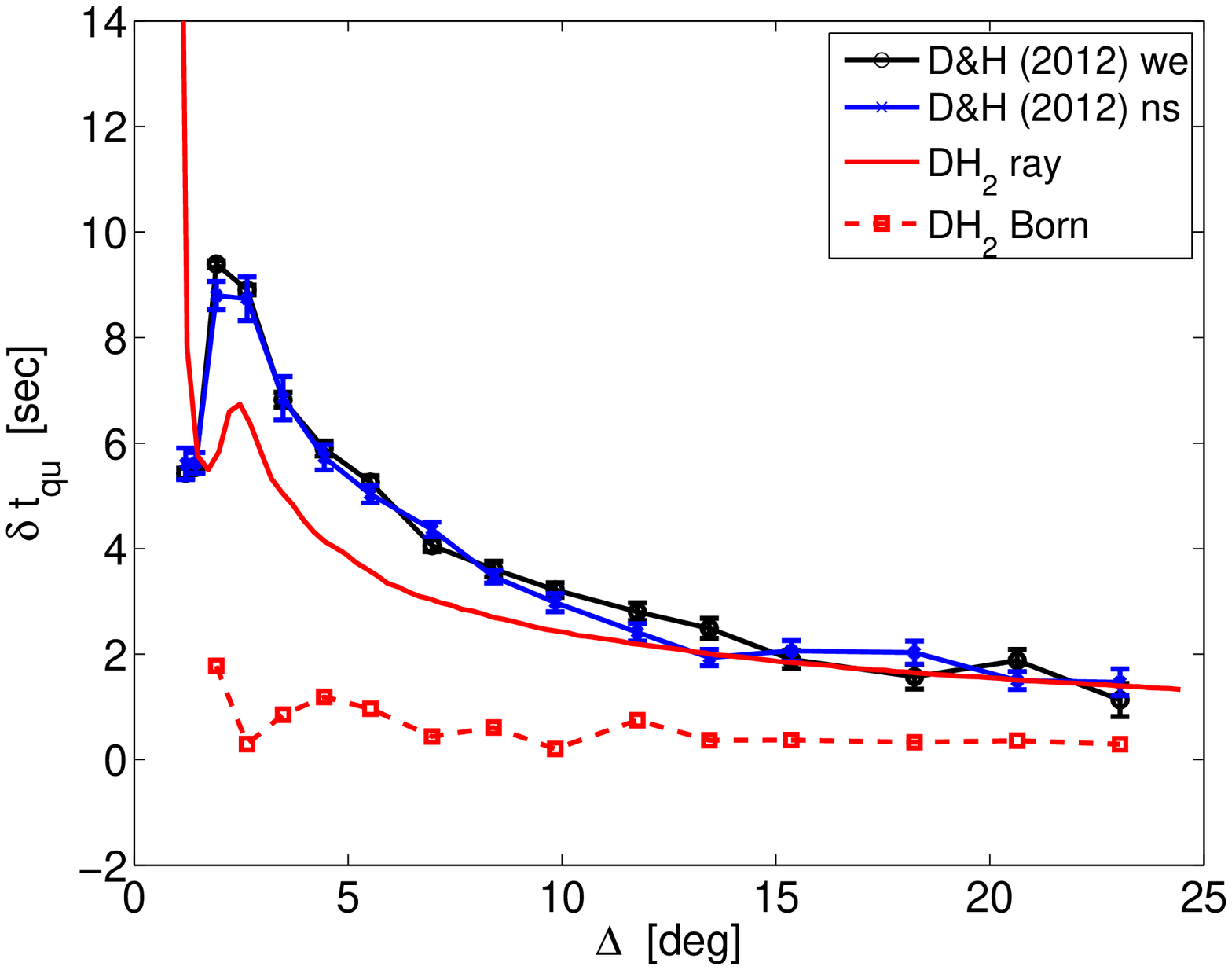}
\end{array}$
\end{center}
\caption{Left: Comparing our large-distance `oi' (left) and quadrant (right) Born-approximation forward-modeled travel times (dashed lines) to the corresponding \citet{duvall2012} measured (D\&H curves) and ray-approximation times (solid lines) for model $\rm{DH}_2$.}
\label{fig:largedist}
\end{figure}

% Wide filter comparisons
\subsection{Large-Distance Born Comparison Using Wide Phase-Speed Kernels}
	\label{largedist}
As a second test, a set of Born-approximation sensitivity kernels for flows \citep{birch2007} was computed over $\Delta_2$ using the same fourteen wide phase-speed filters employed in the \citet{duvall2012} and \citet{duvall2014} studies. Such large-distance, wide-filter Born kernels have not previously been used before in a time-distance analysis. These kernels were used with the $\rm{DH}_2$ model to provide forward-modeled travel-time maps. Figure~\ref{fig:largedist} compares the `oi' (left) and quadrant (right) cell-center values to the \citet{duvall2012} and \citet{duvall2014} measurements and ray-approximation times. Though they differ in amplitude, below $\Delta = 3.5^{\circ}$ the Born and ray-approximation `oi' travel times show similar trends, initially rising sharply and peaking around $\Delta = 2.5^{\circ}$ before falling off. In the range $\Delta = 7$--$8^{\circ}$, something interesting happens: rather than approaching a value $\sim5.1$~sec, the Born travel times exhibit a second large peak (of unknown origin, as visual inspection of the kernels shows no obvious anomalies) before decreasing again and leveling off at values $\sim2$~sec at larger distances. Unlike the ray travel times, it is apparent that these Born travel times do not closely match the data at any distance. Examination of the quadrant travel times also shows large discrepancies. Here, we find that the Born travel times are small and relatively constant over all distances, differing from both the measurements and ray-approximation times by an order of magnitude at small $\Delta$, and a factor of $4.5$ at larger distances.

\subsection{Disagreement Between Measured and Forward-Modeled Travel Times}
	\label{kernels}
% Figure
\begin{figure}[t!]
\begin{center}$
\begin{array}{cc}
\includegraphics[width=0.48\linewidth,clip=]{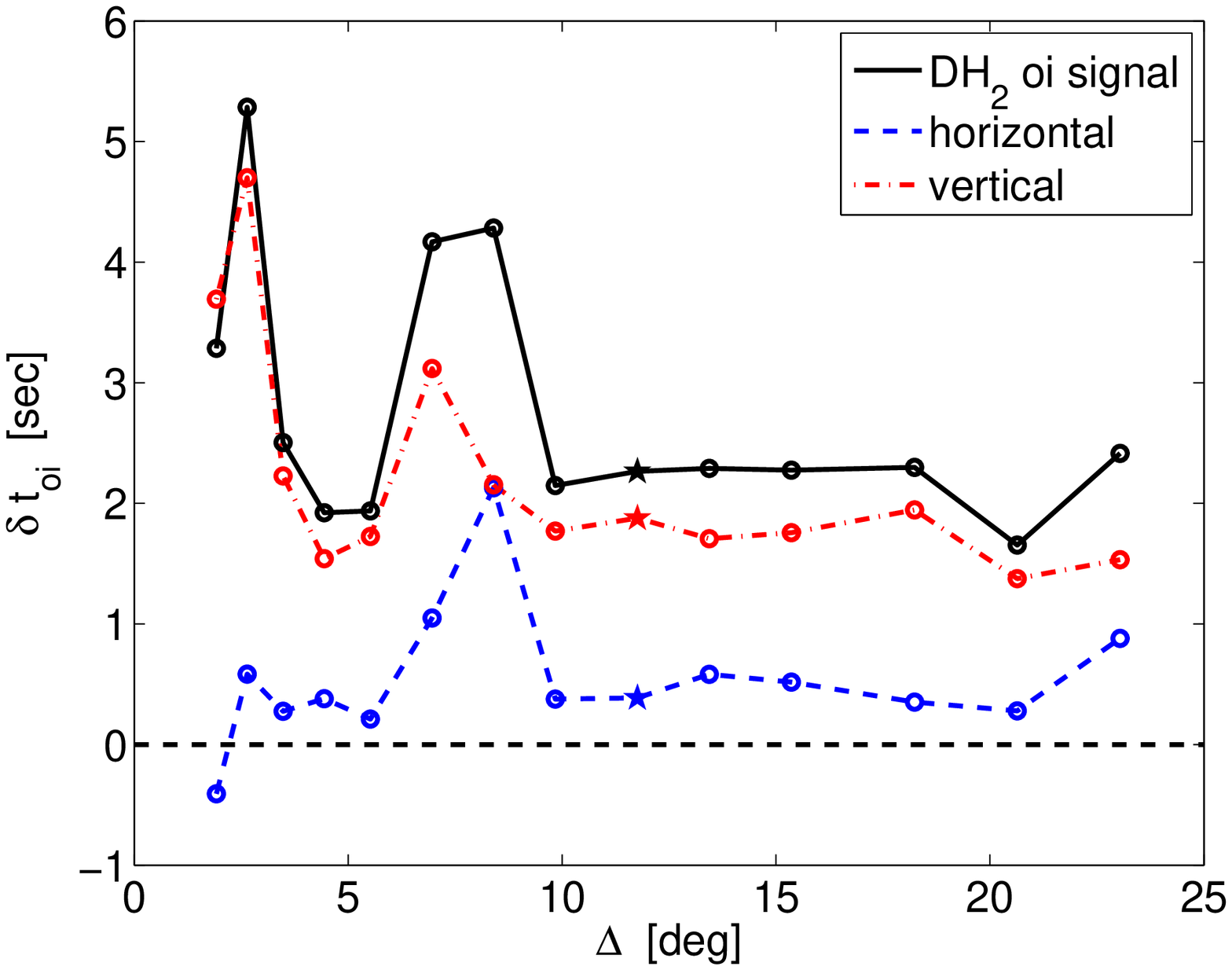} &
\includegraphics[width=0.48\linewidth,clip=]{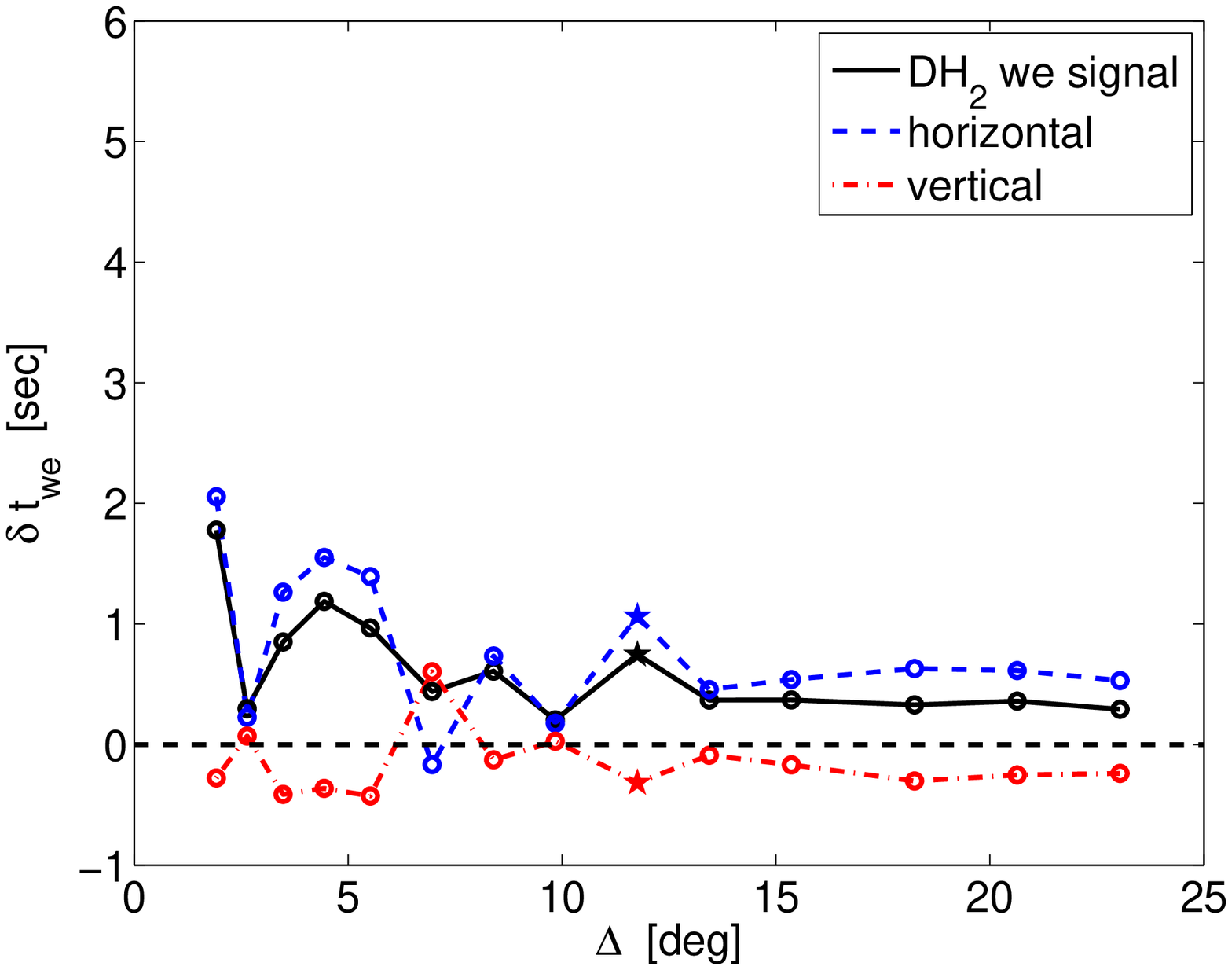} \\
\includegraphics[width=0.48\linewidth,clip=]{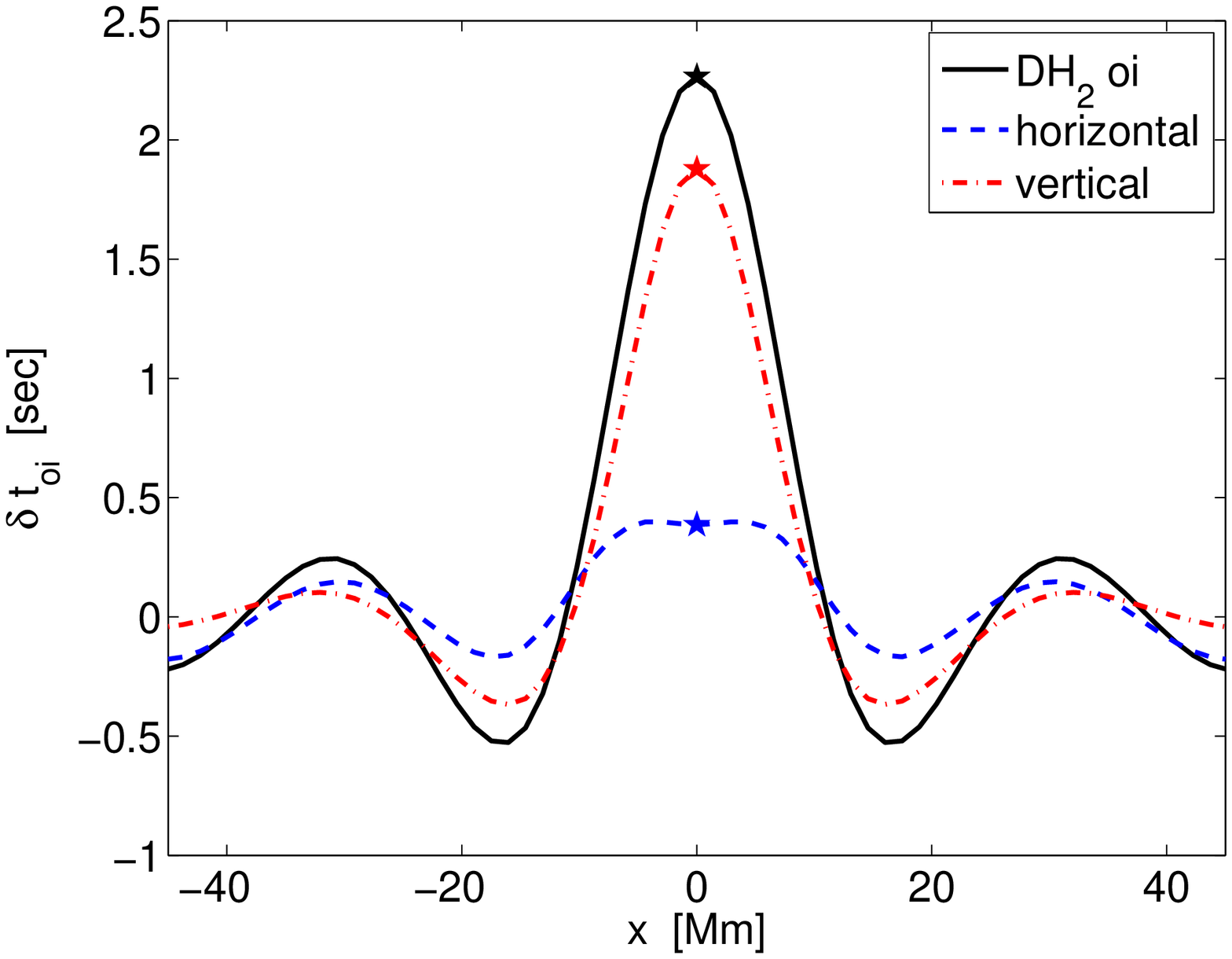} &
\includegraphics[width=0.48\linewidth,clip=]{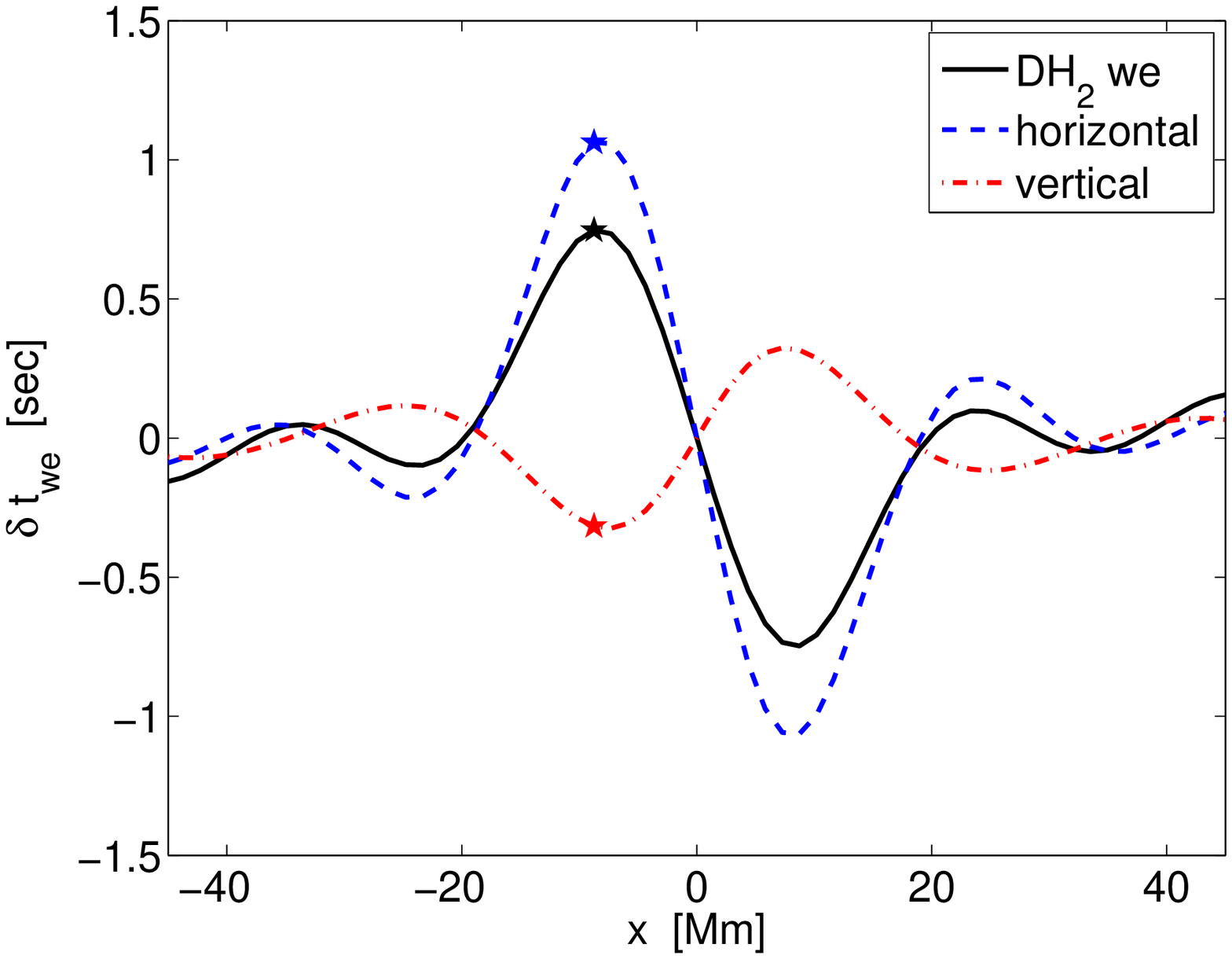}
\end{array}$
\end{center}
\caption{Top row: Horizontal and vertical contributions to all large-distance $\rm{DH}_2$ Born-approximation forward-modeled `oi' (left) and quadrant (right) travel times shown in Figure~\ref{fig:largedist}. Bottom row: Cuts at $y=0$ in the east-west direction through the $\Delta = 11.76^{\circ}$ $\rm{DH}_2$ `oi' (left) and `we' (right) Born-approximation travel-time maps with the individual horizontal and vertical contributions to these times included. These can be directly compared to Figure~$2$ of \citet{duvall2014}. The starred values in the bottom row figures correspond to the starred points in the top row figures.}
\label{fig:largedist_comp}
\end{figure}

Based on these comparisons, it is difficult to draw a firm conclusion about whether or not $\rm{DH}_2$ is indeed a reasonable representation of an average supergranule. It could be that the model is inadequate in light of the disagreement between measured and Born travel times over these distance regimes, or it could be that the Born kernels used in this analysis (those obtained directly from the pipeline, as well as the newly-computed large-distance ones) are not correct.

In terms of the wide-filter results over $\Delta_2$, some disagreement between the measurements and our forward-modeled travel times may be expected, as the computation of these kernels assumes a plane-parallel geometry, neglecting the curvature of the solar surface -- an assumption which admittedly begins to break down at large distances. It is surprising, however, that the Born travel times do not match the data well even at relatively short distances where the deviation from plane-parallel geometry is still quite small.

It is also possible that some mismatch is due to the inability of these kernels to adequately separate the horizontal and vertical flow components at these distances, introducing a type of `cross talk.' To test this, we computed the horizontal and vertical contributions to the `oi' and `we' forward-modeled travel times for each distance. These are shown in the top left and top right panels of Figure~\ref{fig:largedist_comp} respectively. These figures show that the kernels are in fact able to separate the contributions fairly well - in other words, one component contributes to the signal significantly more than the other - particularly for the `oi' times, except in the range $\Delta = 7$--$8^{\circ}$. The bottom left and bottom right panels of Figure~\ref{fig:largedist_comp} show east-west cuts at $y=0$ through the `oi' and `we' maps for one particular distance ($\Delta = 11.76^{\circ}$), along with the individual horizontal and vertical contributions to these travel times. The starred points in the bottom row figures correspond to the starred points in the top row figures. We find that the ray kernels used in Figure~2 of \citet{duvall2014} separate the contributions more fully at these distances than these particular Born kernels do.

\begin{figure}[H]
\centerline{
\includegraphics[width=0.9\textwidth,clip=]{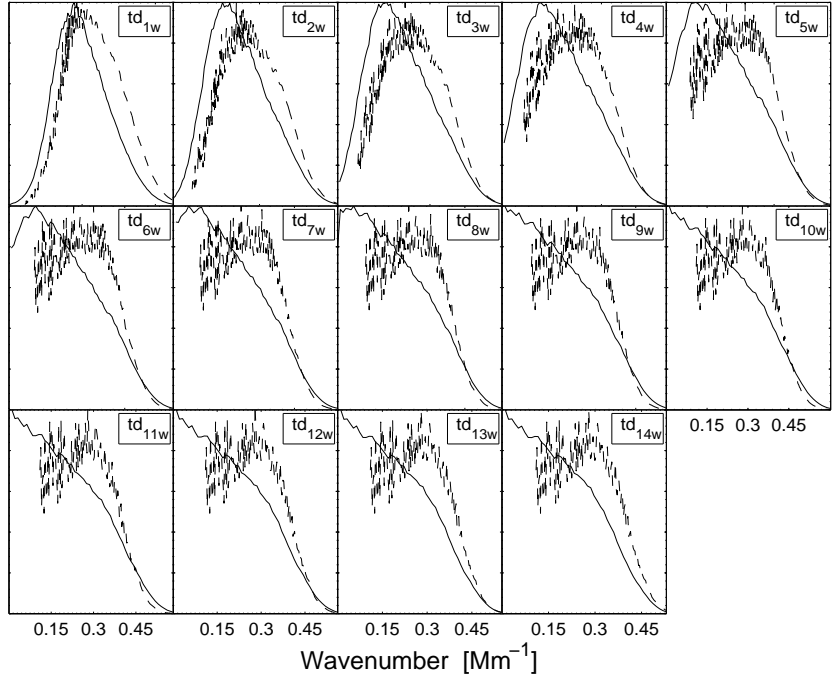}}
\centerline{\small
\hspace{0.48 \textwidth} \color{black}{(a)} \hfill}
\vspace{0.05\textwidth}
\centerline{
\includegraphics[width=0.9\textwidth,clip=]{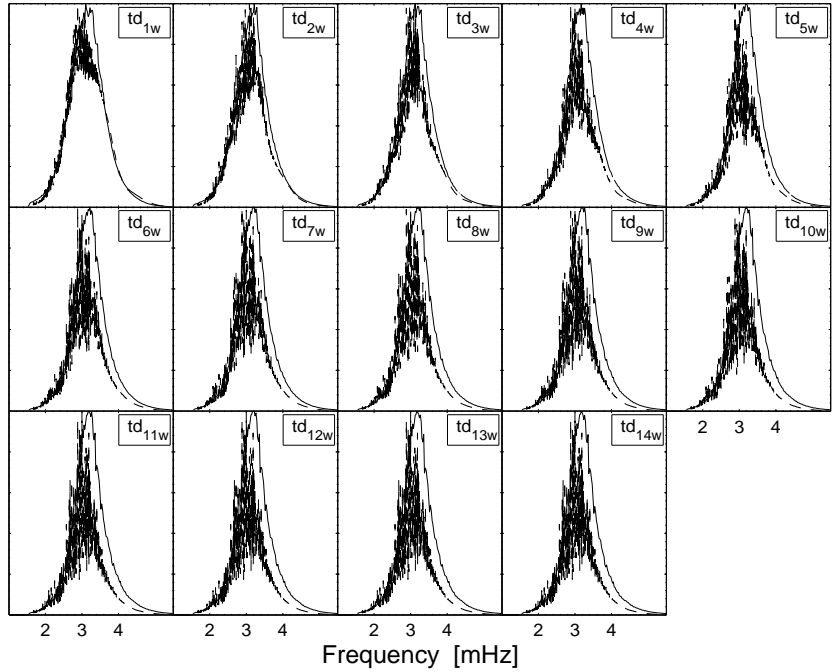}}
\centerline{\small
\hspace{0.48 \textwidth} \color{black}{(b)} \hfill}
\vspace{0.03\textwidth}  % Shift back to the panel bottom
\caption{Filtered data (solid lines) and model (dashed lines) power spectra after integrating over frequency (a) and wavenumber (b) for all large-distance Born kernels used in this analysis. The power has been normalized to unity in each case for easier comparison. Filter number is shown in the upper right-hand corner of each panel. These are the same fourteen filters used in the \citet{duvall2012} study, with larger filter numbers correspond to larger phase-speed values. They correspond to wave travel distances of about $2-24^\circ$, in approximately $2^\circ$ bins.}
\label{fig:intarray}
\end{figure}

The computation of the Born kernels relies on an accurate modeling of the HMI acoustic power spectrum \citep{birch2004a}. This can be quite challenging in some cases, particularly when using filters of higher phase-speed, and the wide filters of \citet{duvall2012}. It is therefore possible that some discrepancy, at least over $\Delta_2$, is coming from an insufficient modeling of this data power. To get a sense for how well the power has been modeled in each case, we compared the filtered data and model power spectra after integrating separately along frequency and wavenumber. Figure~\ref{fig:intarray} shows the integrated data power (solid lines) and model power (dashed lines) plotted together for each kernel. Filter number is given in the upper right-hand corner of each panel. The power has been normalized to unity in each case for easier comparison. We note that the same phase-speed filters are used for both the model and data power spectra. To explore this idea a bit further, we examined our set of wide-filter kernels more closely.

% Figure

%after first applying our $f$-mode and $p_1$ ridge filters (panels [a] and [c] respectively), as well as the standard HMI and wide phase-speed filters (labeled $\rm{td_{n}}$ and $\rm{td_{w}}$, respectively, in panels [b] and [d]). The power has been normalized to unity in each case for easier comparison.

%In most cases, it is usually not too difficult to model the $f$-mode and $p_{1}$ power, and indeed we see that the data and model power spectra agree reasonably well here. It is more difficult to do, however, when using the standard HMI phase-speed filters.

%though is somewhat worse when integrated over wavenumber (panel [d]). The wide phase-speed filters are more difficult still, with a majority of the model power spectra providing a less than ideal match to the data.

In terms of power integrated over frequency (Figure~\ref{fig:intarray}a), the model power is found to deviate most significantly from the data at high wavenumber for the lowest phase-speed filter, $\rm{td}_{1}$. As we look to filters of higher phase-speed, the model power appears to quickly ``pull away" from the data, with the fit becoming progressively worse at nearly all wavenumber values. Beyond filter $\rm{td}_{4}$, the model profiles generally are indistinguishable from one another. In most cases, we are unable to model the power at low wavenumber when using these wide filters.

To some extent, the agreement between data and model power appears to be slightly better when integrated over wavenumber (Figure~\ref{fig:intarray}b), though we do have difficulty getting good agreement between the spectra at higher frequency in most cases. We find that the profiles for each of the filters exhibit a strong oscillatory behavior, which is also present in Figure~\ref{fig:intarray}a, though to a somewhat lesser degree. Increasing the number of frequency and wavenumber model grid points usually helps to counteract this, however this can slow kernel computation time down considerably, and, after a certain point, starts to become impractical. We have currently not tested what overall effect this behavior has on the computation of forward-modeled travel times.

These types of figures are useful for understanding modeling uncertainties and are not always provided in publications. Unfortunately, we do not have available the HMI pipeline model power spectra with which to compare.

\subsection{The Best-Fit Flow Models}
  \label{bestfit}
In light of the fact that the best \citet{duvall2012} flow model is in disagreement with the forward travel times in the Born approximation, we were interested in seeing what kind of supergranule features could provide the best fit to the measurements made over the separate $\Delta_1$ and $\Delta_2$ regimes, under the assumption that both sets of Born kernels are accurate. Using the continuity equation and a Model~S \citep{jcd1996} density profile, $512,000$ Duvall \& Hanasoge-style flow models (i.e. mass-conserving models with a Gaussian $v_z$ flow profile) of varying peak $v_z$ amplitude, peak $v_z$ depth, and $v_z$ FWHM were produced in a Monte Carlo-like fashion. Imposing observational constrains on these models to have a surface $v_z(z=0) \approx 10~\rm{m\,s^{-1}}$ and a surface $v_{x,y}^{\rm{RMS}}(z=0) = 100$--$500~\rm{m\,s^{-1}}$ resulted in a total of 590 ``viable" flow models.

% Figure
\begin{figure}[t!]
\begin{center}$
\begin{array}{cc}
\includegraphics[width=0.48\linewidth,clip=]{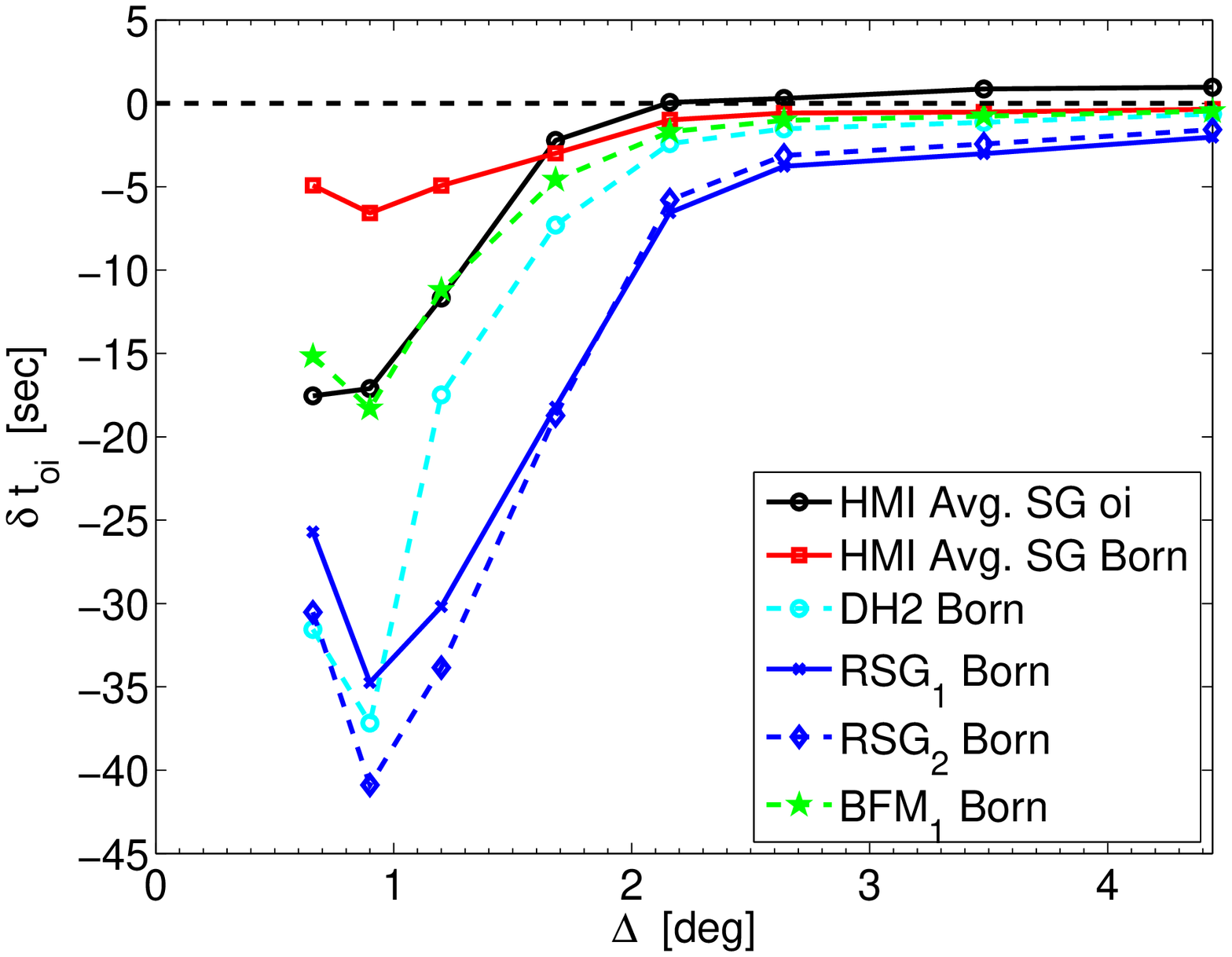}
\includegraphics[width=0.48\linewidth,clip=]{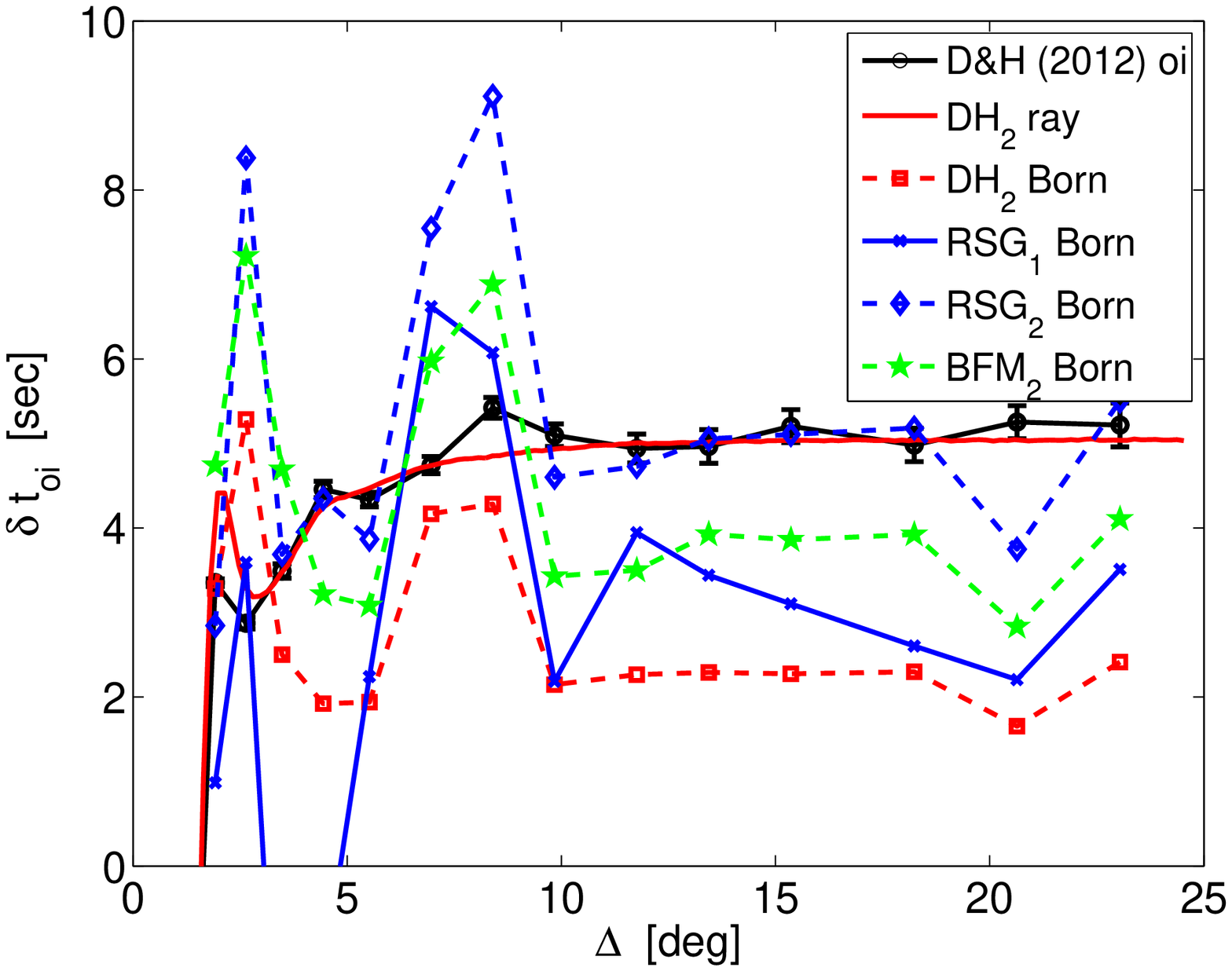}
\end{array}$
\end{center}
%\caption{Top row: Short-distance $f$-mode (left) and $p_1$ (right) `oi' travel-time comparison for all flow features. The top row figures share the same legend. Bottom row: Large-distance `oi' (left) and quadrant (right) travel-time comparison for all flow features. These plots are essentially identical to Figures~\ref{fig:largedist} and \ref{fig:shortdist}, only they now include the BFM (dashed-star lines) and $\rm{RSG}_2$ (dashed-diamond lines) values for comparison. Measured travel times are denoted by the D\&H curves.}
\caption{Short (left) and large-distance (right) `oi' travel-time comparison for all flow features. These plots are identical to Figures~\ref{fig:hmifigure} and \ref{fig:largedist}, only they now include the BFMs (dashed-star lines), $\rm{RSG}_1$, and $\rm{RSG}_2$ (these features are discussed in Section~\ref{rempelflow}) values for comparison. The measured travel times in the left and right-hand figures are denoted by the `HMI Avg. SG oi' and `D\&H (2012) oi' curves respectively. Recall the left (right) plot implies the use of narrow (wide) phase-speed filters.}
\label{fig:best_plots}
\end{figure}

Forward-modeled Born travel times were computed for these flow models over $\Delta_1$ and $\Delta_2$. The best-fit models to all short and large-distance `oi' measurements (in a least-squares sense, referred to as $\rm{BFM_1}$ and $\rm{BFM_2}$, respectively, hereafter) were identified, and have the following parameters:
\\\\
$\rm{BFM_1}$:\\
$v_z(z=0)=9~\mathrm{m\,s^{-1}}$ and $v_z^{\mathrm{max}}(z=-1.7~\mathrm{Mm})=154~\rm{m\,s^{-1}}$, $\mathrm{FWHM}_z=1.7~\mathrm{Mm}$\\
$v_{\mathrm{h}}(z=0)=148~\mathrm{m\,s^{-1}}$ and $v_{\mathrm{h}}^{\mathrm{max}}(z=-1.2~\mathrm{Mm})=559~\mathrm{m\,s^{-1}}$
\\\\
$\rm{BFM_2}$:\\
$v_z(z=0)=14~\mathrm{m\,s^{-1}}$ and $v_z^{\mathrm{max}}(z=-3.6~\mathrm{Mm})=352~\mathrm{m\,s^{-1}}$, $\mathrm{FWHM}_z=3.3~\mathrm{Mm}$\\
$v_{\mathrm{h}}(z=0)=135~\mathrm{m\,s^{-1}}$ and $v_{\mathrm{h}}^{\rm{max}}(z=-2.5~\mathrm{Mm})=786~\mathrm{m\,s^{-1}}$
\\\\
Depth cuts through the two models are shown in Figure~\ref{fig:1Dprofile} along with $\rm{DH}_2$ and the average HMI supergranule. $\rm{BFM}_1$ is shallower and of lower amplitude than $\rm{DH}_2$, while $\rm{BFM}_2$ is of larger amplitude and is much more extended in depth.

% Figure
\begin{figure}[t!]
\begin{center}$
\begin{array}{c}
\includegraphics[width=0.98\linewidth,clip=]{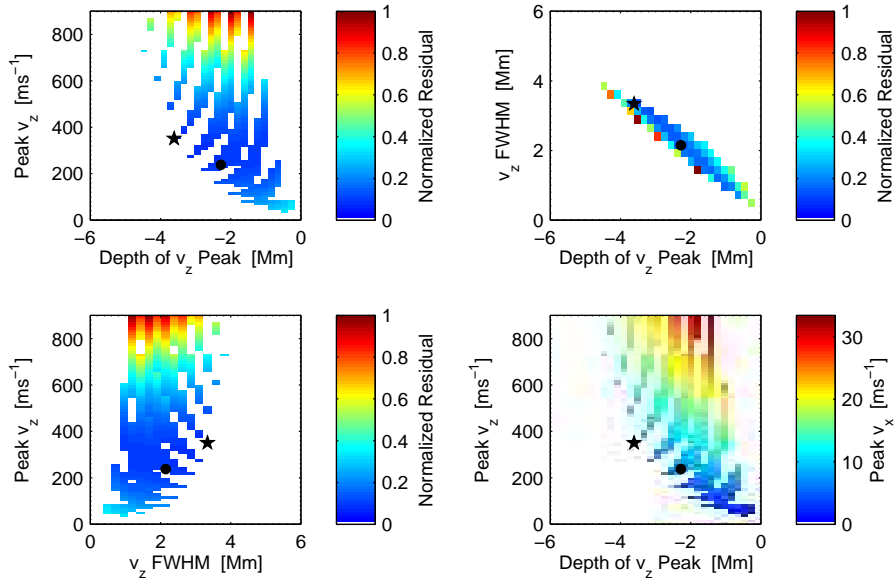}
\end{array}$
\end{center}
\caption{Figures (a), (b), and (c) show the sum of the residuals between the $\Delta_2$ measured and forward-modeled `oi' travel times for each of the $590$ flow models as a function of the model parameters. These have been normalized to the largest residual value in each panel. Small residuals denote flow models that more closely approximate the measurements. These figures serve to illustrate the degeneracies between many of the models. Panel (d) shows the peak $v_x$ velocity for each of the models, with values corresponding to the color bar. The color bar values have been divided by a factor of $100$ for easier viewing. White areas in each panel represent regions where no viable models (i.e. ones obeying the surface flow constraints) were identified. The circle and star points mark the locations of $\rm{DH}_2$ and $\rm{BFM_2}$, respectively, in this parameter space.}
\label{fig:paramspace}
\end{figure}
\label{fig:paramspace}

The resulting forward travel times from these models are shown in Figure~\ref{fig:best_plots}. These plots are identical to Figures~\ref{fig:hmifigure} and \ref{fig:largedist}, only they now have the additional BFM values included for comparison (dashed-star lines). We find that the models do in fact provide better fits to the measurements when compared to the $\rm{DH}_2$ Born-approximation travel times, especially with regard to the $\Delta_1$ measurements (left). The fit over $\Delta_2$ (right), however, is still quite poor. Rather than comparing only cell-center `oi' travel time values, we also compared the full measured and modeled travel-time maps when available. This was only possible over $\Delta_1$, as we did not have available the full Duvall \& Hanasoge travel-time maps with which to compare. The travel-time maps from each model were compared separately to those of the averaged HMI measurements for each filter. For each comparison, the relative error was computed by finding the RMS difference between maps.
%\begin{equation}
%  \mathrm{Error}=\mathrm{RMS}[\tau_{\rm model}-\tau_{\rm measured}]
%\label{RMS} 
%\end{equation}
Overall, we find no difference in terms of which model best fits the data when comparing the full travel-time maps versus using only the cell-center values as shown in Figure~\ref{fig:best_plots} (left).

Experimenting with many different flow models like these provides a few interesting conclusions. We find that the model that best approximates the measured travel times over $\Delta_1$ does not fit the $\Delta_2$, wide phase-speed measured times well, and vice versa. It is very difficult to simultaneously get a reasonable fit over both distance regimes with a single model, which is not surprising given the different filtering. It also appears that the parameter space is somewhat degenerate in the sense that there are many flow models for which the fit to the measured travel times is nearly equally good. These models are often quite dissimilar in peak $v_z$ amplitude, peak depth, and $\mathrm{FWHM}_z$. To illustrate this, Figure~\ref{fig:paramspace} shows the sum of the residuals between measured and forward-modeled `oi' travel times over $\Delta_2$ for each of the $590$ flow models as a function of these parameters. These have been normalized to the largest residual value in each panel, with the smallest residual values representing models that are able to more closely match the measurements. White areas represent regions where no viable models (i.e. ones obeying the surface flow constraints) were identified. The circle and star points mark the locations of $\rm{DH}_2$ and $\rm{BFM}_2$, respectively, in this parameter space. The overall difference between them is not substantial.

% Figure
\begin{figure}[t!]
\begin{center}$
\begin{array}{cc}
\includegraphics[width=0.49\linewidth,clip=]{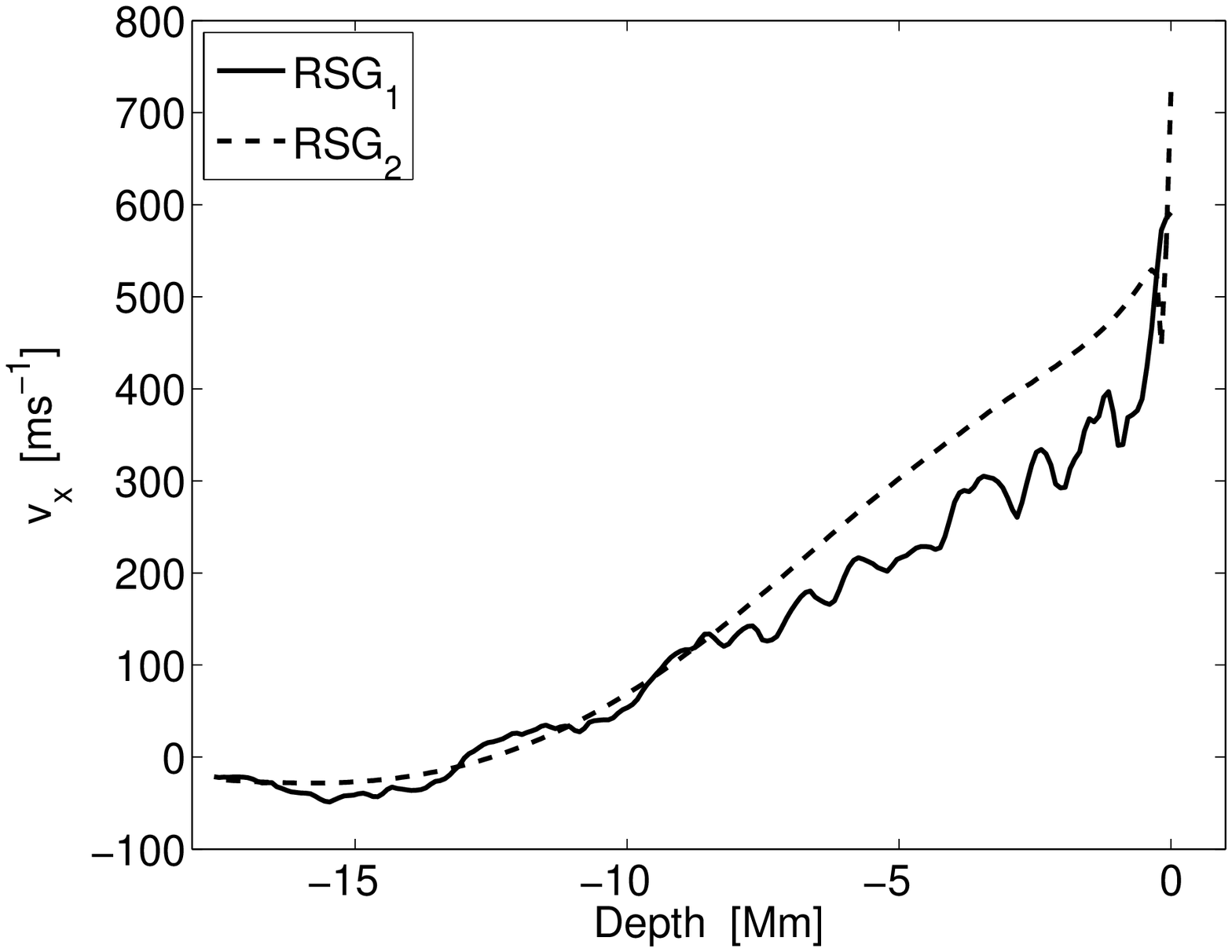}
\includegraphics[width=0.49\linewidth,clip=]{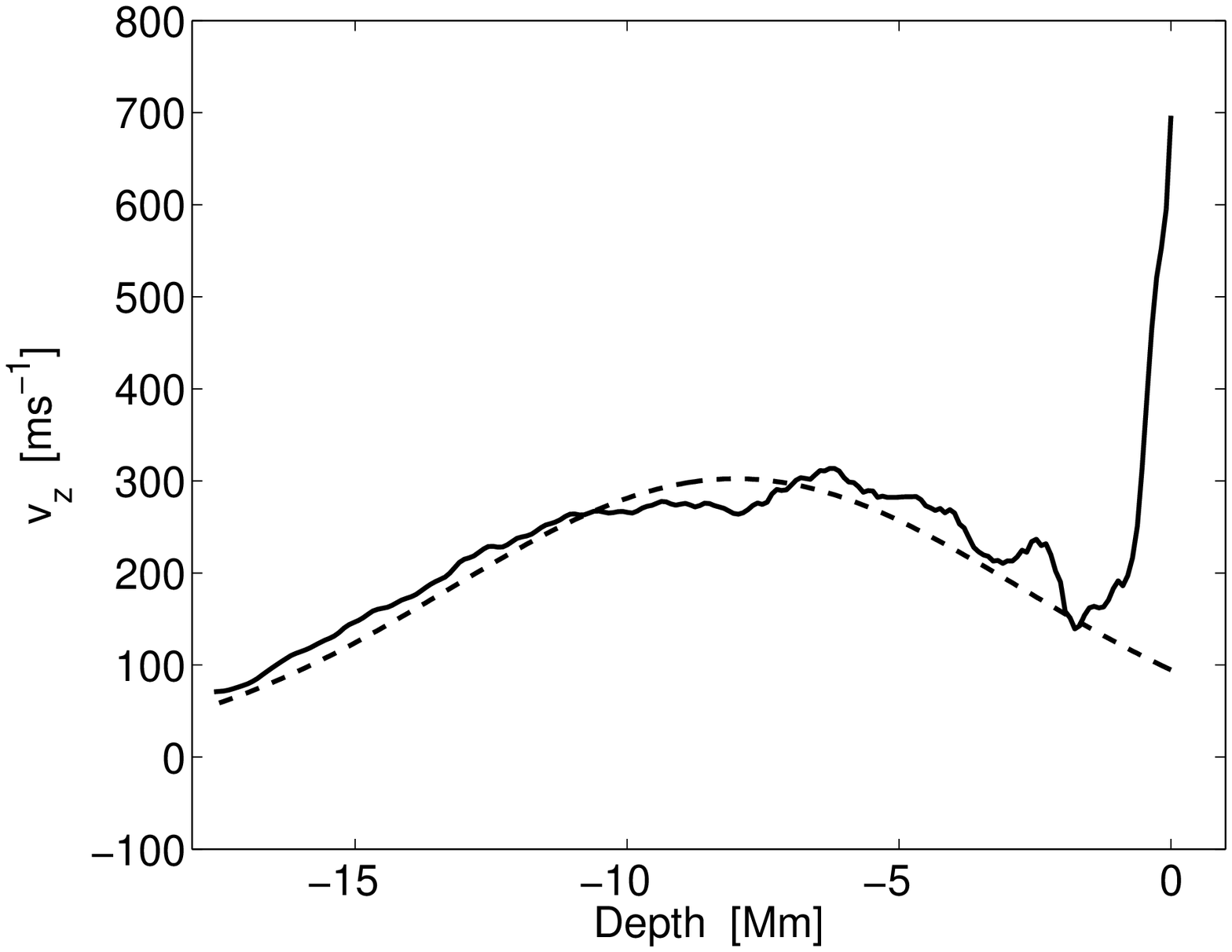}
\end{array}$
\end{center}
\caption{One-dimensional cuts in depth through the $\rm{RSG}_1$ and $\rm{RSG}_2$ $v_x$ (left) and $v_z$ (right) flow components. These profiles are taken through the points of maximum surface $v_x$ and $v_z$, respectively, for the two flow fields.}
\label{fig:fp1}
\end{figure}

\section{Simulated Supergranule-Sized Flow Feature}
	\label{rempelflow}
It is possible to derive an alternative model of supergranulation from the large-scale flow structures present in recent realistic magnetohydrodyamic quiet-Sun simulations \citep[e.g.][]{rempel2014}. Having already analyzed one such simulation previously \citep[QS2 in][]{degrave2014}, these data were already available. The near-surface flow field of this simulation was smoothed with a two-dimensional Gaussian function of $\sigma=4$~Mm to help remove small-scale fluctuations and accentuate the regions of large-scale outflow. Points of maximum divergence were then located from the flows, thereby marking the positions of outflow centers. Averages were taken about seven of these features to produce a single average ``supergranule" (referred to as $\rm{RSG}_1$ hereafter). Cuts through the $v_x$ and $v_z$ flow components of $\rm{RSG}_1$ are shown in Figure~\ref{fig:fp1} (solid lines). These can be directly compared to the HMI, $\rm{DH}_2$, and BFM profiles in Figure~\ref{fig:1Dprofile}. The $\rm{RSG}_1$ feature is clearly very different from the others in its overall flow structure, particularly its deeper extent in depth to at least 15~Mm.

%As a test, we wanted to see if it was possible to closely reproduce the $\rm{RSG}_1$ feature with a mass-conserving flow model derived using a variation of the \citet{duvall2012} models. A model (referred to as $\rm{RSG}_2$ hereafter) was chosen to give a satisfactory match to the $\rm{RSG}_1$ $v_z$ profile in depth, with the $v_x$ profile being determined completely by the choice of $v_z$. Figure~\ref{fig:fp1} shows cuts through the $v_x$ and $v_z$ flow components of $\rm{RSG}_2$ (dashed lines). This model does a reasonably good job of reproducing the overall amplitude of the $\rm{RSG}_1$ flow components as they vary in depth, though is unable to reproduce the large-amplitude spike in $v_z$ that the feature exhibits near the surface, and also tends to overestimate the near-surface $v_x$ flows to some degree.

The supergranule obtained this way does not adequately satisfy the continuity equation (see Table~\ref{mctab}, RSG$_1$). To determine if this is one of the reasons for the different flow profile, we used RSG$_1$ as a starting point to derive a fully mass-conserving flow model like in \citet{duvall2012}. A model (referred to as $\rm{RSG}_2$ hereafter) was chosen to give a satisfactory match to the $\rm{RSG}_1$ $v_z$ profile in depth, with the $v_x$ profile being determined completely by the choice of $v_z$. Figure~\ref{fig:fp1} shows cuts through the $v_x$ and $v_z$ flow components of $\rm{RSG}_2$ (dashed lines). The model matches reasonably well the overall amplitude of the $\rm{RSG}_1$ flow components as they vary in depth, though does not have the large-amplitude spike in $v_z$ that $\rm{RSG}_1$ exhibits near the surface. It also has stronger near-surface horizontal flows.

Forward-modeled travel times were computed for $\rm{RSG}_1$ and $\rm{RSG}_2$ in the Born approximation over both $\Delta_1$ and $\Delta_2$ distance ranges, with the results plotted in Figure~\ref{fig:best_plots}. Over $\Delta_1$, the two features show similar values at all distances. However, they are generally unable to approximate the HMI measurements well, being larger in amplitude by a factor of $1.5$--$2.5$ at small $\Delta$. This is not too surprising given the very strong near-surface horizontal flows. We note that unlike the statistically-averaged HMI measurements, none of the models produce positive `oi' cell-center travel-time differences over the larger distances of $\Delta_1$.

Over $\Delta_2$ (Figure~\ref{fig:best_plots}, right), $\rm{RSG}_1$ and $\rm{RSG}_2$ travel times show larger disagreement between each other, however, $\rm{RSG}_2$ matches the measured travel times at nearly all distances. Like $\rm{BFM}_2$, $\rm{RSG}_1$ and $\rm{RSG}_2$ tend to overshoot the measurements at shorter distances, and also exhibit the anomalous peak between $\Delta = 7$--$8^{\circ}$ that the other models also show. Though not shown in Figure~\ref{fig:best_plots}, the `we' and `ns' quadrant forward-modeled times for $\rm{RSG}_1$, $\rm{RSG}_2$ and $\rm{BFM}_2$ show no significant improvement over the $\rm{DH}_2$ Born values in Figure~\ref{fig:largedist}.

% CONCLUSION
\section{Conclusions}
	\label{conclusions}
HMI time-distance pipeline velocity data products suggest that the ``average" supergranule exhibits the following structure:

\begin{itemize}
\item A radial outflow down to a depth of $7$--$10$~Mm
\item An transition from outflow to inflow at a depth of $7$--$10$~Mm
\item A weak inflow flow extending down to a depth of $\sim20$~Mm
\end{itemize}
These properties differ significantly from the shallow model of supergranulation proposed by \citet{duvall2012}, and further supported by \citet{duvall2014}. The near-surface RMS velocity of the average HMI feature is also found to be considerably smaller (by an order of magnitude) than the values typically associated with solar supergranulation. Much of the overall HMI supergranule structure and its weak flow amplitudes can be explained by examining the HMI pipeline averaging kernels for the near-surface inversions, which are found to be very broad in depth, and nearly identical in terms of sensitivity along the $z$-direction. Additional reduction in flow amplitude could be attributed, at least in part, to the use of narrow standard phase-speed filters in the analysis, along with the smoothing of recovered flows inherent in time-distance inversions. A convolution between the HMI near-surface $v_x$ inversion averaging kernel with model $\rm{DH_2}$ resulted in a $50\%$ reduction in $v_x$ RMS flow amplitude.

To identify possible sources of discrepancy between the data and model, we studied forward-modeled travel-time differences in the Born approximation to compare with the HMI measurements. Using kernels computed with the standard (narrow) phase-speed filters in the HMI pipeline over distances $\Delta_1 = 0.5$--$4.5^{\circ}$, the model from \cite{duvall2012}, which was originally derived in a ray theory approach, does not match the measured HMI travel times for an average supergranule (Figure~\ref{fig:hmifigure}).

At larger distances ($\Delta_2 = 2$--$24^{\circ}$), travel-time measurements exist only from \cite{duvall2012} who employed wider phase-speed filters. Born kernels computed using these filters also do not reproduce what ray theory predicts for the shallow supergranule model (Figure~\ref{fig:largedist}). It is unclear if it is appropriate to compute Born kernels with such filters, due to potential modeling inaccuracies over such a large region of the power spectrum. For these rather small travel-time differences, corrections due to the curvature of the Sun that are not taken into account here could become important too. 

When we instead assume that the Born kernels computed with short-distance, standard phase-speed filters and the large-distance, wide phase-speed filters are correct, we determined a best-fit supergranulation model to the measurements over each of those distance ranges. The resulting supergranule is different in each case (Figure~\ref{fig:vxvz}). The \citet{duvall2012} model falls somewhere between these two solutions in terms of amplitude and depth variations.

Another representation of a supergranule was derived by averaging over large-scale flow features in a realistic 3D numerical simulation. Minor modifications to the flow profile to make it fully satisfy the continuity equation were made. The structure has very strong flows that extend to much larger depths than the HMI or \citet{duvall2012} supergranules. In the Born approximation, travel times computed from this supergranule only match the data reasonably well over a small range at large distances. 

%This feature was also modeled with a mass-conserving \citet{duvall2012}-style flow model. Despite strong similarities in their overall flow structure, some disagreement was found between their $\Delta_2$ forward-modeled travel times, though they agreed well over $\Delta_1$. Overall, they did not approximate the measurements as well as the best-fit models.

%As a final comment, it is important to realize when carrying out the kind of statistical averaging scheme presented in this work that helioseismic inversions for subsurface flows can be very unreliable at depth. This is especially true when inverting for the vertical velocity component, $v_z$. As mentioned in Section~\ref{intro}, for example, time-distance inversions for flows in a quiet-Sun simulation by \citet{degrave2014} reported the existence of spurious flows at depths below $7$~Mm which did not actually exist in the data. If a similar problem exists with the HMI data products, it is possible that averages have been taken over flow signal that is not real, but is instead a product of flow inversions being carried out where signal-to-noise is low, and where wave coverage is poor.

It is quite possible that neither the \citet{duvall2012} flow model nor the average HMI feature are accurate representations of an actual supergranule. The latter may be especially true in light of the recent work by \citet{hanasoge2014} where it was found that, even in the case of an ideal inversion scenario, model supergranule flows could not be accurately recovered. \citet{svanda2015} also suggested that inaccuracies in the forward problem may be amplified in the inversions process, leading to erroneous subsurface flows recovered through time-distance helioseismology. In any case, we find that the supergranules implied from three different approaches are not consistent with each other. The one most closely in agreement with past surface measurements (by construction), as well as large-distance travel-time differences interpreted from ray theory, is the model proposed by \citet{duvall2012}. More work needs to be done to show that the Born approximation can be used to determine this structure.

\begin{acks}
The authors gratefully acknowledge past support by the NASA SDO Science Center through contract NNH09CE41C awarded to NWRA, and helpful discussions with Junwei Zhao, Tom Duvall~J.r., and Matthias Rempel. The data used here are courtesy of NASA/SDO and the HMI Science Team, whom we thank for their dedicated work. K.D. also acknowledge funding from NSF award NSF/AGS-1351311 and sub-award AURA/NSO No. N06504C-N.
\end{acks}

% MASS CONSERVATION
\appendix
\section{Mass Conservation of the Various Flow Features}
\label{mass}
The average HMI and $\rm{RSG}_1$ features were checked to see how well their flows conserve mass relative to the various models described throughout this work. We do this by expanding the continuity equation (term [A] in Equation~\ref{eq:cont}) into its separate horizontal and vertical components (terms [B] and [C] respectively) assuming a time-independent Model~S density profile \citep{jcd1996} which varies only in the $z$-direction:
\begin{equation}
\underbrace{\nabla \cdot (\rho \vec{v}) \vphantom{\frac{\partial (\rho v_z)}{\partial z}}}_{[\rm{A}]} \equiv \underbrace{\nabla_{\rm{h}} \cdot (\rho v_{\rm{h}}) \vphantom{\frac{\partial (\rho v_z)}{\partial z}}}_{[\rm{B}]} + \underbrace{\frac{\partial (\rho v_z)}{\partial z}}_{[\rm{C}]} = 0.
\label{eq:cont}
\end{equation}
The model supergranules are constructed analytically to be mass conserving, though evaluate numerically to some small non-zero value due to numerical gradients and machine precision. As such, we check mass conservation by calculating terms [B] and [C] within a $30$~Mm radius of cell center, and integrating them along all three spatial dimensions. Their ratios are then taken with respect to [$\rm{B}+\rm{C}$]. In the case of a mass-conserving flow, the ratios ($\rm{B}+\rm{C}$)/B and ($\rm{B}+\rm{C}$)/C should be of the same magnitude and of opposite sign. The mass conservation ratios for each feature are presented in Table~\ref{mctab}. As expected, $\rm{DH}_2$, $\rm{RSG}_2$, and the BFM all satisfy continuity to a high precision. The HMI and $\rm{RSG}_1$ supergranules, to varying degrees, do not.

\begin{table}[t!]
\caption{Mass conservation ratios for the HMI average supergranule, $\rm{DH}_2$, $\rm{RSG}_1$, $\rm{RSG}_2$, and the BFM. Parameters B and C are the spatially-integrated values computed from the expanded continuity equation in Equation~\ref{eq:cont}.} % and have units of $\rm{kg}\,m^{-3}\,s^{-1}$.}
\label{mctab}
\begin{tabular}{lccc} % centered columns (4 columns)
\hline
Feature & ($\rm{B}+\rm{C}$)/B & ($\rm{B}+\rm{C}$)/C\\
\hline
\multicolumn{1}{l}{HMI}              & $-3.55\times10^{-1}$ & $2.62\times10^{-1}$\\
\multicolumn{1}{l}{$\rm{DH}_2$}       & $1.54\times10^{-4}$ & $-1.54\times10^{-4}$\\
\multicolumn{1}{l}{$\rm{BFM}_1$}       & $1.36\times10^{-4}$ & $-1.36\times10^{-4}$\\
\multicolumn{1}{l}{$\rm{BFM}_2$}       & $-1.55\times10^{-4}$ & $1.55\times10^{-4}$\\
\multicolumn{1}{l}{$\rm{RSG}_1$}       & $-3.04\times10^{-2}$ & $2.95\times10^{-2}$\\
\multicolumn{1}{l}{$\rm{RSG}_2$}       & $1.12\times10^{-5}$ & $-1.12\times10^{-5}$\\
\hline
\end{tabular}
\end{table}

\clearpage
% format of references provided by the journal (.bst)
%\bibliographystyle{spr-mp-sola}
\bibliographystyle{spr-mp-sola-limited} %% Alternative style: no title,

   % name your Bibtex file containing your references (.bib)
\tracingmacros=2
\bibliography{sola_bibliography_example}

\begin{thebibliography}{33}
% BibTex style file: spr-mp-sola-limited.bst (nameyear,cnd), 2014-01-28
\ifx\bisbn     \undefined \def\bisbn  #1{ISBN #1}\fi
\ifx\binits    \undefined \def\binits#1{#1}\fi
\ifx\bauthor   \undefined \def\bauthor#1{#1}\fi
\ifx\batitle   \undefined \def\batitle#1{#1}\fi
\ifx\bjtitle   \undefined \def\bjtitle#1{\textit{#1}}\fi
\ifx\bvolume   \undefined \def\bvolume#1{\textbf{#1}}\fi
\ifx\byear     \undefined \def\byear#1{#1}\fi
\ifx\bissue    \undefined \def\bissue#1{#1}\fi
\ifx\bfpage    \undefined \def\bfpage#1{#1}\fi
\ifx\blpage    \undefined \def\blpage #1{#1}\fi
\ifx\burl      \undefined \def\burl#1{\textsf{#1}}\fi
\ifx\href      \undefined \def\href#1#2{\textsf{#2}}\fi
\ifx\betal     \undefined \def\betal{\textit{et al.}}\fi
\ifx\bctitle   \undefined \def\bctitle#1{#1}\fi
\ifx\beditor   \undefined \def\beditor#1{#1}\fi
\ifx\bbtitle   \undefined \def\bbtitle#1{\textit{#1}}\fi
\ifx\bedition  \undefined \def\bedition#1{#1}\fi
\ifx\bseriesno \undefined \def\bseriesno#1{\textbf{#1}}\fi
\ifx\blocation \undefined \def\blocation#1{#1}\fi
\ifx\bsertitle \undefined \def\bsertitle#1{\textit{#1}}\fi
\ifx\bsnm      \undefined \def\bsnm#1{#1}\fi
\ifx\bsuffix   \undefined \def\bsuffix#1{#1}\fi
\ifx\bparticle \undefined \def\bparticle#1{#1}\fi
\ifx\barticle  \undefined \def\barticle#1{}\fi
\ifx\binstitute  \undefined \def\binstitute#1{#1}\fi
\ifx\bpublisher  \undefined \def\bpublisher#1{#1}\fi
\ifx\doiurl    \undefined
  \def\doiurl#1{\href{http://dx.doi.org/#1}{\textsf{DOI}}}\fi
\ifx\arxivurl  \undefined
  \def\arxivurl#1{\href{http://arxiv.org/abs/#1}{\textsf{arXiv}}}\fi
\ifx\adsurl    \undefined
  \def\adsurl#1{\href{http://adsabs.harvard.edu/abs/#1}{\textsf{ADS}}}\fi
\ifx\botherref \undefined \def\botherref#1{}\fi
\ifx\url       \undefined \def\url#1{\textsf{#1}}\fi
\ifx\bchapter  \undefined \def\bchapter#1{}\fi
\ifx\bbook     \undefined \def\bbook#1{}\fi
\ifx\bcomment  \undefined \def\bcomment#1{#1}\fi
\ifx\oauthor   \undefined \def\oauthor#1{#1}\fi
\ifx\citeauthoryear \undefined\def \citeauthoryear#1{#1}\fi
\def\endbibitem {}
\ifx\bconflocation  \undefined \def\bconflocation#1{#1} \fi

\bibitem[\protect\citeauthoryear{{Benson}, {Stein}, and
  {Nordlund}}{2006}]{benson2006}
\begin{bchapter}
\bauthor{\bsnm{{Benson}}, \binits{D.}},
\bauthor{\bsnm{{Stein}}, \binits{R.}},
\bauthor{\bsnm{{Nordlund}}, \binits{{\AA}.}}:
\byear{2006},
In: \beditor{\bsnm{{Leibacher}}, \binits{J.}},
\beditor{\bsnm{{Stein}}, \binits{R.F.}},
\beditor{\bsnm{{Uitenbroek}}, \binits{H.}} (eds.)
\bbtitle{Solar MHD Theory and Observations: A High Spatial Resolution
  Perspective},
\bsertitle{Astronomical Society of the Pacific Conference Series}
\bseriesno{354},
\bfpage{92}.
\adsurl{2006ASPC..354...92B}.
\end{bchapter}
\endbibitem

\bibitem[\protect\citeauthoryear{{Birch} and {Gizon}}{2007}]{birch2007}
\begin{barticle}
\bauthor{\bsnm{{Birch}}, \binits{A.C.}},
\bauthor{\bsnm{{Gizon}}, \binits{L.}}:
\byear{2007},
\bjtitle{Astronomische Nachrichten}
\bvolume{328},
\bfpage{228}.
\doiurl{10.1002/asna.200610724}.
\adsurl{2007AN....328..228B}.
\end{barticle}
\endbibitem

\bibitem[\protect\citeauthoryear{{Birch}, {Kosovichev}, and
  {Duvall}}{2004}]{birch2004a}
\begin{barticle}
\bauthor{\bsnm{{Birch}}, \binits{A.C.}},
\bauthor{\bsnm{{Kosovichev}}, \binits{A.G.}},
\bauthor{\bsnm{{Duvall}}, \binits{T.L.} \bsuffix{Jr.}}:
\byear{2004},
\bjtitle{\apj}
\bvolume{608},
\bfpage{580}.
\doiurl{10.1086/386361}.
\adsurl{cgi-bin/nph-bib_query?bibcode=2004ApJ...608..580B&db_key=AST}.
\end{barticle}
\endbibitem

\bibitem[\protect\citeauthoryear{{Birch} \textit{et~al.}}{2006}]{birch2006}
\begin{bchapter}
\bauthor{\bsnm{{Birch}}, \binits{A.}},
\bauthor{\bsnm{{Duvall}}, \binits{T.L.}},
\bauthor{\bsnm{{Gizon}}, \binits{L.}},
\bauthor{\bsnm{{Jackiewicz}}, \binits{J.}}:
\byear{2006},
In: \bbtitle{AAS/Solar Physics Division Meeting \#37},
\bsertitle{Bulletin of the American Astronomical Society}
\bseriesno{38},
\bfpage{224}.
\adsurl{2006SPD....37.0505B}.
\end{bchapter}
\endbibitem

\bibitem[\protect\citeauthoryear{{Christensen-Dalsgaard}
  \textit{et~al.}}{1996}]{jcd1996}
\begin{barticle}
\bauthor{\bsnm{{Christensen-Dalsgaard}}, \binits{J.}},
\bauthor{\bsnm{{Dappen}}, \binits{W.}},
\bauthor{\bsnm{{Ajukov}}, \binits{S.V.}},
\bauthor{\bsnm{{Anderson}}, \binits{E.R.}},
\bauthor{\bsnm{{Antia}}, \binits{H.M.}},
\bauthor{\bsnm{{Basu}}, \binits{S.}},
\bauthor{\bsnm{{Baturin}}, \binits{V.A.}},
\bauthor{\bsnm{{Berthomieu}}, \binits{G.}},
\bauthor{\bsnm{{Chaboyer}}, \binits{B.}},
\bauthor{\bsnm{{Chitre}}, \binits{S.M.}},
\bauthor{\bsnm{{Cox}}, \binits{A.N.}},
\bauthor{\bsnm{{Demarque}}, \binits{P.}},
\bauthor{\bsnm{{Donatowicz}}, \binits{J.}},
\bauthor{\bsnm{{Dziembowski}}, \binits{W.A.}},
\bauthor{\bsnm{{Gabriel}}, \binits{M.}},
\bauthor{\bsnm{{Gough}}, \binits{D.O.}},
\bauthor{\bsnm{{Guenther}}, \binits{D.B.}},
\bauthor{\bsnm{{Guzik}}, \binits{J.A.}},
\bauthor{\bsnm{{Harvey}}, \binits{J.W.}},
\bauthor{\bsnm{{Hill}}, \binits{F.}},
\bauthor{\bsnm{{Houdek}}, \binits{G.}},
\bauthor{\bsnm{{Iglesias}}, \binits{C.A.}},
\bauthor{\bsnm{{Kosovichev}}, \binits{A.G.}},
\bauthor{\bsnm{{Leibacher}}, \binits{J.W.}},
\bauthor{\bsnm{{Morel}}, \binits{P.}},
\bauthor{\bsnm{{Proffitt}}, \binits{C.R.}},
\bauthor{\bsnm{{Provost}}, \binits{J.}},
\bauthor{\bsnm{{Reiter}}, \binits{J.}},
\bauthor{\bsnm{{Rhodes}}, \binits{E.J.} \bsuffix{Jr.}},
\bauthor{\bsnm{{Rogers}}, \binits{F.J.}},
\bauthor{\bsnm{{Roxburgh}}, \binits{I.W.}},
\bauthor{\bsnm{{Thompson}}, \binits{M.J.}},
\bauthor{\bsnm{{Ulrich}}, \binits{R.K.}}:
\byear{1996},
\bjtitle{Science}
\bvolume{272},
\bfpage{1286}.
\adsurl{cgi-bin/nph-bib_query?bibcode=1996Sci...272.1286C&db_key=AST}.
\end{barticle}
\endbibitem

\bibitem[\protect\citeauthoryear{{Couvidat}
  \textit{et~al.}}{2012}]{couvidat2012}
\begin{barticle}
\bauthor{\bsnm{{Couvidat}}, \binits{S.}},
\bauthor{\bsnm{{Zhao}}, \binits{J.}},
\bauthor{\bsnm{{Birch}}, \binits{A.C.}},
\bauthor{\bsnm{{Kosovichev}}, \binits{A.G.}},
\bauthor{\bsnm{{Duvall}}, \binits{T.L.}},
\bauthor{\bsnm{{Parchevsky}}, \binits{K.}},
\bauthor{\bsnm{{Scherrer}}, \binits{P.H.}}:
\byear{2012},
\bjtitle{\solphys}
\bvolume{275},
\bfpage{357}.
\doiurl{10.1007/s11207-010-9652-y}.
\adsurl{2012SoPh..275..357C}.
\end{barticle}
\endbibitem

\bibitem[\protect\citeauthoryear{{DeGrave}, {Jackiewicz}, and
  {Rempel}}{2014}]{degrave2014}
\begin{barticle}
\bauthor{\bsnm{{DeGrave}}, \binits{K.}},
\bauthor{\bsnm{{Jackiewicz}}, \binits{J.}},
\bauthor{\bsnm{{Rempel}}, \binits{M.}}:
\byear{2014},
\bjtitle{\apj}
\bvolume{788},
\bfpage{127}.
\doiurl{10.1088/0004-637X/788/2/127}.
\adsurl{2014ApJ...788..127D}.
\end{barticle}
\endbibitem

\bibitem[\protect\citeauthoryear{{Duvall} and {Hanasoge}}{2013}]{duvall2012}
\begin{barticle}
\bauthor{\bsnm{{Duvall}}, \binits{T.L.}},
\bauthor{\bsnm{{Hanasoge}}, \binits{S.M.}}:
\byear{2013},
\bjtitle{\solphys}
\bvolume{287},
\bfpage{71}.
\doiurl{10.1007/s11207-012-0010-0}.
\adsurl{2013SoPh..287...71D}.
\end{barticle}
\endbibitem

\bibitem[\protect\citeauthoryear{{Duvall}, {Hanasoge}, and
  {Chakraborty}}{2014}]{duvall2014}
\begin{barticle}
\bauthor{\bsnm{{Duvall}}, \binits{T.L.}},
\bauthor{\bsnm{{Hanasoge}}, \binits{S.M.}},
\bauthor{\bsnm{{Chakraborty}}, \binits{S.}}:
\byear{2014},
\bjtitle{\solphys}
\bvolume{289},
\bfpage{3421}.
\doiurl{10.1007/s11207-014-0537-3}.
\adsurl{2014SoPh..289.3421D}.
\end{barticle}
\endbibitem

\bibitem[\protect\citeauthoryear{{Duvall}}{1998}]{duvall1998}
\begin{bchapter}
\bauthor{\bsnm{{Duvall}}, \binits{T.L.} \bsuffix{Jr.}}:
\byear{1998},
In: \beditor{\bsnm{{Korzennik}}, \binits{S.}} (ed.)
\bbtitle{Structure and Dynamics of the Interior of the Sun and Sun-like Stars},
\bsertitle{ESA Special Publication}
\bseriesno{418},
\bfpage{581}.
\adsurl{1998ESASP.418..581D}.
\end{bchapter}
\endbibitem

\bibitem[\protect\citeauthoryear{{Duvall} and {Birch}}{2010}]{duvall2010}
\begin{barticle}
\bauthor{\bsnm{{Duvall}}, \binits{T.L.} \bsuffix{Jr.}},
\bauthor{\bsnm{{Birch}}, \binits{A.C.}}:
\byear{2010},
\bjtitle{\apjl}
\bvolume{725},
\bfpage{L47}.
\doiurl{10.1088/2041-8205/725/1/L47}.
\adsurl{2010ApJ...725L..47D}.
\end{barticle}
\endbibitem

\bibitem[\protect\citeauthoryear{{Duvall} \textit{et~al.}}{1997}]{duvall1997}
\begin{barticle}
\bauthor{\bsnm{{Duvall}}, \binits{T.L.} \bsuffix{Jr.}},
\bauthor{\bsnm{{Kosovichev}}, \binits{A.G.}},
\bauthor{\bsnm{{Scherrer}}, \binits{P.H.}},
\bauthor{\bsnm{{Bogart}}, \binits{R.S.}},
\bauthor{\bsnm{{Bush}}, \binits{R.I.}},
\bauthor{\bsnm{{de Forest}}, \binits{C.}},
\bauthor{\bsnm{{Hoeksema}}, \binits{J.T.}},
\bauthor{\bsnm{{Schou}}, \binits{J.}},
\bauthor{\bsnm{{Saba}}, \binits{J.L.R.}},
\bauthor{\bsnm{{Tarbell}}, \binits{T.D.}},
\bauthor{\bsnm{{Title}}, \binits{A.M.}},
\bauthor{\bsnm{{Wolfson}}, \binits{C.J.}},
\bauthor{\bsnm{{Milford}}, \binits{P.N.}}:
\byear{1997},
\bjtitle{\solphys}
\bvolume{170},
\bfpage{63}.
\adsurl{cgi-bin/nph-bib_query?bibcode=1997SoPh..170...63D&db_key=AST}.
\end{barticle}
\endbibitem

\bibitem[\protect\citeauthoryear{{Gizon} and {Birch}}{2002}]{gb02}
\begin{barticle}
\bauthor{\bsnm{{Gizon}}, \binits{L.}},
\bauthor{\bsnm{{Birch}}, \binits{A.C.}}:
\byear{2002},
\bjtitle{\apj}
\bvolume{571},
\bfpage{966}.
\doiurl{10.1086/340015}.
\adsurl{2002ApJ...571..966G}.
\end{barticle}
\endbibitem

\bibitem[\protect\citeauthoryear{{Hanasoge}}{2014}]{hanasoge2014}
\begin{barticle}
\bauthor{\bsnm{{Hanasoge}}, \binits{S.M.}}:
\byear{2014},
\bjtitle{\apj}
\bvolume{797},
\bfpage{23}.
\doiurl{10.1088/0004-637X/797/1/23}.
\adsurl{2014ApJ...797...23H}.
\end{barticle}
\endbibitem

\bibitem[\protect\citeauthoryear{{Hart}}{1954}]{hart1954}
\begin{barticle}
\bauthor{\bsnm{{Hart}}, \binits{A.B.}}:
\byear{1954},
\bjtitle{\mnras}
\bvolume{114},
\bfpage{17}.
\adsurl{1954MNRAS.114...17H}.
\end{barticle}
\endbibitem

\bibitem[\protect\citeauthoryear{{Hart}}{1956}]{hart1956}
\begin{barticle}
\bauthor{\bsnm{{Hart}}, \binits{A.B.}}:
\byear{1956},
\bjtitle{\mnras}
\bvolume{116},
\bfpage{38}.
\adsurl{1956MNRAS.116...38H}.
\end{barticle}
\endbibitem

\bibitem[\protect\citeauthoryear{{Hathaway}
  \textit{et~al.}}{2000}]{hathaway2000}
\begin{barticle}
\bauthor{\bsnm{{Hathaway}}, \binits{D.H.}},
\bauthor{\bsnm{{Beck}}, \binits{J.G.}},
\bauthor{\bsnm{{Bogart}}, \binits{R.S.}},
\bauthor{\bsnm{{Bachmann}}, \binits{K.T.}},
\bauthor{\bsnm{{Khatri}}, \binits{G.}},
\bauthor{\bsnm{{Petitto}}, \binits{J.M.}},
\bauthor{\bsnm{{Han}}, \binits{S.}},
\bauthor{\bsnm{{Raymond}}, \binits{J.}}:
\byear{2000},
\bjtitle{\solphys}
\bvolume{193},
\bfpage{299}.
\doiurl{10.1023/A:1005200809766}.
\adsurl{2000SoPh..193..299H}.
\end{barticle}
\endbibitem

\bibitem[\protect\citeauthoryear{{Hathaway}
  \textit{et~al.}}{2002}]{hathaway2002}
\begin{barticle}
\bauthor{\bsnm{{Hathaway}}, \binits{D.H.}},
\bauthor{\bsnm{{Beck}}, \binits{J.G.}},
\bauthor{\bsnm{{Han}}, \binits{S.}},
\bauthor{\bsnm{{Raymond}}, \binits{J.}}:
\byear{2002},
\bjtitle{\solphys}
\bvolume{205},
\bfpage{25}.
\adsurl{2002SoPh..205...25H}.
\end{barticle}
\endbibitem

\bibitem[\protect\citeauthoryear{{Hirzberger}
  \textit{et~al.}}{2008}]{hirzberger2008}
\begin{barticle}
\bauthor{\bsnm{{Hirzberger}}, \binits{J.}},
\bauthor{\bsnm{{Gizon}}, \binits{L.}},
\bauthor{\bsnm{{Solanki}}, \binits{S.K.}},
\bauthor{\bsnm{{Duvall}}, \binits{T.L.}}:
\byear{2008},
\bjtitle{\solphys}
\bvolume{251},
\bfpage{417}.
\doiurl{10.1007/s11207-008-9206-8}.
\adsurl{2008SoPh..251..417H}.
\end{barticle}
\endbibitem

\bibitem[\protect\citeauthoryear{{Jackiewicz}, {Gizon}, and
  {Birch}}{2008}]{jackiewicz2008}
\begin{barticle}
\bauthor{\bsnm{{Jackiewicz}}, \binits{J.}},
\bauthor{\bsnm{{Gizon}}, \binits{L.}},
\bauthor{\bsnm{{Birch}}, \binits{A.C.}}:
\byear{2008},
\bjtitle{\solphys}
\bvolume{251},
\bfpage{381}.
\doiurl{10.1007/s11207-008-9158-z}.
\adsurl{2008SoPh..251..381J}.
\end{barticle}
\endbibitem

\bibitem[\protect\citeauthoryear{{Kosovichev} and
  {Duvall}}{1997}]{kosovichev1997}
\begin{bchapter}
\bauthor{\bsnm{{Kosovichev}}, \binits{A.G.}},
\bauthor{\bsnm{{Duvall}}, \binits{T.L.} \bsuffix{Jr.}}:
\byear{1997},
In: \beditor{\bsnm{{Pijpers}}, \binits{F.P.}},
\beditor{\bsnm{{Christensen-Dalsgaard}}, \binits{J.}},
\beditor{\bsnm{{Rosenthal}}, \binits{C.S.}} (eds.)
\bbtitle{SCORe'96 : Solar Convection and Oscillations and their Relationship},
\bsertitle{Astrophysics and Space Science Library}
\bseriesno{225},
\bfpage{241}.
\adsurl{1997ASSL..225..241K}.
\end{bchapter}
\endbibitem

\bibitem[\protect\citeauthoryear{{Rempel}}{2014}]{rempel2014}
\begin{barticle}
\bauthor{\bsnm{{Rempel}}, \binits{M.}}:
\byear{2014},
\bjtitle{\apj}
\bvolume{789},
\bfpage{132}.
\doiurl{10.1088/0004-637X/789/2/132}.
\adsurl{2014ApJ...789..132R}.
\end{barticle}
\endbibitem

\bibitem[\protect\citeauthoryear{{Rieutord}
  \textit{et~al.}}{2008}]{rieutord2008}
\begin{barticle}
\bauthor{\bsnm{{Rieutord}}, \binits{M.}},
\bauthor{\bsnm{{Meunier}}, \binits{N.}},
\bauthor{\bsnm{{Roudier}}, \binits{T.}},
\bauthor{\bsnm{{Rondi}}, \binits{S.}},
\bauthor{\bsnm{{Beigbeder}}, \binits{F.}},
\bauthor{\bsnm{{Par{\`e}s}}, \binits{L.}}:
\byear{2008},
\bjtitle{\aap}
\bvolume{479},
\bfpage{L17}.
\doiurl{10.1051/0004-6361:20079077}.
\adsurl{2008A\%26A...479L..17R}.
\end{barticle}
\endbibitem

\bibitem[\protect\citeauthoryear{{Roudier} \textit{et~al.}}{2014}]{roudier2014}
\begin{barticle}
\bauthor{\bsnm{{Roudier}}, \binits{T.}},
\bauthor{\bsnm{{{\v S}vanda}}, \binits{M.}},
\bauthor{\bsnm{{Rieutord}}, \binits{M.}},
\bauthor{\bsnm{{Malherbe}}, \binits{J.M.}},
\bauthor{\bsnm{{Burston}}, \binits{R.}},
\bauthor{\bsnm{{Gizon}}, \binits{L.}}:
\byear{2014},
\bjtitle{\aap}
\bvolume{567},
\bfpage{A138}.
\doiurl{10.1051/0004-6361/201423577}.
\adsurl{2014A\%26A...567A.138R}.
\end{barticle}
\endbibitem

\bibitem[\protect\citeauthoryear{{Simon} and {Leighton}}{1964}]{simon1964}
\begin{barticle}
\bauthor{\bsnm{{Simon}}, \binits{G.W.}},
\bauthor{\bsnm{{Leighton}}, \binits{R.B.}}:
\byear{1964},
\bjtitle{\apj}
\bvolume{140},
\bfpage{1120}.
\doiurl{10.1086/148010}.
\adsurl{1964ApJ...140.1120S}.
\end{barticle}
\endbibitem

\bibitem[\protect\citeauthoryear{{{\v S}vanda}}{2012}]{svanda2012}
\begin{barticle}
\bauthor{\bsnm{{{\v S}vanda}}, \binits{M.}}:
\byear{2012},
\bjtitle{\apjl}
\bvolume{759},
\bfpage{L29}.
\doiurl{10.1088/2041-8205/759/2/L29}.
\adsurl{2012ApJ...759L..29S}.
\end{barticle}
\endbibitem

\bibitem[\protect\citeauthoryear{{{\v S}vanda}}{2015}]{svanda2015}
\begin{barticle}
\bauthor{\bsnm{{{\v S}vanda}}, \binits{M.}}:
\byear{2015},
\bjtitle{\aap}
\bvolume{575},
\bfpage{A122}.
\doiurl{10.1051/0004-6361/201425203}.
\adsurl{2015A\%26A...575A.122S}.
\end{barticle}
\endbibitem

\bibitem[\protect\citeauthoryear{{Williams}
  \textit{et~al.}}{2014}]{williams2014}
\begin{barticle}
\bauthor{\bsnm{{Williams}}, \binits{P.E.}},
\bauthor{\bsnm{{Pesnell}}, \binits{W.D.}},
\bauthor{\bsnm{{Beck}}, \binits{J.G.}},
\bauthor{\bsnm{{Lee}}, \binits{S.}}:
\byear{2014},
\bjtitle{\solphys}
\bvolume{289},
\bfpage{11}.
\doiurl{10.1007/s11207-013-0330-8}.
\adsurl{2014SoPh..289...11W}.
\end{barticle}
\endbibitem

\bibitem[\protect\citeauthoryear{{Woodard}}{2007}]{woodard2007}
\begin{barticle}
\bauthor{\bsnm{{Woodard}}, \binits{M.F.}}:
\byear{2007},
\bjtitle{\apj}
\bvolume{668},
\bfpage{1189}.
\doiurl{10.1086/521391}.
\adsurl{2007ApJ...668.1189W}.
\end{barticle}
\endbibitem

\bibitem[\protect\citeauthoryear{{Worden} and {Simon}}{1976}]{worden1976}
\begin{barticle}
\bauthor{\bsnm{{Worden}}, \binits{S.P.}},
\bauthor{\bsnm{{Simon}}, \binits{G.W.}}:
\byear{1976},
\bjtitle{\solphys}
\bvolume{46},
\bfpage{73}.
\doiurl{10.1007/BF00157555}.
\adsurl{1976SoPh...46...73W}.
\end{barticle}
\endbibitem

\bibitem[\protect\citeauthoryear{{Zhao} and {Kosovichev}}{2003}]{zhao2003}
\begin{bchapter}
\bauthor{\bsnm{{Zhao}}, \binits{J.}},
\bauthor{\bsnm{{Kosovichev}}, \binits{A.G.}}:
\byear{2003},
In: \beditor{\bsnm{{Sawaya-Lacoste}}, \binits{H.}} (ed.)
\bbtitle{GONG+ 2002. Local and Global Helioseismology: the Present and Future},
\bsertitle{ESA Special Publication}
\bseriesno{517},
\bfpage{417}.
\adsurl{2003ESASP.517..417Z}.
\end{bchapter}
\endbibitem

\bibitem[\protect\citeauthoryear{{Zhao} \textit{et~al.}}{2007}]{zhao2007}
\begin{barticle}
\bauthor{\bsnm{{Zhao}}, \binits{J.}},
\bauthor{\bsnm{{Georgobiani}}, \binits{D.}},
\bauthor{\bsnm{{Kosovichev}}, \binits{A.G.}},
\bauthor{\bsnm{{Benson}}, \binits{D.}},
\bauthor{\bsnm{{Stein}}, \binits{R.F.}},
\bauthor{\bsnm{{Nordlund}}, \binits{{\AA}.}}:
\byear{2007},
\bjtitle{\apj}
\bvolume{659},
\bfpage{848}.
\doiurl{10.1086/512009}.
\adsurl{2007ApJ...659..848Z}.
\end{barticle}
\endbibitem

\bibitem[\protect\citeauthoryear{{Zhao} \textit{et~al.}}{2012}]{zhao2012}
\begin{barticle}
\bauthor{\bsnm{{Zhao}}, \binits{J.}},
\bauthor{\bsnm{{Couvidat}}, \binits{S.}},
\bauthor{\bsnm{{Bogart}}, \binits{R.S.}},
\bauthor{\bsnm{{Parchevsky}}, \binits{K.V.}},
\bauthor{\bsnm{{Birch}}, \binits{A.C.}},
\bauthor{\bsnm{{Duvall}}, \binits{T.L.}},
\bauthor{\bsnm{{Beck}}, \binits{J.G.}},
\bauthor{\bsnm{{Kosovichev}}, \binits{A.G.}},
\bauthor{\bsnm{{Scherrer}}, \binits{P.H.}}:
\byear{2012},
\bjtitle{\solphys}
\bvolume{275},
\bfpage{375}.
\doiurl{10.1007/s11207-011-9757-y}.
\adsurl{2012SoPh..275..375Z}.
\end{barticle}
\endbibitem

\end{thebibliography}

\end{article} 

\end{document}